\documentclass[aps,prd,notitlepage]{revtex4-1}

\usepackage{pstricks}
\usepackage{pst-coil}
\usepackage{amsmath,amssymb,graphicx,graphics,color}
\usepackage{natbib,hyperref}
\usepackage{stackengine}
\usepackage{url}

\usepackage{etoolbox}
\usepackage{stackrel}

\usepackage{lipsum}

\makeatletter
\def\@mkboth#1#2{}
\newlength\appendixwidth
\preto\appendix{\addtocontents{toc}{\protect\patchl@section}}
\newcommand{\patchl@section}{%
  \settowidth{\appendixwidth}{\textbf{Appendix }}%
  \addtolength{\appendixwidth}{1.5em}%
  \patchcmd{\l@section}{1.5em}{\appendixwidth}{}{\ddt}%
}
\makeatother

\newtheorem{comments}{Comment}
\newtheorem{notation}{Notation}
\newtheorem{theorem}{Theorem}
\newtheorem{proposition}{Proposition}
\newtheorem{lemma}{Lemma}
\newtheorem{corollary}{Corollary}
\newtheorem{definition}{Definition}

\newcommand{\be}{\begin{equation}}
\newcommand{\ee}{\end{equation}}
\newcommand{\bi}{\begin{itemize}}
\newcommand{\ei}{\end{itemize}}
\newcommand{\bes}{\begin{eqnarray*}}
\newcommand{\ees}{\end{eqnarray*}}
\newcommand{\beq}{\begin{eqnarray}}
\newcommand{\eeq}{\end{eqnarray}}

\newcommand{\pd}[2]{\frac{\partial #1}{\partial #2}}

\newcommand{\co}{{\cal{O}}}

\newcommand{\cm}{{\mathcal{M}}}
\newcommand{\sig}{\sigma}

\newcommand{\hor}{{\mathcal{H}}}

\newcommand{\qed}{{\hfill$\blacksquare$}\vskip10pt}

\newcommand{\startproof}{\noindent\textbf{Proof: }}

\newcommand{\godel}{{G\"{o}del}}
\newcommand{\al}{\alpha}

\newcommand{\ax}{\mathcal{A}}
\newcommand{\ch}{\mathcal{H}}
\newcommand{\floor}[1]{\left\lfloor #1 \right\rfloor}
\newcommand{\backc}{$\supset\!-$shaped }
\newcommand{\env}{{\mathcal{E}}}
\newcommand{\fp}{{\mathcal{F}}}

\newcommand{\ce}{{\mathcal{E}}}

\newcommand{\ese}[1]{\env_{{#1},\rm{SE}}}
\newcommand{\se}{{\rm{SE}}}
\newcommand{\ueone}{u_{\env_1}}

\begin{document}

\title[Causality violating signals in  G\"{o}del's universe]{Causality violation without time-travel: closed lightlike paths in  G\"{o}del's universe }
\author{Brien C. Nolan}
\address{Centre for Astrophysics and Relativity, School of Mathematical Sciences, Dublin City University, Glasnevin, Dublin 9, Ireland.}
\email{brien.nolan@dcu.ie}

\begin{abstract}
We revisit the issue of causality violations in {\godel}'s universe, restricting to geodesic motions. It is well-known that while there are closed timelike curves in this spacetime, there are no closed causal geodesics. We show further that no observer can communicate directly (i.e.\ using a single causal geodesic) with their own past. However, we show that this type of causality violation can be achieved by a system of relays: we prove that from any event $P$ in {\godel}'s universe, there is a future-directed \textit{lightlike path} - a sequence of future-directed null geodesic segments, laid end to end - which has $P$ as its past and future endpoints. By analysing the envelope of the family of future directed null geodesics emanating from a point of the spacetime, we show that this lightlike path must contain a minimum of eight geodesic segments, and show further that this bound is attained. We prove a related general result, that events of a time orientable spacetime are connected by a (closed) timelike curve if and only if they are connected by a (closed) lightlike path. This suggests a means of violating causality in {\godel}'s universe without the need for unfeasibly large accelerations, using instead a sequence of light signals reflected by a suitably located system of mirrors. 
 \end{abstract}

\maketitle

\section{Introduction: G\"odel's Universe}

In 1949, Kurt \godel \cite{godel1949example} published a solution of Einstein's equations which provides what appears to be the first example of a spacetime containing closed timelike curves (CTCs). Van Stockum's solution \cite{van1937gravitational}, published in 1937, also contains CTCs, but their presence was not recognised until the 1960's: see \cite{maitra1966stationary} and \cite{tipler1974rotating}. As such, \godel's solution has played a significant role in the study of causality violations and the associated concept of time-travel in relativity theory, and in theoretical speculations more generally (see \cite{gleick2017time} for a recent account of the concept of time-travel and its history). 


The question we address here is whether or not there are causality violations in \godel's universe that rely only on geodesic motion. This is motivated by the fact that the CTC's found in \cite{godel1949example} are accelerated world-lines. The magnitude of the acceleration of these world lines is constant and  has the same order of magnitude as the local energy density of the spacetime (considering the case of a dust-filled universe). Furthermore, an observer travelling on such a CTC must move at a speed at least $c/\sqrt{2}$ relative to the matter in the universe, and must have access to vast quantities of fuel (see footnote 11 of \cite{godel1949remark}: the numbers provided by \godel\ indicate that a rocket with a mass of 1kg would require fuel with a mass approximately that of the moon). This mass is essentially the exponential of the total integrated acceleration $TA$ along the CTC. Malament \cite{malament1985minimal} proved that $TA\geq \ln(2+\sqrt{5})\simeq1.4436$ along any CTC in \godel's universe, and has conjectured that  $TA\geq2\pi(9+6\sqrt{3})^{1/2}\simeq 27.6691$ along any CTC in the spacetime \cite{malament1987note}. Manchak \cite{manchak2011efficient} subsequently showed that Malament's conjectured lower bound can be violated. Natario \cite{natario2012optimal} presented a candidate for the optimal CTC - i.e.\ that with the least $TA$: this has $TA\simeq 24.9947$. Natario also provides a strong case for the optimal nature of this CTC, and argued that Malament's bound does indeed hold for periodic CTCs, where the tangent vectors to the closed curve at the (coincident) initial and terminal points of the closed curve are equal. So in all cases, time travel in \godel's universe requires unfeasible accelerations and vast quantities of fuel. This is also the case if the CTC is replaced by a sequence of timelike geodesics (see Definition \ref{def:trip} and Proposition \ref{prop:Penrose-2.23} below, which come from \cite{penrose1972techniques}; see also Comment \ref{comm:geo-segs} below). So we ask: is there an alternative means of violating causality in \godel's universe?

We consider this question from a few different perspectives. From the simplest perspective, we can ask if there are closed causal geodesics in \godel's universe. It has been known for some time that there are not \cite{kundt1956tr}. For clarity and completeness, we provide a proof of this result: see Proposition \ref{prop:no-ccg} below. So we consider the possibility of \textit{communicating} with one's own past: this is the essence of causality violation, as it creates the same opportunities (and paradoxes) as \textit{travelling} to one's own past. From the spacetime perspective, the key question is this: can we find a future-pointing, timelike geodesic $\gamma$, parametrised by proper time $s$ (which increases into the future) - an observer - with the property that there exists a future-pointing causal geodesic $\mu$ that extends from an event $P\in\gamma$ to an event $Q\in\gamma$ for which $s|_P>s|_Q$?  ($P$ is the older version of the observer, who communicates with their younger self $Q$ via the \textit{future-pointing} causal geodesic $\mu$.) The answer to this question is no: no such pair of causal geodesics exists in \godel's universe (Proposition \ref{prop:no-direct-comm}). 

Keeping our focus on the notion of communicating with the past we ask: can we find a timelike geodesic $\gamma$ (with future-increasing proper time $s$) and a finite sequence of events $P_1,\dots,P_n$ of \godel's universe with the property that there is a future-pointing null geodesic from $P_i$ to $P_{i+1}, 1\leq i\leq n-1$, and such that $P=P_1$ and $Q=P_n$ lie on $\gamma$ with $s|_P>s|_Q$? The interpretation here is that the observer at the event $P$ can commmunicate with their past self $Q$ via a system of geodesic relays. (Adapting the terminology of \cite{penrose1972techniques}, we refer to this sequence of null geodesic segments as a \textbf{lightlike path} from $P$ to $Q$.) Our main results comprise an affirmative answer to this question. We prove the following results:

First, we present a general result, applicable to any time orientable spacetime, that relates the existence of lightlike paths connecting two events to the existence of a timelike curve connecting those events (no causality violation need be implied). Then we show that the minimum number of future pointing null geodesic segments required to construct a closed lightlike path in \godel's spacetime is $N=8$, and that this bound is attained. (An 8-segment lightlike path can also be used to send a signal to an observer's own past.) 

The absence of the causality violation ruled out in Proposition 2 as described above is essentially contained in the results of Novello \textit{et al.} \cite{novello1983geodesic}, where the authors study geodesics of \godel's universe using an effective potential approach. This paper also highlights a key property of the spacetime: the existence of a coordinate $\tau$ which is a time coordinate within a certain distance of a given observer, but which becomes spacelike beyond this distance - beyond the so-called the \godel\ horizon. Outside the horizon, the coordinate may have \textit{decreasing} values along future-pointing causal geodesics. The existence of such a coordinate is implicit in \cite{chandrasekhar1961geodesics}, and first appears to have been mentioned explicitly in \cite{pfarr1981time}, where it is noted that ``this running backwards of [the coordinate $\tau$] has nothing to do with a possible going backward in time or time travel" (\cite{pfarr1981time}, p.\ 1078). Likewise, Novello \textit{et al.} \cite{novello1983geodesic} and Grave \textit{et al.} \cite{grave2009godel} study future-pointing geodesics along which $\tau$ may decrease, but both conclude that no violation of causality occurs on the basis of this phenomenon. In fact we can show that this feature of \godel's universe may be exploited to generate the causality violations described above.

In the following section, we review the metric, the isometries and the geodesics of {\godel}'s universe. We revisit the result that there are no closed causal geodesics in \godel's spacetime, and we show further that no observer in the spacetime can send a signal directly (i.e.\ via a single causal geodesic) to their own past. In Section III, we present the general result relating lightlike paths to causal curves in a time orientable spacetime (Theorem 1). The long Section IV contains our main result (Theorem 2 below) on optimal closed lightlike paths in {\godel}'s universe - i.e.\ on closed lightlike paths that contain the least number of segments. Some basic properties of null geodesic segments in \godel's universe are described in Section IV-A, and we derive the form of the first segment of an optimal path. In Section IV-B, we give a sequence of results that narrows down the possibilities for those segments (of future pointing null geodesics) that together form the optimal path. The results show that the search for segments of the optimal path may be reduced from a three-parameter set (plus a choice of a sign) to a one parameter set (and no choice of sign remaining). We identify the central role of the envelope of a certain family of null geodesic segments emanating from a fixed point of the spacetime. In Section IV-C, we complete the proof of Theorem 2. We use the conventions of \cite{wald1984general}, and set $G=c=1$. Throughout the paper, a \textbf{curve} is a $C^1$ mapping from an interval (of non-zero measure) to the spacetime with nowhere vanishing derivative. For ease of reading, where a proof does not introduce a concept or quantity required later in the paper, it is given in the appendix.  We use the symbol $\blacksquare$ to indicate the end of a proof (or the statement of a result for which the proof is immediate or implicit in the preceding text).

\section{Metric, isometries and geodesics}


In cylindrical coordinates $x^{}=(\tau,r,\phi,\zeta)\in M=\mathbb{R}\times[0,+\infty)\times(-\pi,\pi]\times\mathbb{R}$, the line element of {\godel}'s universe reads (\cite{godel1949example}, \cite{grave2009godel})
\be ds^2=-d\tau^2-2\sqrt{2}\al r^2d\phi d\tau + r^2(1-\al^2 r^2)d\phi^2 + (1+\al^2r^2)^{-1}dr^2+d\zeta^2, \label{metric-cyl1}\ee
where $2\pi-$periodic identification of the coordinate $\phi$ applies. (\godel\ uses the coordinate $\rho$ where $ \al r=\sinh\al\rho$.) The parameter $\al$ plays effectively no role in the geometry of the spacetime other than setting the scale of the density and pressure, which are given by (respectively)
\begin{eqnarray} 8\pi\rho &=& 2\al^2-\Lambda,\label{density}\\
8\pi P &=& 2\al^2+\Lambda,\label{pressure}
\end{eqnarray}
and are therefore constant ($\Lambda$ is the cosmological constant). In these coordinates, the fluid flow vector is $\vec{u}=\pd{}{\tau}$. The parameter $\alpha$ can be absorbed into the coordinates $\tau,r,\zeta$ (i.e.\ by defining $T=\al\tau$, and then renaming $T$ as $\tau$, and similar for $r$ and $\zeta$). Then the line element satisfies 
\be \al^2ds^2=-d\tau^2-2\sqrt{2}r^2d\phi d\tau + r^2(1- r^2)d\phi^2 + (1+r^2)^{-1}dr^2+d\zeta^2. \label{metric-cyl}\ee 
We will work on the conformal spacetime with line element $\al^2ds^2$. Since the conformal factor is a constant, all results relating to geodesics and global structure carry over to the physical spacetime with line element $ds^2$. 

\begin{figure}
\centering
\includegraphics[width=0.9\textwidth]{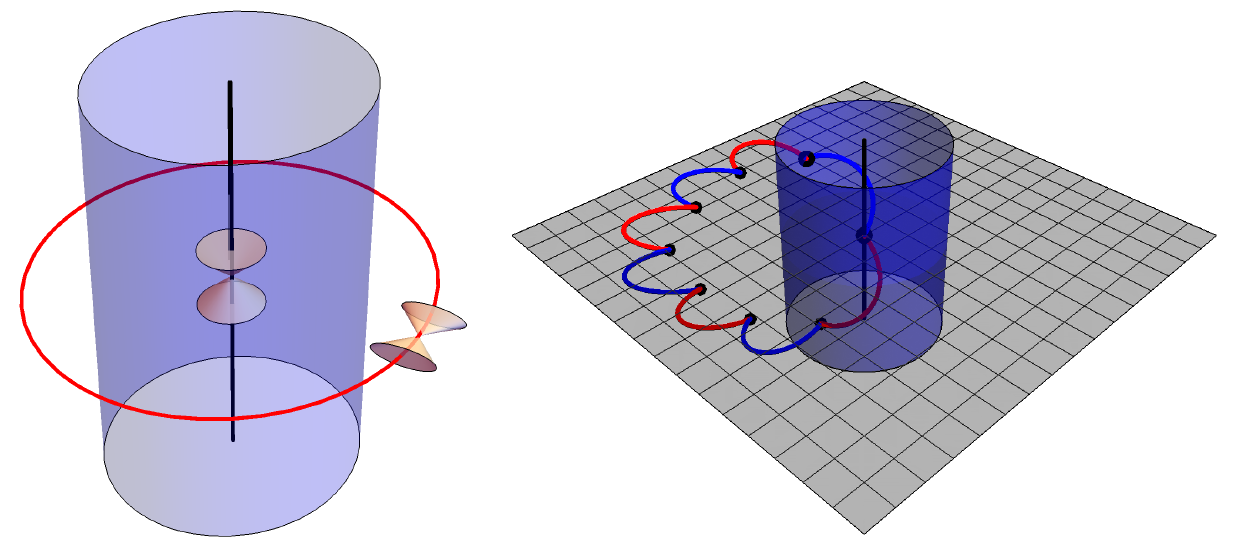}
\caption{Representation of a closed timelike curve (left) and a closed lightlike path (right) in \godel's spacetime. The figures show spacetime diagrams in the coordinates of (\ref{metric-cyl}), with the $z$ coordinate suppressed and with the coordinate $\tau$ increasing along the vertical axis. The axis $\{r=0\}$ is shown bold, and the \godel\ horizon is shaded (blue). The image on the left shows that outside the horizon, the light cones tip over, allowing for the existence of closed timelike curves. In the image on the right, the closed lightlike path has initial point $\co$ on the axis with $\tau(\co)=0$. The first segment terminates on the horizon, and the fourth segment terminate on $\Sigma_0=\{\tau=0\}$, which is shown in grey. The fifth to eighth segments sit below $\Sigma_0$, with the eighth and last segment terminates at $\co$. These segments are numerical plots of future pointing null geodesics as identified in the proof of Theorem \ref{thm-main}: see Figure \ref{fig:construction} and Section \ref{subsect:optimal-path} below.}
\label{fig:CTC}
\end{figure}

In these coordinates, (some) closed timelike curves are relatively easy to identify. Consider the 3-parameter family of curves with $(\tau,r,\phi,\zeta)=(\tau_0,r_0,\Phi(s),\zeta_0)$ where $\tau_0,r_0$ and $\zeta_0$ are constant and where $r_0>1$. Then (\ref{metric-cyl}) shows that these curves are timelike, and the periodicity of the coordinate $\phi$ shows that they are closed (see Figure \ref{fig:CTC}). Taking the parameter $s$ to be proper time gives
\be \Phi'(s) = \pm r_0^{-1}(r_0^2-1)^{-1/2}.\label{vel-ctc}\ee The 4-acceleration is 
\be \vec{a} =r_0^{-1}(1-r_0^2)^{-1}(1+r_0^2)(1-2r_0^2)\pd{}{r},\label{4-acc-ctc}\ee
with norm-squared
\be g(\vec{a},\vec{a}) =  q(\beta), \label{acc-norm}\ee
where $\beta = r_0^2>1$ and $q(x) = (1+x)(1-2x)^2/(x(1-x)^2)$. The function $q$ is monotone decreasing on $(1,+\infty)$ with $q(x)\to+\infty$ as $x\to 1^+$ and $q(x)\to4$ as $x\to+\infty$. Thus there is a minimum value of the acceleration on these CTCs: $g(\vec{a},\vec{a})>4$. We can also calculate the 3-velocity of a observer travelling on the CTC relative to the matter at rest on the fluid flow lines. The norm $v$ of this velocity - the speed - satisfies 
\be v^2 = \frac{1+r_0^2}{2r_0^2} \geq \frac12. \label{speed} \ee
This is the origin of \godel's statement that an observer on these CTCs must travel at a minimum speed of $c/\sqrt{2}$.

The geodesics of \godel's spacetime were first solved by Kundt \cite{kundt1956tr}, and have been considered on numerous occasions since then: see e.g.\ \cite{chandrasekhar1961geodesics,stein1970paradoxical,pfarr1981time,novello1983geodesic,chicone2006explicit, franchi2009relativistic,grave2009godel,buser2013visualization,bini2019godel}. The analysis of the geodesics is greatly facilitated by the high degree of symmetry present. The full set of five linearly independent Killing vector fields of \godel\ spacetime is relatively straightforward to calculate. To investigate geodesics of the spacetime in these coordinates, we make use only of the Killing vector fields
\be \vec{\eta}_1=-\pd{}{\tau},\quad \vec{\eta}_2=-\pd{}{\phi},\quad \vec{\eta}_3=\pd{}{\zeta}. \label{kvfs-cyl} \ee
The Killing vector field $\vec{\eta}_2$ is the generator of an axial symmetry of the spacetime, with axis given by the set $\ax:=\{r=0\}$. The axis is a world-line of the fluid, and so by homogeneity, the spacetime is axially symmetric about any fluid flow world-line.

For each geodesic with tangent vector $\vec{v}$, we have the constants of motion $L_i = g(\vec{v},\vec{\eta}_i), i=1,2,3$. Then the geodesic equations may be reduced to 
\begin{eqnarray} 
\dot{\tau} &=& (1+r^2)^{-1}\left((1-r^2)L_1+\sqrt{2} L_2\right), \label{tau-dot} \\
\dot{\phi} &=&  (1+r^2)^{-1}\left( \sqrt{2} L_1 - r^{-2}L_2\right), \label{phi-dot} \\
\dot{r}^2 &=& - \frac{L_2^2}{r^2} + 2\sqrt{2} L_1L_2+(1-r^2)L_1^2 - (1+r^2)(L_3^2-\epsilon),\label{r-dot} \\
\dot{\zeta} &=& L_3. \label{zeta-dot}
\end{eqnarray}
Here, $\epsilon=-1$ for timelike geodesics, $0$ for null geodesics and $+1$ for spacelike geodesics. 

 As \godel\ \cite{godel1949example} pointed out, the existence of the fluid flow vector field means that a continuous choice of past and future may be made throughout the spacetime - i.e.\ the spacetime is time-orientable \cite{penrose1972techniques}. This is crucial in what follows, and so we make a key observation about this issue. In the coordinates of (\ref{metric-cyl}), the fluid flow vector is  $\vec{u}=\pd{}{\tau}$. In fact, this form does not arise uniquely, but only up to a sign. So we adopt the convention that $\tau$ increases into the future along the fluid flow lines. Then $\vec{u}$ as given defines the future half of the light cone at each point, and yields the following useful observation: a causal curve with tangent vector field $\vec{v}$ is future-directed at a point $P$ if and only if $g(\vec{u},\vec{v})|_P < 0$. Since $\vec{u}$ is a Killing field, this translates conveniently into a statement about constants of motion in the case of geodesics \cite{grave2009godel}:

\begin{lemma}\label{lem:fp-causal} Let $\gamma$ be a causal geodesic of \godel's universe with tangent vector $\vec{v}$. Then $L_1\neq 0$, and $\gamma$ is future-pointing if and only if $L_1>0$. \qed
\end{lemma}

The constant of motion $L_2$ also carries useful information about the nature of geodesics. It is immediate from (\ref{r-dot}) that if $L_2\neq 0$ for a geodesic $\gamma$, then that geodesic cannot reach the axis. When $L_2=0$, a further condition on the other constants of motion must be satisfied in order that the geodesic exists. Thus we have:

\begin{lemma}\label{lem:meets-axis} Let $\gamma$ be a geodesic of \godel's universe. Then $\gamma$ meets $\ax$ if and only if $L_2=0$ and $L_1^2-L_3^2+\epsilon\geq0$. \qed
\end{lemma}

We can now exploit homogeneity to prove the absence of closed causal geodesics in \godel's universe, and to prove the impossibility of an observer sending a signal directly to their own past. 

\begin{proposition}\label{prop:no-ccg}
There are no closed causal geodesics in \godel's universe. 
\end{proposition}

\noindent\textbf{Proof:} Let $\gamma$ be a causal geodesic of \godel's universe and let $P$ be any event on $\gamma$. Then by homogeneity, we can assume that $P$ lies on the axis, and so the equations of the geodesic are given by (\ref{tau-dot})-(\ref{zeta-dot}) with $L_2=0$ and $L_1^2-L_3^2+\epsilon\geq 0$. In particular, 
\be \dot{r}^2 = (L_1^2-L_3^2+\epsilon) -  r^2(L_1^2+L_3^2-\epsilon)\geq0, \label{rdot-axis} \ee
so that along the geodesic,
\be  r^2 \leq \frac{L_1^2-L_3^2+\epsilon}{L_1^2+L_3^2-\epsilon}\leq 1. \label{r-bound-acis} \ee  
We note that the denominator in the rational term is strictly positive, and that the upper bound $r=1$ is attained if and only if $L_3=\epsilon=0$ (we will refer to geodesics with $L_3=0$ as \textbf{planar} geodesics: the coordinate $\zeta$ is constant along these geodesics). Then (\ref{tau-dot}) with $L_2=0$ shows that $\tau$ is either non-decreasing or non-increasing along the geodesic, and is strictly increasing or strictly decreasing except at isolated points $r=1$ of the geodesic (recall that $L_1\neq 0$ by Lemma 1). This proves that $\gamma$ cannot be closed, as we cannot have $\tau(s_2)=\tau(s_1)$ for different values $s_1, s_2$ of the parameter $s$ on the geodesic. 
\qed

This well-known result is implicit in the work of Kundt \cite{kundt1956tr}. Working in a quasi-rectangular coordinate system $\{w,x,y,z\}$ (with $z=\zeta$), Kundt derives a `spatially bound' feature of the geodesics: the projection of the geodesics into the $x-y$ plane marks out a closed curved. He then calculates the elapse of the the coordinate $w$ along a complete circuit of this closed curve (see equation (12) of \cite{kundt1956tr}), and obtains a positive result (equation (15) of \cite{kundt1956tr}). This is sufficient to demonstrate the absence of closed causal geodesics in the spacetime. The geodesics of this spacetime were also considered in a paper of Chandrasekhar and Wright \cite{chandrasekhar1961geodesics}. 
The authors show that the particular closed timelike curves identified by \godel\ in \cite{godel1949example} are not geodesics - and (erroneously) state that their own conclusions on geodesic motion are ``contrary to some statements of \godel" (\cite{chandrasekhar1961geodesics}, p. 347). This discrepancy appears to have been noticed first by Stein \cite{stein1970paradoxical}. 

Without any further analysis of the solutions of the geodesic equations, we can state the following result that further limits the possibility of violating causality in {\godel}'s universe using geodesic motions. This result shows that an observer cannot send a signal directly to their own past.  

\begin{proposition}\label{prop:no-direct-comm}
Let $\gamma$ be a future pointing causal geodesic of \godel's universe, and let $s$ be a parameter along the geodesic that increases into the future. Then there cannot exist a future pointing causal geodesic $\mu$, with future-increasing parameter $u$, with the property that $\mu(u_1)=\gamma(s_2)$ and $\mu(u_2)=\gamma(s_1)$ where $s_1<s_2$ and $u_1<u_2$. 
\end{proposition}

\startproof Let $\gamma,\mu,s$ and $u$ be as in the statement, and let $P=\mu(u_1)=\gamma(s_2)$. By homogeneity, we can assume that $P$ lies on the axis $\ax$. Then both $\gamma$ and $\mu$ are future pointing causal geodesics that pass through the origin, and so are both described by   (\ref{tau-dot})-(\ref{zeta-dot}) with $L_2=0$ but with different values of the constants $L_1, L_3$. As both are future pointing, we have $L_1|_\gamma>0$ and $L_1|_\mu>0$, and $\tau$ is non-decreasing, and increasing almost everywhere, along both geodesics. Therefore no point $Q$ that lies to the future of $P$ on $\mu$ can lie to the past of $P$ on $\gamma$, proving the proposition. \qed  

The causal geodesics emanating from any given event $P$ of \godel's universe are \textit{confined} \cite{novello1983geodesic} in the following sense. By homogeneity, $P$ is a point on the axis $\ax$, and the geodesics are subject to the bound $r\leq 1 $, which we refer to as the \textit{\godel\ radius}. Planar null geodesics attain this bound, and these generate an envelope containing all future-pointing causal geodesics emanating from $P$. In this sense, the hypersurface $r=1$ forms a horizon relative to the axis $\ax$: events of the spacetime can communicate directly (via a causal geodesics) with events on $\ax$ only if they lie within the interior region. For this reason, the hypersurface $r=1$ is referred to as the \textit{\godel\ horizon} $\ch$ relative to the axis $\ax$. This confinement property is the spatial boundedness derived by Kundt \cite{kundt1956tr} (as mentioned above), and is elaborated explicitly in \cite{novello1983geodesic}. The fact that $\tau$ may not decrease along these geodesics is readily understood as a metric property of this coordinate: we find 
\be g^{-1}(d\tau,d\tau) = -\frac{1-r^2}{1+r^2},\label{tau-temp}\ee
and so the surfaces $\tau=$constant are spacelike in the \textit{interior region}, $r< 1$. We note that the monotone nature of $\tau$ in the interior region is flagged in both \cite{novello1983geodesic} and \cite{grave2009godel}, and the preservation of causality along geodesics in this region is stated explicitly. Our Proposition 2 attempts to clarify a key aspect of this preservation. As (\ref{tau-temp}) shows, $\tau$ fails to be a time coordinate beyond the \godel\ radius - i.e.\ in the \textit{exterior region}, $r>1$. In fact Novello et al. show that ``the time coordinate" $\tau$ may decrease along future-pointing causal geodesics that extend into the exterior region - but they conclude that this ``does not represent a direct violation of causality with geodesics" (\cite{novello1983geodesic}, pp. 786-787). (A necessary and sufficient condition for these geodesics to extend into the exterior region is that $L_2\neq0$.) This feature of the geodesics of \godel's universe is further studied in \cite{grave2009godel}: these authors also conclude that ``[in] all cases, causality is not violated". We now proceed to show that by exploiting this feature of the geodesics identified in \cite{novello1983geodesic}, we can indeed violate causality in \godel's universe without the need for the extravagant speeds associated with the CTCs described above - i.e.\ using only geodesic motions. Before giving the detailed results on this, we consider some general causality issues in time orientable spacetimes.    

\section{Timelike curves and lightlike paths}

In this section, we present a general result (Theorem \ref{thm-lightlike-trips}) that applies to \godel's spacetime to show that the closed timelike curve may be replaced by a causality violating chain of null geodesic segments - a lightlike path. The result applies generally to timelike curves, and not just closed timelike curves. This brief section is mostly technical, with just the statement of Theorem \ref{thm-lightlike-trips} and Corollary \ref{cor:clt-exist} (and associated definitions) being required in the remainder of the paper. We need to recall certain results of \cite{penrose1972techniques}. We begin the discussion with this definition:

\begin{definition} Let $(M,g)$ be a time orientable spacetime and let $A,B \in M$. A \textbf{lightlike path} from $A$ to $B$ is a curve which is piecewise a future-pointing null geodesic, with past endpoint $A$ and future endpoint $B$. We write $A \lll B$ to indicate the existence of a lightlike path from $A$ to $B$. Thus the statement $A \lll B$ is equivalent to the statement that there exists a finite set of points $A_0=A, A_1,A_2,\dots,A_n=B$ and a set of $n$  future-pointing null geodesics $\gamma_i$ from $A_{i-1}$ to $A_i$, $1\leq i\leq n$.
\end{definition}

This copies directly Penrose's definition of a \textit{trip} (\cite{penrose1972techniques} and Definition \ref{def:trip} below), but with timelike geodesic segments replaced by null geodesic segments. We recall the following definitions and results of \cite{penrose1972techniques}. We work throughout in a time orientable spacetime $(M,g)$. 

\begin{definition} [\cite{penrose1972techniques}, Definition 2.1]\label{def:trip}  A \textbf{trip} from $A$ to $B$ is a curve which is piecewise a future-pointing timelike geodesic, with past endpoint $A$ and future endpoint $B$. We write $A \ll B$ to indicate the existence of a  trip from $A$ to $B$. Thus the statement $A \ll B$ is equivalent to the statement that there exists a finite set of points $A_0=A, A_1,A_2,\dots,A_n=B$ and a set of $n$  future-pointing timelike geodesics $\gamma_i$ from $A_{i-1}$ to $A_i$, $1\leq i\leq n$.
\end{definition}

\begin{definition} [\cite{penrose1972techniques}, Definition 2.3]\label{def:causal-trip}  A \textbf{causal trip} from $A$ to $B$ is a curve which is piecewise a future-pointing causal geodesic, with past endpoint $A$ and future endpoint $B$. We write $A \prec B$ to indicate the existence of a causal trip from $A$ to $B$. Thus the statement $A \prec B$ is equivalent to the statement that there exists a finite set of points $A_0=A, A_1,A_2,\dots,A_n=b$ and a set of $n$  future-pointing causal geodesics $\gamma_i$ from $A_{i-1}$ to $A_i$, $1\leq i\leq n$.
\end{definition}

\begin{proposition}[\cite{penrose1972techniques}, Proposition 2.20]\label{prop:Penrose-2.20}

If $A\prec B$ but $A \not\ll B$, then there is a null geodesic from $A$ to $B$. 
\end{proposition}

\begin{proposition}[\cite{penrose1972techniques}, Proposition 2.23]\label{prop:Penrose-2.23}
 There exists a future directed timelike curve from $A$ to $B$ if and only if $A \ll B$. 
\end{proposition}

We now state and prove our main result, which shows that Proposition 2.23 of \cite{penrose1972techniques} carries over from the case of (timelike) trips to lightlike paths.

\begin{theorem}\label{thm-lightlike-trips} Let $(M,g)$ be a time orientable spacetime and let $A,B\in M$. 
\begin{itemize}
\item[(i)] If $A\lll B$, then either there is a future-pointing timelike curve from $A$ to $B$, or there is a future-pointing null geodesic from $A$ to $B$. 
\item[(ii)] If there exists a future-pointing timelike curve from $A$ to $B$, then $A\lll B$. \qed
\end{itemize}
\end{theorem}

 
Following the rule set down in the introduction, the proof is given in the appendix. This comprises an application of techniques from \cite{penrose1972techniques}. 
An immediate consequence of Theorem \ref{thm-lightlike-trips} is this (recall that by homogeneity, there are CTCs through every point of {\godel}'s universe):

\begin{corollary}\label{cor:clt-exist} Let $\co$ be any event in {\godel}'s universe. Then there is a closed lightlike path from $\co$ to $\co$. \qed
\end{corollary}

\begin{comments}\label{comm:geo-segs}
We will discuss below the extent to which this provides an alternative means of violating causality. We note also that Proposition \ref{prop:Penrose-2.23} is of immediate interest to the topic of the present paper, as it indicates that \godel's CTC may be replaced by a piecewise $C^1$ curve comprising a sequence of timelike geodesic segments. But there is a heavy fuel cost (i.e.\ a large contribution to the total integrated acceleration) at each junction of successive segments: see for example Natario's calculation \cite{natario2012optimal}, which shows a contribution $\Delta TA\simeq 3.6158$ to the total integrated acceleration to provide the boost required at the end point of his closed timelike curve to make the CTC periodic: i.e.\ to make the tangent continuous at the initial and final points of the curve. 
\end{comments}

\section{Closed lightlike paths}

Corollary \ref{cor:clt-exist} is the basis for the construction of causality violations based on a chain of null geodesic segments. This allows for causality violations without the need for unfeasibly high speeds or extravagant amounts of fuel. However, to construct the lightlike path, we need (e.g.) a sequence of mirrors to deflect the trajectory of each incoming null geodesic onwards to the next mirror, and ultimately, back to the observer at $P$. Alternatively, this process could be carried out by a network of cooperative agents, each passing on the signal/message (which would include details of the required trajectory of the onward message) to the next agent. Either way, it is clear that the fewer the null geodesic segments in the lightlike path, the better. Thus we see this as an optimisation problem, and so we ask the question which is answered in the statement of the following theorem:

\begin{theorem}\label{thm-main} Let $\co$ be an event of {\godel}'s universe. Then a closed lightlike path from $\co$ to $\co$ contains at least $N=8$ future pointing null geodesic segments. Furthermore, this bound is attained. 
\end{theorem}

The remainder of this section (and the paper) provides a proof of this statement. We take $\co$ to be an arbitrary point of the spacetime, which (by homogeneity) we may choose to be located on the axis (with coordinate values $r=\tau=\zeta=0$). We use $N$ to refer to the minimum number of future pointing null geodesic segments required to construct the lightlike path from $\co$ to $\co$. The closed lightlike path that we construct relies on the fact that the coordinate $\tau$ may \textit{decrease} along future-pointing causal geodesics outside the horizon \cite{novello1983geodesic,grave2009godel}.

\begin{notation}
Given a complete geodesic $\gamma:\mathbb{R}\to M:s\mapsto\gamma(s)\in M$, we use the notation $\gamma_{[P,Q]}$ to refer to the segment of the geodesic with $s\in[s_1,s_2]$ where $\gamma(s_1)=P$ and $\gamma(s_2)=Q$, and extend the notation in the obvious way using interval notation, so that the set of points $\gamma_{(P,Q]}$ includes $Q$ but not $P$. A segment from $P$ to $Q$ on a geodesic $\gamma_a$ will be denoted $\gamma_{a,[P,Q]}$. 
\end{notation} 

We break up the discussion into three subsections. In the first of these, we discuss some basic properties of null geodesics in \godel's spacetime, and determine the structure of the first segment of the optimal path, taking us from the axis to the horizon. Here and below, an optimal path refers to a closed lightlike path from $\co$ to $\co$ comprising the minimum number of null geodesic segments. Optimal paths exist, and Proposition \ref{prop:no-direct-comm} proves that $N\geq2$. 
In subsection \ref{subsect:segments}, we prove a number of results relating to null geodesic segments exterior to the horizon. We establish the key result that all segments of the optimal path must be planar.  In Subsection \ref{subsect:optimal-path}, we construct the optimal path, and complete the proof of Theorem \ref{thm-main}.

We note that $\tau$ must increase along any future pointing causal geodesic through $\co$. Along with the fact that $\tau$ can decrease along future pointing causal geodesics outside the horizon specifies the overall strategy: we seek segments of future pointing null geodesics that extend from $\co$ to the horizon $\hor$. We then identify a path with the least number of segments that returns to $\hor$ at a sufficiently lower value of $\tau$ so that a final segment can be found from $\hor$ to $\co$: $\tau$ must increase along this segment. 

\subsection{Basic properties of the null geodesics}\label{subsect:basics}
Our first task is to find the optimal path from $\co$ (located on the axis $\{r=0\}$) to the horizon $\hor=\{r=1\}$. In this instance, `optimal' means the path along which the increase of $\tau$ is minimised. We settle this question as follows (see Figure \ref{fig:interior}:

\begin{figure}
\centering
{\includegraphics[width=0.4\linewidth]{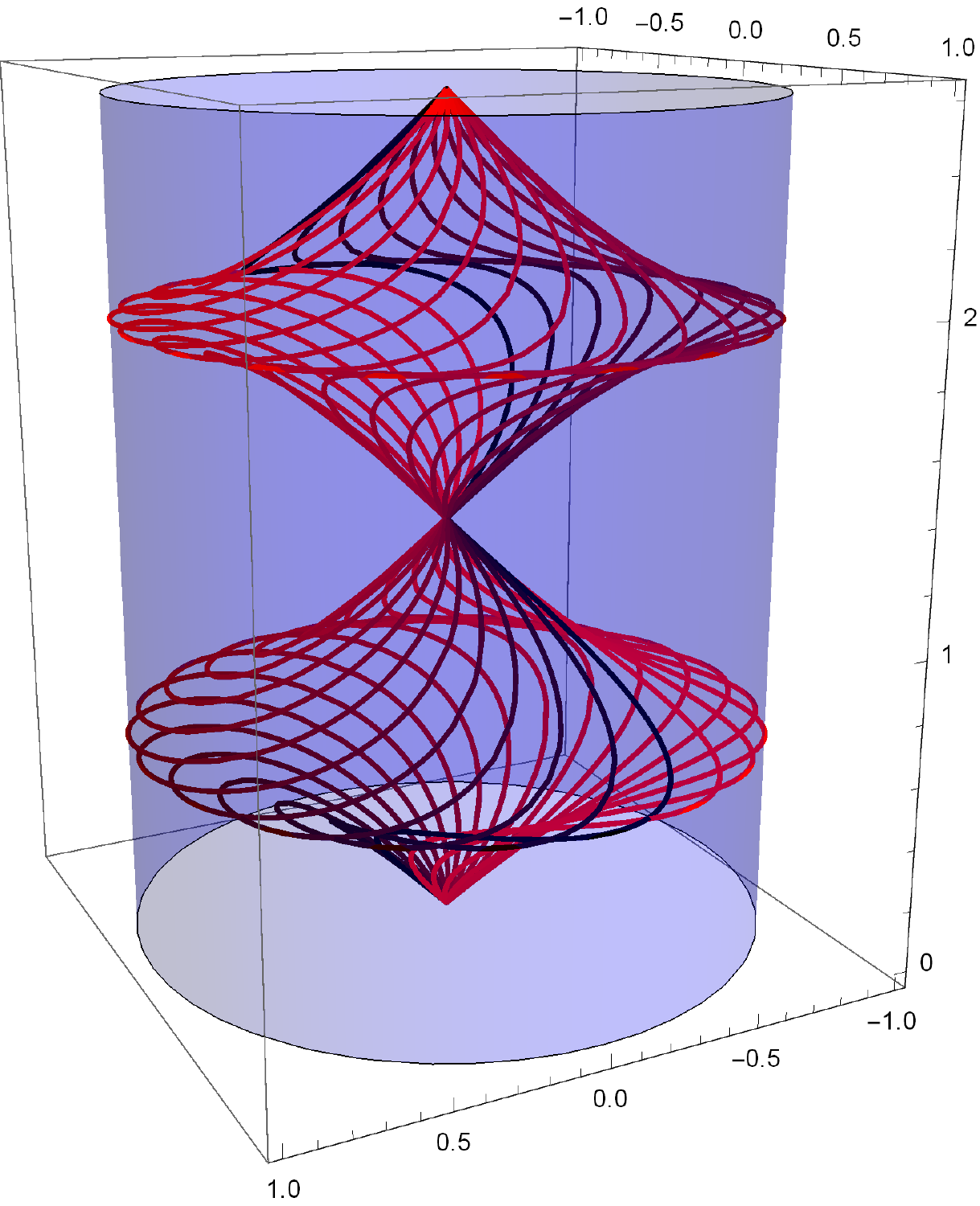},\includegraphics[width=0.4\linewidth]{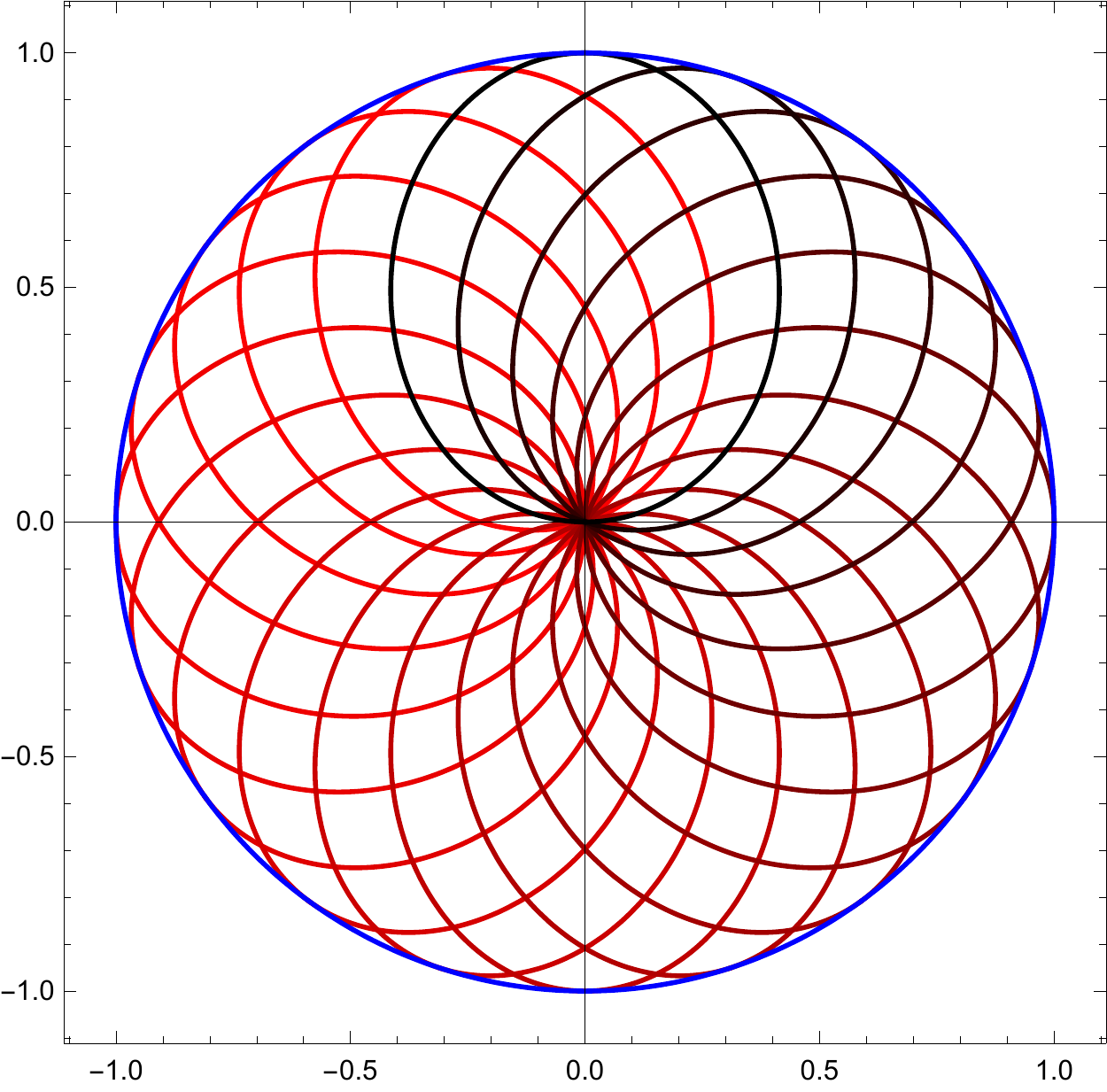}}
\caption{Planar null geodesics inside the horizon. The image on the left shows a spacetime diagram of a selection of future pointing null geodesics emerging from a point of the axis, in the cylindrical coordinates of (\ref{metric-cyl}) and with the coordinate $z$ suppressed. The geodesics pass through the axis, and so have $L_2=0$. The vertical axis represents $\tau$. The geodesics emerge from a point on the axis with $\tau=0$, reach the horizon $\{r=1\}$ at $\tau=\tau_*=(\sqrt{2}-1)\pi/2\simeq 0.6506$, and refocus at the axis at $\tau=2\tau_*$.  Two cycles are shown. The image on the right shows the projection of one cycle of the geodesics into the $x-y$ plane ($x=r\cos\phi, y=r\sin\phi$). Shading is used to distinguish the individual geodesics. The horizon is shown shaded (blue) in both images.}
\label{fig:interior}
\end{figure}

\begin{lemma}\label{lem:axis-to-hor} Let $\gamma$ be a causal trip from $\co\in\ax$ to an event $P\in\hor$. Then 
\be \Delta\tau_{\co P}=\tau(P)-\tau(\co)\geq \tau_*:=\frac{\pi}{2}(\sqrt{2}-1),\label{del-tau-A-H} \ee
with equality if and only if $\gamma$ comprises a single planar null geodesic with $L_2=0$.
\end{lemma}

\startproof Let $A$ be any point inside the horizon, so that $r(A)<1$, let $\mu$ be a future-pointing causal geodesic through $A$ and let $\nu$ be a future-pointing radial ($L_2=0$), planar ($L_3=0$), null ($\epsilon=0$) geodesic through $A$, both outward directed in the sense that $\dot{r}>0$ at $A$ for both geodesics. Using (\ref{tau-dot}) and (\ref{r-dot}), we have
\be \left.\frac{\dot{\tau}}{\dot{r}}\right|_{\nu} = \frac{(1-r^2)^{1/2}}{1+r^2}, \label{ratio1}
\ee
whereas on $\mu$,
\begin{eqnarray}
\dot{r}^2 &=& -\frac{L_2^2}{r^2}+2\sqrt{2} L_1L_2+(1-r^2)L_1^2-(1+r^2)(L_3^2-\epsilon) \nonumber \\
&\leq & (1-r^2)^{-1}\left((1-r^2)^2L_1^2+2\sqrt{2}(1-r^2) L_2L_3-(1-r^2)\frac{L_2^2}{r^2}\right) \nonumber \\
&=& (1-r^2)^{-1}\left(((1-r^2)L_1+\sqrt{2} L_2)^2-(1+r^2)\frac{L_2^2}{r^2}\right)
\end{eqnarray}
which gives
\be 0<\dot{r}<(1-r^2)^{-1/2}(\sqrt{2} L_2+(1-r^2)L_1). \ee
Then using (\ref{tau-dot}), we see that 
\be \left.\frac{\dot{\tau}}{\dot{r}}\right|_\mu>\left.\frac{\dot{\tau}}{\dot{r}}\right|_{\nu}. \ee
Thus at any event $A$ of the causal trip $\gamma$ at which $\dot{r}>0$, $\frac{d\tau}{dr}$ has its positive minimum along a segment which is a radial planar null geodesic. It follows that the minimum elapse of $\tau$ on a future-pointing causal trip from the axis to the horizon is attained along the future-pointing, radial, planar null geodesic $\nu$. Integrating (\ref{ratio1}) (writing the left hand side as $d\tau/dr$) yields \be \Delta\tau_{\co P}=\frac{\pi}{2}(\sqrt{2}-1).\ee\qed

\begin{comments}\label{comm:N3}
It follows from Lemma \ref{lem:axis-to-hor} that the \textit{last} of the $N$ segments that form the closed lightlike path from $\co$ to $\co$ must have $L_2=0$ and so must lie within the horizon, with at most one point on the horizon (see (\ref{r-bound-acis})). The geodesic equations (\ref{tau-dot}) and (\ref{r-dot}) have time-reversal and time-translation invariance: it follows from this that Lemma 4 applies also to the last segment of the path, and hence the greatest value that $\tau$ may have at the initial point of the last null geodesic segment of the path is $\tau=-\tau_*=-\frac{\pi}{2}(\sqrt{2}-1)$. Hence the optimal choice for the last segment is an ingoing radial, planar null geodesic from $\hor$ to $\ax$. Notice that this establishes that $N\geq3$: we need the outgoing planar null geodesic from the axis to the horizon, at least one segment on which $\tau$ decreases, and the ingoing planar null geodesic from the horizon to the axis.  
\end{comments}

\begin{comments}\label{comm:toS0}
So at this stage, our problem is the following: find the minimum number of future-pointing null geodesic segments that connect a point on the horizon with $\tau=\frac{\pi}{2}(\sqrt{2}-1)$ to an earlier point on the horizon with $\tau=-\frac{\pi}{2}(\sqrt{2}-1)$. The null geodesic equations contain three parameters - the conserved quantities $L_i, i=1,2,3$ - so we are seeking to optimize over a multidimensional parameter space. This is not ideal, but there are two strategies that help simplify the problem. The first is to note that our problem is effectively to reach the hypersurface $\Sigma_0=\{x^\alpha\in M: \tau=0\}$ using the least possible number of null geodesic segments. Suppose we produce a lightlike path from $P\in\hor$ at which $\tau=\tau_*$ to $Q\in\Sigma_0$. The null geodesic equations possess reflection symmetries that allow us to follow a lightlike path (constructed by reflection of the $P-Q$ path about $\tau=0$ in the $r-\tau$ plane) from $Q$ to a point $R\in\hor$ which has $\tau=-\tau_*$. This is achieved by application of Lemma \ref{lem:retrace} below to each null geodesic segment of $P-Q$. The second strategy involves reducing the number of free parameters to one. The angular momentum constant $L_2$ can easily be set aside (Proposition \ref{prop:key-ngs}). We can also set $L_3=0$. Proving this requires considerably more effort: see Subsection \ref{subsect:segments}.
\end{comments}

\begin{lemma}\label{lem:retrace}
Let $\gamma_a:[0,1]\to M$ be a segment of a future-pointing null geodesic with parameters $(L_1,L_2,L_3)$ along which $(\tau,r,\phi,\zeta)=(\tau_a(s),r_a(s),\phi_a(s),\zeta_a(s))$ and with initial and terminal points 
\be A_0=(r_a(0),\tau_a(0),\phi_a(0),\zeta_a(0))=(r_0,\tau_0,\phi_0,\zeta_0),\quad A_1=(r_a(1),\tau_a(1),\phi_a(1),\zeta_a(1))=(r_1,\tau_1,\phi_1,\zeta_1). \label{gama-endpoints} \ee
Then the equations $(\tau,r,\phi,\zeta)=(\tau_b(s),r_b(s),\phi_b(s),\zeta_b(s)), s\in[0,1]$ where
\begin{eqnarray}
\tau_b(s) &=& \tau_2+\tau_1-\tau_a(1-s), \label{tau-rev} \\
r_b(s) &=& r_a(1-s), \label{r-rev} \\
\phi_b(s) &=& \phi_2+\phi_1-\phi_a(1-s),\label{phi-rev} \\
\zeta_b(s) &=& \zeta_a(1-s) \label{zeta-rev} 
\end{eqnarray}
describe a future-pointing null geodesic segment $\gamma_b:[0,1]\to M$ with parameters $(L_1,L_2,-L_3)$ and with initial and terminal points
\be B_0=(r_b(0),\tau_b(0),\phi_b(0),\zeta_b(0))=(r_1,\tau_2,\phi_2,\zeta_1),\quad B_1=(r_b(1),\tau_b(1),\phi_b(1),\zeta_b(1))=(r_0,\tau_3,\phi_3,\zeta_0) \label{gamb-endpoints} \ee
and where 
\begin{eqnarray}
\tau_3-\tau_2 &=& \tau_1-\tau_0,\label{tau-diff-rev} \\
\phi_3-\phi_2 &=& \phi_1-\phi_0. \label{phi-diff-rev}
\end{eqnarray}
Thus $\gamma_b$ retraces the path of $\gamma_a$ in the $r-\tau$ plane with an overall translation of $\tau$ and with the same net elapse of $\tau$. The segment is also subject to an overall rotation in $\phi$, but returns to the same $\zeta$= constant hypersurface on which $\gamma_a$ originated.
\end{lemma}

\startproof The conclusions follow immediately by substitution of (\ref{tau-rev})-(\ref{zeta-rev}) into the geodesic equations (\ref{tau-dot})-(\ref{zeta-dot}) and by relevant evaluations.\qed

\begin{lemma}\label{lem:params-L}
\begin{enumerate}
    \item[(i)] A causal geodesic with parameters $L_1,L_2,L_3,\epsilon$ exists if and only if
    \be 2\sqrt{2} L_1L_2+L_1^2-L_3^2\geq 0 \label{para-cond0} \ee
    and 
    \be (2\sqrt{2} L_1L_2+L_1^2-L_3^2)^2-4L_2^2(L_1^2+L_3^2-\epsilon)\geq 0. \label{para-cond}\ee
    \item[(ii)]
Along a causal geodesic, the coordinate $r$ satisfies
$r_1\leq r\leq r_2$, where
\be r_{1,2}^2 = \frac{1}{2(L_1^2+L_3^2-\epsilon)}\left[2\sqrt{2} L_1L_2+L_1^2-L_3^2+\epsilon\pm\left((2\sqrt{2} L_1L_2+L_1^2-L_3^2+\epsilon)^2-4L_2^2(L_1^2+L_3^2-\epsilon)\right)^{1/2}\right],\label{r12-def}\ee
with $r_1$ corresponding to the lower sign and $r_2$ to the upper.
\item[(iii)] If $L_2<0$, then $r_2<1$.
\end{enumerate}
\qed
\end{lemma}

The proof of this lemma follows more or less immediately from (\ref{r-dot}), the right hand side of which must be non-negative on an interval of $r$ values of positive measure. The upper and lower bounds for $r$ play a crucial role in the analysis below. Part (iii) of this lemma, in combination with Lemma 2, indicates that we must have $L_2>0$ along the segments that traverse the region exterior to the horizon, along which we can bring about the required decrease in $\tau$. We can then absorb $L_2$ into the affine parameter along the geodesic. This enables the following convenient description of the null geodesic equations and their solutions. See also \cite{kundt1956tr,chandrasekhar1961geodesics,novello1983geodesic,grave2009godel}. 

\begin{proposition}\label{prop:key-ngs}
 The geodesic equations for a null geodesic with $L_2>0$ may be written as 
\begin{eqnarray}
\dot{\tau} &=& (1+r^2)^{-1}((1-r^2)\kappa+\sqrt{2}),\label{tau-d} \\
\dot{\phi}&=&(1+r^2)^{-1}(\sqrt{2}\kappa-r^{-2}),\label{phi-d}\\
\dot{r}^2 &=& -\frac{1}{r^2}+(2\sqrt{2}\kappa+\kappa^2-\lambda^2)-(\kappa^2+\lambda^2)r^2,\label{r-d}\\
\dot{\zeta} &=& \lambda, \label{z-d}
\end{eqnarray}
where the overdot represents differentiation with respect to the parameter $s=\frac{s'}{L_2}$, and where $\kappa=L_1/L_2, \lambda=L_3/L_2$ and $s'$ is the affine parameter of (\ref{tau-dot}-\ref{zeta-dot}). The parameter $s$ increases into the future, and the geodesic is future-pointing if and only if $\kappa>0$. Furthermore:
\begin{enumerate}
    \item[(i)] \textbf{Existence of solutions:} The necessary and sufficient condition for existence of a solution with $\kappa>0$ is 
    \be |\lambda|\leq \kappa, \label{geo-exist} \ee
and when solutions exist, they exist globally and are smooth in $s$. 
\item[(ii)] \textbf{Global behaviour of $r$:} $r$ is bounded along the geodesic, and the minimum and maximum of $r$ along the geodesic are $r_1, r_2$ respectively, where
\be r_{1,2}^2 = \frac{1}{2(\kappa^2+\lambda^2)}\left[2\sqrt{2}\kappa+\kappa^2-\lambda^2\pm\left((2\sqrt{2}\kappa+\kappa^2-\lambda^2)^2-4(\kappa^2+\lambda^2)\right)^{1/2}\right]. \label{r12-new}\ee
\item[(iii)] \textbf{Global behaviour of $\tau$:} 
\begin{enumerate}
    \item [(a)] Local minimum and maximum points of the coordinate $\tau$ exist on the geodesic if and only if 
\be |\lambda|\leq \overline{\lambda}(\kappa):= \frac{\kappa}{(\sqrt{2}\kappa+2)^{1/2}}, \label{l3-bound}\ee and occur at $r=r_3\in(r_1,r_2)$ where 
\be r_3^2 = 1+\frac{\sqrt{2}}{\kappa}. \label{r3-def}\ee
\item[(b)] If (\ref{l3-bound}) holds, then as the parameter $s$ increases the geodesic repeatedly passes through sequences of four points which correspond to the global minimum of $r$; a local maximum of $\tau$; the global maximum of $r$ and a local minimum of $\tau$.
\item[(c)] If (\ref{l3-bound}) does not hold, then $\tau$ is monotone along the geodesic (monotone increasing for future-pointing null geodesics).
\end{enumerate}
\item[(iv)] \textbf{Closed form of the solutions:} The solutions of (\ref{tau-d})-(\ref{z-d}) are given by 
\begin{eqnarray}
u &=& \frac{u_1+u_2}{2}+\frac{u_2-u_1}{2}\sin\sigma, \label{u-sol}\\
\tau &=& \sqrt{2}\arctan\left(\frac{c+b\tan\frac{\sigma}{2}}{\sqrt{b^2-c^2}}\right)-\frac{\kappa}{2(\kappa^2+\lambda^2)^{1/2}}\sigma+\sqrt{2}\pi\left(1+\floor{\frac{\sigma-\pi}{2\pi}}\right)+k_\tau, \label{tau-sol}\\
\phi &=& \arctan\left(\frac{c+b\tan\frac{\sigma}{2}}{\sqrt{b^2-c^2}}\right)-\arctan\left(\frac{c+(b-2)\tan\frac{\sigma}{2}}{\sqrt{(b-2)^2-c^2}}\right)+k_\phi,\label{phi-sol}\\
\zeta &=& \lambda s + k_\zeta,\label{z-sol}
\end{eqnarray}
where
\begin{eqnarray}
u &=& r^2, \label{u-def}\\
u_i&=&r_i^2,\quad i=1,2, \label{ui-def}\\
b&=& 2+u_1+u_2, \label{b-def}\\
c&=& u_2-u_1, \label{c-def}\\
\sigma &=& 2(\kappa^2+\lambda^2)^{1/2}s+\sigma_0,\label{sig-def}
\end{eqnarray}
and 
\be \sigma_0,k_\tau,k_\phi,k_\zeta \label{const}\ee
are constants of integration. In (\ref{tau-sol}), the floor function has the usual definition:
\be \floor{x} = \max\{y\in\mathbb{Z}:y\leq x\}. \ee

\end{enumerate}

\end{proposition}

\startproof The preamble follows immediately from the definitions and from Lemma \ref{lem:params-L}. For parts (i) and (ii), we note that a necessary and sufficient condition for existence is that there is a non-empty interval of values of $r$ for which the right hand side of (\ref{r-d}) is non-negative. It follows that we must have
\be 2\sqrt{2}\kappa+\kappa^2-\lambda^2 \geq 0, \label{ell-ineq1} \ee
and 
\begin{eqnarray} 
0 &\leq & (2\sqrt{2}\kappa+\kappa^2-\lambda^2)^2-4(\kappa^2+\lambda^2) \nonumber \\
&=& (\kappa^2-\lambda^2)(4+4\sqrt{2}\kappa+\kappa^2-\lambda^2).
\label{ell-ineq2}
\end{eqnarray}
From (\ref{ell-ineq1}) and positivity of $\kappa$, we see that the second factor in (\ref{ell-ineq2}) is positive. We then see that the three conditions $\kappa>0$, (\ref{ell-ineq1}) and (\ref{ell-ineq2}) are equivalent to the two conditions $\kappa>0$ and $\kappa^2-\lambda^2\geq$ as required. 

Part (ii) follows immediately from (\ref{r12-def}). 

For part (iii-a), the max/min existence condition follows by solving (\ref{tau-d}) with $\dot{\tau}=0$ for $r$, and checking the sign of $\dot{r}^2$ at this value of $r$. To prove (iii-b), note that differentiating (\ref{tau-d}) and evaluating at a local extremum of $\tau$, we see that local maxima (respectively minima) of $\tau$ occur at points where $\dot{r}>0$ (respectively $\dot{r}<0$). In the case where (\ref{l3-bound}) does not hold, $\dot{\tau}$ remains non-zero for $r\in[r_1,r_2]$. We find that $\dot{\tau}>0$ at $r=r_1$, and so $\dot{\tau}$ remains positive along the geodesic, establishing (iii-c). The solutions in part (iv) are obtained as follows. We make the change of variable $u=r^2$ and rewrite (\ref{r-d}) as an equation in $u$. This is readily solved to yield (\ref{u-sol}) (a negative root arises: this may be absorbed into the constant of integration). Equation (\ref{tau-d}) can then be integrated to yield the local solution. The global solution is found by adding the floor function to obtain the unique $C^1$ continuation of the local solution (which in fact yields a smooth solution). Equation (\ref{phi-d}) is solved by direct integration: the local solution yields the global solution. The solution for $\zeta$ arises trivially. \qed

The following corollary could not be more simple, but calls attention to a fact that is used repeatedly below: 

\begin{corollary} We can write
\be \dot{u}=c(\kappa^2+\lambda^2)^{1/2}\cos\sigma. \label{u-dot-sol} \ee
\qed
\end{corollary}

The quantities $r_1$ and $r_2$ play an important role in what follows, and they satisfy the following properties. 

\begin{lemma}\label{lem:ri-l3-dep}
Let $\kappa,\lambda$ both be positive and satisfy (\ref{geo-exist}). Then
\be \pd{r_1}{\lambda}>0, \quad \pd{r_2}{\lambda}<0.\label{r1-inc-r2-dec}\ee
\end{lemma}

\startproof We can prove these inequalities as follows. Solve the equations (\ref{r12-new}) for $\kappa$ and $\lambda$ in terms of $u_1=r_1^2$ and $u_2=r_2^2$. Calculate the derivatives of $u_1,u_2$ with respect to 
\be \Lambda=\lambda^2 \label{cap-lam-def}\ee 
and substitute for $\kappa$ and $\lambda$ in terms of $u_1$ and $u_2$. This yields 
\begin{eqnarray}
\pd{u_1}{\Lambda}&=&\frac{u_1^2(1+u_1)u_2}{u_2-u_1}>0,\label{u1-dlambda}\\
\pd{u_2}{\Lambda}&=&-\frac{u_1u_2^2(1+u_2)}{u_2-u_1}<0,\label{u2-dlambda}
\end{eqnarray}
from which the result follows as $\lambda>0$. (As well as furnishing this proof, the derivatives (\ref{u1-dlambda}) and (\ref{u2-dlambda}) will be useful below.) \qed

The following corollary introduces some quantities that will be important below.

\begin{corollary}\label{corr:lambda-bounds}
\begin{itemize}
\item[(i)]
A future pointing null geodesic with parameters $(\kappa,\lambda)$ meets the cylinder  with radius $R$ if and only if $r_1(\kappa,\lambda)\leq R\leq r_2(\kappa,\lambda)$. If the future pointing null geodesic with parameters $(\kappa,\lambda)$ with $\lambda>0$ meets this cylinder, then so too does every other future pointing null geodesic with parameters $(\kappa,\lambda')$ for all $0\leq\lambda'<\lambda$.
\item[(ii)] 
A future pointing null geodesic $\gamma$ with parameters $\kappa>0$ and $\lambda=0$ reaches the cylinder with radius $R=\sqrt{u}\geq1$ if and only if 
\begin{eqnarray}
\bar{u}_1(\kappa) \leq u \leq \bar{u}_2(\kappa), \label{ubar-bounds}
\end{eqnarray}
where $\bar{u}_i(\kappa)=u_i(\kappa,0), i=1,2$ (\textit{cf.} (\ref{ui-def}), (\ref{r12-def})) so that 
\begin{eqnarray}
\bar{u}_1(\kappa) &=& \frac{2\sqrt{2}+\kappa-(\kappa^2+4\sqrt{2}\kappa+4)^{1/2}}{2\kappa},\label{u1bar-def-1} \\ 
\bar{u}_2(\kappa) &=& \frac{2\sqrt{2}+\kappa+(\kappa^2+4\sqrt{2}\kappa+4)^{1/2}}{2\kappa}.\label{u2bar-def-1}
\end{eqnarray}
These satisfy
\be \bar{u}_1'(\kappa)<0,\quad\bar{u}_2'(\kappa)<0,\quad \kappa>0. \label{ubar-decr}\ee
\item[(iii)]
The bounds (\ref{ubar-bounds}) are equivalent to 
\be
\kappa_1(u) \leq \kappa \leq \kappa_2(u), \label{kap-bounds}
\ee
where for $u=1$ we have
\be \kappa_1(1) = \frac{1}{2\sqrt{2}},\quad \kappa_2(1) = +\infty, \label{eq:kap-1-2-1} \ee
and for $u>1$, 
\begin{eqnarray}
\kappa_1(u) &=& \frac{\sqrt{2}u-(u^2+u)^{1/2}}{u(u-1)},\label{kap1-def}\\
\kappa_2(u) &=& \frac{\sqrt{2}u+(u^2+u)^{1/2}}{u(u-1)}.\label{kap2-def}
\end{eqnarray}
For $u>1$, these quantities satisfy
\be \kappa_1(u) < \kappa_+(u) <\kappa_2(u), \label{kap1+2} \ee
where 
\be \kappa_+(u)=\frac{\sqrt{2}}{u-1},\label{kpm-def} \ee
which is the unique value of $\kappa$ for which $\dot{\tau}$ vanishes at a given value of $u$.
\end{itemize}
\end{corollary}

\startproof Part (i) follows immediately from Proposition \ref{prop:key-ngs}. The bounds (\ref{ubar-bounds}) of part (ii) then follow by setting $\lambda=0$. These can be inverted to produce (\ref{kap-bounds}), but in fact these are more easily obtained by `solving' $\left.\dot{r}^2\right|_{\lambda=0}=f(u;\kappa)\geq0$ for $\kappa$ using (\ref{r-dot}), proving part (iii). The bounds (\ref{kap1+2}) are easily verified. \qed  

\subsection{Segments of the optimal path}\label{subsect:segments}

In the remainder of the paper, we will use the formulation of the null geodesic equations and their solutions given in Proposition \ref{prop:key-ngs}. We will assume that the existence conditions of part (i) of the Proposition 5 hold for all $\kappa,\lambda$ values we encounter. Bearing in mind that our aim is to drive $\tau$ down to zero using the least number of segments, we note that for a fixed value of $\kappa$, segments with large values of $r$ are more favourable (the right hand side of (\ref{tau-d}) is a decreasing function of $r$; larger $r$ gives a more rapidly decreasing $\tau$). So we like segments that decrease $\tau$ and increase $r$. This loosely stated idea is a helpful guide in constructing the optimal path.  But note also how this observation makes the optimisation problem more complicated: it is not simply a matter of selecting each segment by maximising the decrease of $\tau$ along all available paths: it may be preferrable to select a segment that provides an ultimately more favourable increase in $r$, at the short-term expense of a less pronounced decrease in $\tau$. The problem is also complicated by the fact that we have a two-parameter family of geodesics at each initial point of each segment (along with a choice of sign for $\dot{r}$). The principal result of this subsection is the following.

\begin{proposition}\label{prop:segments-optimal} Without loss of generality, each null geodesic segment $\gamma:[0,s_*]\to M$ of an optimal path from $A_0:(r,\tau)=(1,\tau_*)$ to $A_{N-1}:(r,\tau)=(1,-\tau_*)$ has parameters $(\kappa,\lambda)$ satisfying  $\lambda=0$ and 
\be\kappa_1(u)\leq\kappa\leq\kappa_{\rm{min}}(u),\label{kp-bounds} 
\ee
where $\kappa_1$ is defined in (\ref{kap1-def}) and where
\be \kappa_{\rm{min}}(u) = \frac{\sqrt{2}}{u}. \label{eq:kap-min-def} \ee
Each segment has $\dot{r}(0)\geq0$, where $u=R^2$ and $r(0)=R$ is the value of $r$ at the initial point of the segment. Furthermore, each such segment is of the form $\gamma_{[P,Q]}$ with $Q=\gamma(s_*)\in \ese{P}$, where $P=\gamma(0)$ and $\ese{P}$ is the ``south-east" portion of the envelope of the family of future pointing null geodesics from $P$ (see Definition \ref{def:ese} below).
\end{proposition}

\begin{definition}\label{def:se-seg}
A segment $\gamma_{[P,Q]}$ of a future pointing null geodesic $\gamma$ with $\lambda=0$, $\kappa\in[\kappa_1(u),\kappa_{\rm {min}}(u)]$ and $Q\in\ese{P}$ is called an \textbf{${\rm{SE}}$-segment from $P$}. As in the statement of Proposition \ref{prop:segments-optimal}, $u=(r(P))^2$.
\end{definition}

\begin{figure}
    \centering
    \includegraphics{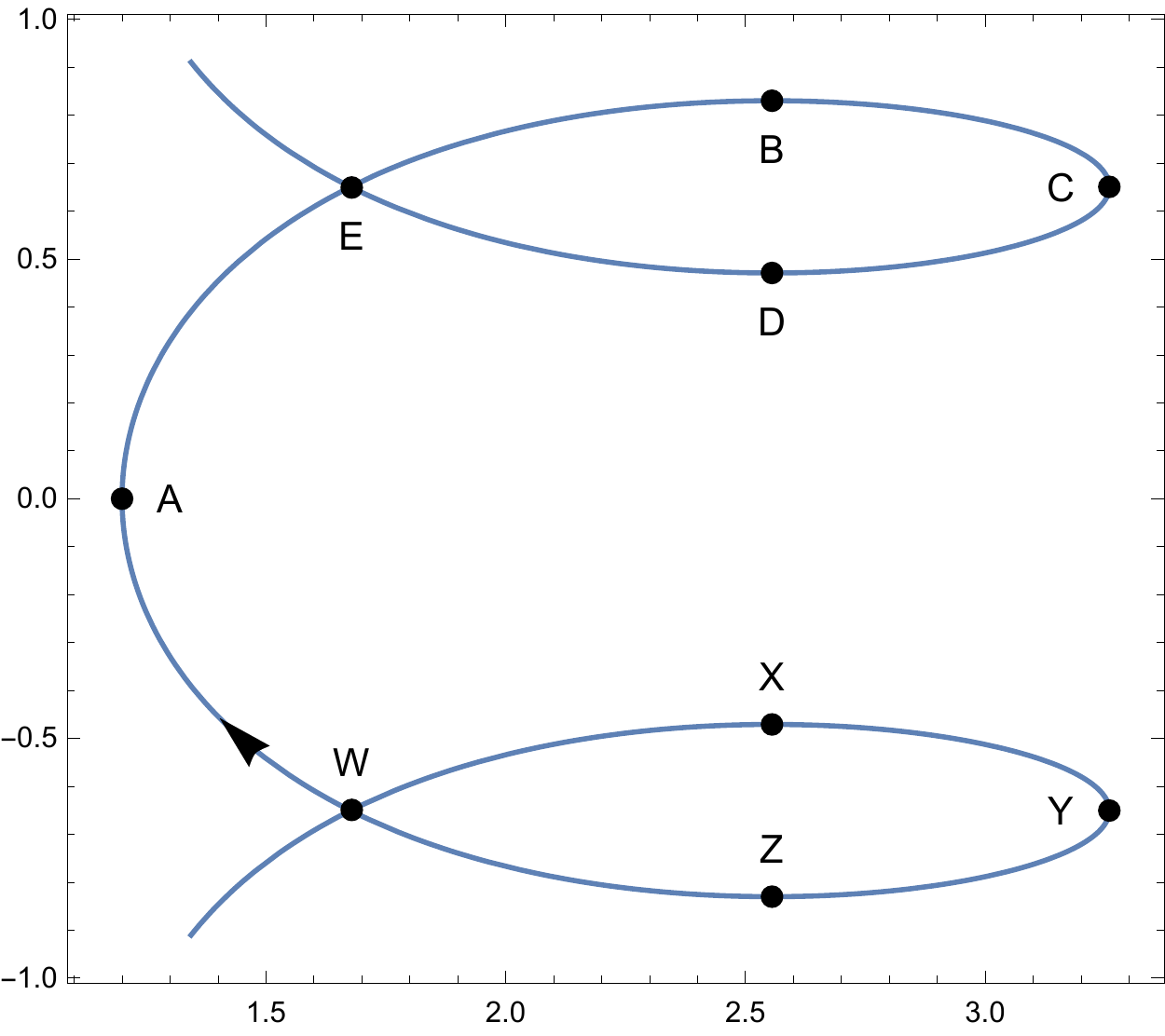}
    \caption{Projection into the $r-\tau$ plane of a typical future pointing null geodesic satisfying (\ref{l3-bound}). Labelling of points follows the definitions of Lemma \ref{lem:tau-seps-generic}. The horizontal and vertical axes represent $r$ and $\tau$ respectively, and (without loss of generality) $\tau$ is set to zero at $A$. Note that the self-intersections $E$ and $W$ of the projection of the geodesic must occur as indicated. $E$ corresponds to a distinct pair of spacetime points on the geodesic, $E_-$ preceding $B$ and $E_+$ succeeding $D$. Similar holds for $W$ ($W_-$ precedes $X$ and $W_+$ succeeds $Z$). The arrowhead indicates the direction of increase of the parameter $s$. This particular geodesic has $\lambda=0,\kappa\simeq0.2557$, yielding minimum and maximum values of $r$ of $r_1=1.2$ and $r_2=3.2591$ respectively, and with  \backc segments at radius $r=2.5556$.}
    \label{fig:lem8fig}
\end{figure}

\begin{comments} The remainder of this section is given over to the proof of this proposition: we outline the structure of the proof here. We begin by identifying an important class of future pointing null geodesic segments which we refer to as \backc transits (Definition \ref{def:transit}). With the aid of (\ref{tau-sol}), we can calculate explicitly, and in a useful form, the elapse of $\tau$ along such segments (Lemma \ref{lem:back-c-shaped}). This result and Lemma \ref{lem:tau-seps-generic} establishes the structure of a generic future pointing null geodesic with parameters $\kappa, \lambda$ satisfying (\ref{kp-bounds}). A generic geodesic is represented in Figure \ref{fig:lem8fig}, which provides a useful reference diagram for the succeeding lemmas. These lemmas will involve comparing two (or more) geodesic segments, and we introduce the concept of one segment being \textbf{better} than the other:
\begin{definition}
Let $\gamma_{a,[P,Q]}$ be a future pointing null geodesic segment. Then the future pointing null geodesic segment $\gamma_{b,[P,Q']}$ is said to be \textbf{better} than $\gamma_{a,[P,Q]}$ if 
\be \tau(Q')<\tau(Q)\quad\hbox{and}\quad r(Q')>r(Q). \label{def:better} \ee
We will say that $\gamma_{b,[P,Q']}$ is \textbf{marginally better} than $\gamma_{a,[P,Q]}$ if one of these strict inequalities is replaced by a non-strict inequality.
\end{definition}

As the name suggests, replacing segments of a closed lightlike path with better segments will decrease (or at least not increase) the number of segments required. The next steps in the proof involve establishing the fact that the segments identified in the statement of Proposition \ref{prop:segments-optimal} are better than all others. We can set aside segments with $\dot{r}(0)<0$ (Lemma \ref{lem:rdot-neg-useless}), those with $\lambda>\overline{\lambda}(\kappa)$ (Lemma \ref{lem:mono-useless-1}; cf. (\ref{l3-bound})), and subsequently those with $\lambda>0$ (Lemma \ref{lem:X}). The final steps require knowledge of the envelope $\env_P$ of the family of future pointing null geodesics emanating from the point $P$. The relevant properties are established in Subsection IV-B-2 below. It is then straightforward to prove Proposition \ref{prop:segments-optimal}, by showing that any given closed lightlike path can be replaced by one constructed using the `better' segments described in the statement of the proposition: there will be fewer (or the same number) of these segments than in the original path.  
\end{comments}

\begin{notation} Given geodesics $\gamma_a$ and $\gamma_b$, we use the notation $q_a(s)$ to represent the functional dependence of the quantity $q$ on the parameter $s$ along $\gamma_a$, and likewise for $\gamma_b$. For a generic geodesic $\gamma$, we will use (e.g.) $r(s)$ to indicate the value of $r$ at the parameter value $s$ on the geodesic. There should be no confusion with the usage $r(P)$ to indicate the value of the coordinate $r$ at the point $P$.
\end{notation}

Proposition \ref{prop:segments-optimal} greatly simplifies the optimisation problem, as it restricts us to a one-parameter problem, with that parameter restricted to a compact set (albeit a different compact set for each segment of the path). The difficulty highlighted above remains: finding the optimal path involves balancing the need to decrease $\tau$ with the desirability of increasing $r$, but Proposition \ref{prop:segments-optimal} makes this more tractable. 

As flagged above, the proof of Proposition \ref{prop:segments-optimal} begins by identifying a class of segments of particular importance:

\begin{definition}\label{def:transit} 
\begin{enumerate}
    \item[(i)]  A \textbf{transit of the cylinder at $r=R$} (a transit at $R$ for short) is defined to be a future-pointing null geodesic segment $\gamma_{[P,Q]}$ whose initial and terminal points $P,Q$ satisfy $r(P)=r(Q)=R$ and with $r>R$ elsewhere on the segment.
    \item[(ii)] A \textbf{$\supset\!-$shaped transit at $R$} has the additional property that $\tau$ has a local maximum at the initial point $P$ and has a local minimum at the terminal point $Q$.
    \end{enumerate}
\end{definition}

\begin{comments}
We note that a transit $\gamma_{[P,Q]}$ at $R$ is a $\supset\!-$shaped transit at $R$ if and only if $\dot{r}|_P>0$ and the parameter $\kappa$ satisfies $\kappa=\kappa_+(\sqrt{R})$, where $\kappa_+$ is defined in (\ref{kpm-def}). The terminology arises from the image formed by the projection of a \backc transit in the $r-\tau$ plane. In Figure \ref{fig:lem8fig}, the segments $\gamma_{[B,D]}$ and $\gamma_{[X,Z]}$ are \backc segments.
\end{comments}

\begin{lemma}\label{lem:back-c-shaped} Let $\gamma_{[B,D]}$ be a $\supset\!-$shaped transit at $R>1$ with parameters $\kappa>0$ and $\lambda\geq0$, so that (\ref{l3-bound}) holds and 
\be R=r(B)=r(D),\quad R^2= 1+\frac{\sqrt{2}}{\kappa}. \label{rA-rB}\ee \begin{itemize}
    \item[(i)] For fixed $R$ and $\kappa$, the elapse of $\tau$ along such a segment, $\Delta\tau_{BD}(\kappa,\lambda)=\tau(D)-\tau(B)$ is negative, and is minimised when $\lambda=0$. 
    \item[(ii)] For $\lambda=0$, the elapse is given by 
\be 
\Delta\tau_{BD}(\kappa)=\tau(D)-\tau(B) = \sqrt{2}\arctan x+\arcsin y-\frac{\pi}{2},\label{del-tau-backC1}
\ee
where
\begin{eqnarray}
x&=&\frac{\sqrt{2}(\sqrt{2}\kappa+1)^{1/2}}{\kappa+\sqrt{2}}=\frac{\sqrt{R^4-1}}{R^2},\label{x-form-backC}\\
y&=&\frac{\kappa}{(\kappa^2+4\sqrt{2}\kappa+4)^{1/2}} =\frac{1}{\sqrt{2R^4-1}}.\label{y-form-backC}
\end{eqnarray}
It follows that $\Delta\tau_{BD}$ is a negative, decreasing function of $R$ on $(1,+\infty)$ with 
\begin{eqnarray}
\lim_{R\to 1^+}\Delta\tau_{BD}&=&0,\label{lim-at-hor}\\
\lim_{R\to \infty}\Delta\tau_{BD}&=&(\sqrt{2}-2)\frac{\pi}{4}=-\frac{\tau_*}{\sqrt{2}},\label{lim-at-inf}
\end{eqnarray}
and so 
\be (\sqrt{2}-2)\frac{\pi}{4} <\Delta\tau_{BD}<0 \label{eq:del-tau-bounds} \ee
for all \backc transits $\gamma_{[B,D]}$.
\end{itemize}
\end{lemma}

\startproof The structure of the proof is straightforward: we evaluate $\tau$ in (\ref{tau-sol}) at the points $B$ and $D$, and apply some elementary calculus. However the calculations are not so straightforward, and so we will give the relevant detail. 

First, we note that the elapse of $\tau$ from $B$ to $D$ is twice the elapse of $\tau$ from $B$ to $C$, where $C$ is the first point to the future of $B$ on $\gamma$ at which $r$ first reaches its global maximum, so that $\dot{r}>0$ on $\gamma_{[B,C)}$ and $\dot{r}<0$ on $\gamma_{(C,D]}$. See Figure \ref{fig:lem8fig}. To see that $\tau(C)-\tau(B)=\tau(D)-\tau(C)$, which is equivalent to the previous claim, we use a change of variable to write 
\be \tau(C)-\tau(B) = \int_{s(B)}^{s(C)} \dot{\tau} ds = \int_{r(B)}^{r(C)} \frac{f(r)}{\sqrt{g(r)}}dr \label{tauAC} \ee
and 
\be \tau(D)-\tau(C) = \int_{s(C)}^{s(D)} \dot{\tau} ds = -\int_{r(C)}^{r(D)} \frac{f(r)}{\sqrt{g(r)}}dr=\int_{r(D)}^{r(C)} \frac{f(r)}{\sqrt{g(r)}}dr. \label{tauCB} \ee
Here, we have used (\ref{tau-dot}) and (\ref{r-dot}) in the forms $\dot{\tau}=f(r(s))$ and $\dot{r}^2=g(r(s))$, paying due attention to the sign of $\dot{r}$ on the relevant segments. Since $r(B)=r(D)$, the claim follows.

So we focus our attention on the elapse of $\tau$ from $B$ to $C$, and to prove part (i), show that this is an increasing function of $\Lambda=\lambda^2$. 

In (\ref{tau-sol}), we can take $k_\tau=0$, and set $s=0$ at $B$. Define $u:=u(0)=R^2$: then 
\be u=1+\frac{\sqrt{2}}{\kappa}=\frac{u_1+u_2}{2}+\frac{u_2-u_1}{2}\sin\sigma_0, \label{u0}\ee
and 
\be \dot{u}(0)=\frac{u_2-u_1}{2}\sqrt{\kappa^2+\lambda^2}\cos\sigma_0. \label{udot0}\ee
We solve (\ref{u0}) to write 
\be \sin\sigma_0=\frac{2u-u_1-u_2}{u_2-u_1}, \label{sin0} \ee
which is readily shown to be positive. Since $r$, and hence $u$, must be increasing at $B$, we must have (without loss of generality) $\sigma_0\in(0,\frac{\pi}{2})$. The point $C$ corresponds to the first zero of $\dot{u}$ on the geodesic segment, so $\sigma=\sigma_1=\frac{\pi}{2}$ at $C$. We can then calculate 
\be \floor{\frac{\sigma_0-\pi}{2\pi}}=\floor{\frac{\sigma_1-\pi}{2\pi}}=-1,\label{floors-equal}\ee
and so 
\begin{eqnarray}
\Delta\tau_{BC}&=&\sqrt{2}(\arctan\alpha - \arctan\beta)-\frac{\kappa}{2(\kappa^2+\lambda^2)^{1/2}}\left(\frac{\pi}{2}-\sigma_0\right). \label{dtau-a-c1} 
\end{eqnarray} 
where 
\be 
\alpha = \frac{b+c}{\sqrt{b^2-c^2}},\quad \beta = \frac{b\tan\sigma_0/2+c}{\sqrt{b^2-c^2}}.
\label{al-be}
\ee 
It is straightforward to see that 
\be 1<\alpha = \left(\frac{b+c}{b-c}\right)^{1/2}<+\infty \label{al-bounds1} \ee
and 
\be 0 < \beta < \alpha, \label{al-bounds2} \ee
and so monotonicity of the arctan function yields
\be 0<\arctan\alpha-\arctan\beta<\frac{\pi}{2}. \label{arctan-bounds} 
\ee
This provides the required information regarding branches of the tan function to apply the arctan addition formula and so obtain (after some manipulations)
\be 
\Delta\tau_{BC}(\kappa,\lambda)=\sqrt{2}\arctan\mu + \frac{\kappa}{2(\kappa^2+\lambda^2)^{1/2}}\left(\sigma_0-\frac{\pi}{2}\right),
\label{dt-a-c-1}
\ee
where 
\be \mu = \left(\frac{1-\tan\frac{\sigma_0}{2}}{1+\tan\frac{\sigma_0}{2}}\right)\left(\frac{1+u_1}{1+u_2}\right)^{1/2}.
\label{mu-def}
\ee
Now take the derivative with respect to $\Lambda$ (\textit{cf.} (\ref{cap-lam-def})) and write
\be \pd{}{\Lambda}\left\{\Delta\tau_{BC}(\kappa,\lambda)\right\}= \frac{\sqrt{2}}{2\mu(1+\mu^2)}\pd{\mu^2}{\Lambda}+\frac{\kappa}{2(\kappa^2+\lambda^2)^{1/2}}\pd{\sigma_0}{\Lambda}-\frac{\kappa}{2(\kappa^2+\lambda^2)^{3/2}}\left(\sigma_0-\frac{\pi}{2}\right). 
\label{dtau-deriv1}
\ee
The last term here is clearly positive since $\sigma_0<\pi/2$, and the first two have the advantage of involving only terms that are algebraic in $\kappa$ and $\lambda$. The derivatives are most readily calculated by writing the relevant functions in terms of $u_1$ and $u_2$:
\begin{eqnarray}
\mu^2 &=& \left(\frac{u_2-u}{u_2+1}\right)\left(\frac{1+u_1}{u-u_1}\right),\label{mu-in-u}\\
\sigma_0 &=& \arcsin\left(\frac{2u-u_1-u_2}{u_2-u_1}\right). \label{arcsin-sig0}
\end{eqnarray}
The $\Lambda$-derivatives are then calculated using (\ref{u1-dlambda},\ref{u2-dlambda}). We then rewrite the sum of the first two terms in (\ref{dtau-deriv1}) in terms of $\kappa$ and $\lambda$. The resulting expression can then be shown to be positive by using the inequality (\ref{l3-bound}). This proves part (i) of the statement. 

To prove part (ii), we calculate $\Delta\tau_{BD}$ directly. As above, we can take $k_\tau=0$, and set $s=s_0=0$ at $B$ and $s=s_1$ at $D$. Since $B$ (respectively $D$) is a local maximum (minimum) of $\tau$, whereat $r$ is increasing (decreasing), we must have 
\begin{eqnarray} u(s_0)&=&\frac{b-2}{2}+\frac{c}{2}\sin\sigma_0=R^2, \label{sigma0-backC-1}\\ 
\dot{u}(s_0)&=&\kappa c \cos\sigma_0 >0, \label{sigma0-backC-2}
\end{eqnarray}
and
\begin{eqnarray} u(s_1)&=&\frac{b-2}{2}+\frac{c}{2}\sin\sigma_1=R^2, \label{sigma1-backC-1} \\ 
\dot{u}(s_1)&=&\kappa c \cos\sigma_1 <0. \label{sigma1-backC-2} 
\end{eqnarray}
It follows that $\sigma_1=\pi-\sigma_0$ and (from above) $0<\sigma_0<\frac{\pi}{2}$. Noting that $\tan\frac{\sigma_1}{2}=\cot\frac{\sigma_0}{2}$, we can simplify by applying the arctan summation formula and thereby obtain the stated formula for $\Delta\tau_{BD}$. The arcsin term arises more readily from (\ref{sigma0-backC-1}) and (\ref{sigma1-backC-1}). The decrease with respect to $R$ arises by a straightforward calculation, as do the limits quoted. The bounds (\ref{eq:del-tau-bounds}) follow from these limits by monotonicity. \qed

\begin{corollary} The elapse of $\tau$ on a $\supset\!-$shaped transit at $R$, $\gamma_{[B,D]}$, with parameters $\kappa>0$ and $\lambda=0$ is a negative, increasing function of $\kappa$ on $(0,+\infty)$ and
\begin{eqnarray}
\lim_{\kappa\to +\infty}\Delta\tau_{BD}&=&0,\label{klim-at-hor}\\
\lim_{\kappa\to 0^+}\Delta\tau_{BD}&=&(\sqrt{2}-2)\frac{\pi}{4}=-\frac{\tau_*}{\sqrt{2}}.\label{klim-at-inf}
\end{eqnarray}
\qed
\end{corollary}


\begin{comments} Lemma \ref{lem:back-c-shaped} shows that we can decrease the value of $\tau$ by following a $\supset\!-$shaped transit at $r=R$ for any value of $R>1$. The greater the value of $R$, the greater the decrease in $\tau$. There is a limiting value for this decrease of $(\sqrt{2}-2)\pi/4\simeq -0.4601$. This provides some quantitative support for the observation that we like segments at large values of $r$. The proposition also provides more or less complete information on $\supset\!-$shaped segments and the elapse of $\tau$ along these segments, with the useful fact that this elapse is minimised along planar segments (i.e.\ those with $\lambda=0$). It is also useful to establish the separation in $\tau$ of the endpoints of successive \backc transits on a single null geodesic. This is the content of the following lemma. Additionally, this lemma provides the last piece of information required to determine a useful picture of generic non-monotone geodesics. See Figure \ref{fig:lem8fig}. \end{comments}

\begin{lemma}\label{lem:tau-seps-generic}
On a null geodesic $\gamma$ satisfying (\ref{l3-bound}), let $X,B$ be successive local maximum points of $\tau$, let $Y,C$ be successive maximum points of $r$, let $Z,D$ be successive local minimum points of $\tau$, let $A$ be the minimum point of $r$ on the segment $\gamma_{[Y,C]}$ and let $F$ be the next minimum point of $r$ to the future of $A$ on $\gamma$  so that 
\be s(X)<s(Y)<s(Z)<s(A)<s(B)<s(C)<s(D)<s(F). \label{eq:successive}\ee
Then:
\begin{itemize}
    \item[(i)]
\be \tau(D) > \tau(X). \label{eq:tau-sep-successive} \ee
\item[(ii)] There exist points $E_-\in\gamma_{(A,B)}$ and $E_+\in\gamma_{(D,F)}$ with $\tau(E_-)=\tau(E_+)$ and $r(E_-)=r(E_+)$. Furthermore, 
\be \tau(E_-)=\tau(E_+)=\tau(C). \label{tauE-tauC} \ee
\end{itemize}
\end{lemma}

\startproof For part (i), our aim is to show that $\Delta\tau_{XD}>0$. We have $\Delta\tau_{XD}=\Delta\tau_{XB}+\Delta\tau_{BD}$, the latter term being the elapse of $\tau$ on a \backc transit. With the obvious meanings of $\sigma_X$ and $\sigma_B$, we must have (\textit{cf.} (\ref{u-dot-sol}))
\be \sin\sigma_X=\sin\sigma_B,\quad \cos\sigma_X=\cos\sigma_B>0, \label{sig-X-B} \ee
where $\sigma_B$ is the minimal value of $\sigma>\sigma_X$ for which these equalities hold. Thus $\sigma_B=\sigma_X+2\pi$, and we can use (\ref{tau-sol}) to calculate
\be \Delta\tau_{XB}=\left(\sqrt{2}-\frac{\kappa}{(\kappa^2+\lambda^2)^{1/2}}\right)\pi>0. \label{tau-XB} \ee
Then 
\begin{eqnarray}
\Delta\tau_{XD} &=& \Delta\tau_{XB}+\Delta\tau_{BD} \nonumber \\
&=& \pi(\sqrt{2}-\frac{\kappa}{(\kappa^2+\lambda^2)^{1/2}}) + \Delta\tau_{BD} \nonumber \\
&\geq& \pi(\sqrt{2}-\frac{\kappa}{(\kappa^2+\lambda^2)^{1/2}}) +\frac{\pi}{4}(\sqrt{2}-2) >0,
\label{del-tau-XD} 
\end{eqnarray}
where we have used the bounds (\ref{eq:del-tau-bounds}) for \backc transits. This completes the proof of part (i).

For part (ii), existence of the points $E_\pm$ which project to the same point in the $r-\tau$ plane follows from a straightforward continuity argument. To see that $\tau(C)=\tau(E_\pm)$, we use a change of variable to show that $\tau(C)-\tau(E_-)=\tau(E_+)-\tau(C)$ (compare the first step in the proof of Lemma \ref{lem:back-c-shaped}). Since $\tau(E_-)=\tau(E_+)$, the equality (\ref{tauE-tauC}) follows.
\qed

\begin{comments} While only some of the points mentioned in Lemma \ref{lem:tau-seps-generic} play a role in the proof, it is convenient to label the other points. The lemma provides the ordering (in $\tau$) of key points on typical null geodesics. The ordering in $s$ of (\ref{eq:successive}) follows from part (iii)-(b) of Proposition \ref{prop:key-ngs}. It follows from this, from Lemma \ref{lem:back-c-shaped} and from Lemma \ref{lem:tau-seps-generic} that Figure \ref{fig:lem8fig} provides an accurate picture of the projection into the $r-\tau$ plane of a generic null geodesic satisfying  (\ref{l3-bound}). This figure provides useful intuition for the results that follow, and we will use it for reference below.
\end{comments}

\subsubsection{Finding better segments}

Our aim is to establish the fact that the segments of Proposition \ref{prop:segments-optimal} are better than other segments. We begin by setting aside: (i) segments that are initially ingoing ($\dot{r}(0)<0$; Lemma \ref{lem:rdot-neg-useless}); (ii) the monotone geodesics described in part (iii)-(c) of Proposition \ref{prop:key-ngs} (Lemma \ref{lem:mono-useless-1}) and (iii) remaining segments with $\lambda\neq 0$ (Lemma \ref{lem:X}). Crucially, this reduces the number of parameters to be considered to just one $(\kappa)$. The proofs of these lemmas are somewhat detailed, but do not introduce any concepts needed in the remainder of the paper. Hence we give the proofs in an appendix. 

\begin{lemma}\label{lem:rdot-neg-useless}
Let $\gamma_a$ be a future pointing null geodesic with $\gamma_a(0)=P$ such that $r_a(0)>1$ and $\dot{r}_a(0)<0$. Then there exists a null geodesic $\gamma_b$ with $\gamma_b(0)=P$, $\dot{r}_b(0)> 0$ such that for any point $Q$ to the future of $P$ on $\gamma_a$ there is a point $Q'$ to the future of $P$ on $\gamma_b$ with $\tau(Q')<\tau(Q)$ and $r(Q') \geq r(Q)$.
\qed
\end{lemma}

At first sight, it seems intuitive that no segment of the optimal path should be of the form described in part (iii)-(c) of Proposition \ref{prop:key-ngs}. On such segments, $\tau$ is monotone increasing. But this increase may come with the benefit of increasing $r$ in such a way as to make a favourable trade-off: the \textit{next} segments, at `large' $r$ may allow for a subsequent decrease of $\tau$ on the next segment substantial enough to out-weigh the increase on the monotone increasing segment. Fortunately, we can rule out the presence of the (iii)-(c) segments on the optimal path without too much difficulty:

\begin{lemma}\label{lem:mono-useless-1} 
Let $\gamma_a$ be a future pointing null geodesic  with $\dot{r}(0)\geq$ and with $\lambda>\bar{\lambda}(\kappa)$, and let $P=\gamma_a(0)$. Let $\gamma_b$ be the unique future pointing null geodesic with $\lambda=0$, $\dot{r}(0)>0$, $\gamma_b(0)=P$ and with $\kappa|_{\gamma_b}=\kappa|_{\gamma_a}$.
Then for every $Q=\gamma_a(s), s> 0$ to the future of $P$ on $\gamma_a$, there exists $s'>0$ and a point $Q'= \gamma_b(s')$ to the future of $P$ on $\gamma_b$ with the property that 
\be \left.r\right|_{P}<\left.r\right|_{Q},\label{r-more}\ee
and
\be \left.\tau\right|_{P}>\left.\tau\right|_{Q}.\label{tau-less}\ee \qed
\end{lemma}

The next result shows that initially outgoing ($\dot{r}(0)\geq0$), non-monotone ($\lambda\leq\overline{\lambda}(\kappa)$) segments with $\lambda>0$ can be replaced with better segments with $\lambda=0$. 

\begin{lemma}\label{lem:X} 
Let $\gamma_a$ be a future pointing null geodesic  with $\dot{r}(0)\geq$ and with $0<\lambda\leq\bar{\lambda}(\kappa)$, and let $P=\gamma_a(0)$. Let $\gamma_b$ be the unique future pointing null geodesic with $\lambda=0$, $\dot{r}(0)>0$, $\gamma_b(0)=P$ and with $\kappa|_{\gamma_b}=\kappa|_{\gamma_a}$.
Then for every $Q=\gamma_a(s), s> 0$ to the future of $P$ on $\gamma_a$, there exists $s'>0$ and a point $Q'= \gamma_b(s')$ to the future of $P$ on $\gamma_b$ with the property that 
\be \left.r\right|_{P}<\left.r\right|_{Q},\label{r-more-x}\ee
and
\be \left.\tau\right|_{P}>\left.\tau\right|_{Q}.\label{tau-less-x}\ee \qed
\end{lemma}

\subsubsection{Properties of the envelope}

At this stage, we need to introduce the envelope $\env_{P}$ of the family ${\mathcal{F}}_P$ of future pointing null geodesic segments from $P$ with $\lambda=0$, and establish some properties of this set. 

\begin{definition}\label{cal-f-p} Let $P$ be a point with $r_0=r|_P\geq 1$ and let $u_0=r_0^2$. Then the family of curves ${\mathcal{F}}_P$ is defined to be the set of all semi-infinite segments $\gamma_{[P,+\infty)}$ of future-pointing null geodesics $\gamma:\mathbb{R}\to M$ for which $\lambda=0$. This family is indexed by the parameter $\kappa\in[\kappa_1(u_0),\kappa_2(u_0)]$ and the choice of the sign of $\dot{r}(0)$. We refer to those segments with $\dot{r}(0)\geq0$ as initially outgoing and those with $\dot{r}(0)<0$ as initially ingoing, and define the corresponding families
\begin{eqnarray}
\fp_P^+=\{\gamma_{[P,+\infty]}\in\fp_P:\dot{r}(0)\geq0\}, \label{fp-pos}\\
\fp_P^-=\{\gamma_{[P,+\infty]}\in\fp_P:\dot{r}(0)<0\}. \label{fp-neg}\\
\end{eqnarray}
\end{definition}

From Proposition \ref{prop:key-ngs}, we have an explicit description of the family $\fp_P$. With $u_0=u(P)$ and $\tau_0=\tau(P)$, we write (\ref{u-sol}) and (\ref{tau-sol}) in the form 
\be u = \alpha(s,k;u_0,\tau_0),\quad \tau=\beta(s,\kappa;u_0,\tau_0) \label{al-be-env} \ee
where we set $\lambda=0$ and where the constants of integration $\sigma_0$ and $k_\tau$ (see (\ref{sig-def}), (\ref{const})) are chosen so that 
\be \alpha(0,\kappa;u_0,\tau_0) = u_0,\quad \beta(0,\kappa;u_0,\tau_0)=\tau_0 \quad\hbox{ for all }\kappa\in[\kappa_1(u_0),\kappa_2(u_0)]. \label{env-initial} \ee
Other restrictions will apply to characterise members of $\fp_P^\pm$: we use the notation $(u,\tau)=(\alpha^\pm,\beta^\pm)$ for members of these families. 
We note that $\alpha,\beta$ are $C^1$ functions of their arguments on the relevant domains. 

Define 
\be \Omega_{u_0} = \{(s,\kappa)\in\mathbb{R}^2: s\in[0,+\infty), \kappa\in[\kappa_1(u_0),\kappa_2(u_0)]\}. 
\label{omega-def}
\ee
Then we can write
\be \fp_P=\{(\alpha(s,\kappa;u_0,\tau_0),\beta(s,\kappa;u_0,\tau_0)):(s,\kappa)\in\Omega_{u_0}
\}. \label{fp-form}\ee
 The envelope of $\fp_P$ is given by \cite{bruce1992curves}
\be \env_{P}=\{(u,\tau)\in\mathbb{R}^2:u=\alpha(s,\kappa), \tau=\beta(s,\kappa), \Delta(s,\kappa)=0, (s,\kappa)\in\Omega_{u_0}\}, \label{env-def}
\ee
where 
\be \Delta(s,\kappa) =
\pd{\alpha}{s}\pd{\beta}{\kappa}-\pd{\alpha}{\kappa}\pd{\beta}{s},
\label{eq:delta-def} 
\ee
and where for convenience we omit the functional dependence on $u_0,\tau_0$. Using the notation $(u,\tau)=(\alpha^\pm,\beta^\pm)$ to represent initially outgoing ($+$) and initially ingoing ($-$) geodesics, we will write (\ref{fp-pos}) and (\ref{fp-neg}) as 
\be \fp_P^\pm=\{(\alpha^\pm(s,\kappa;u_0,\tau_0),\beta^\pm(s,\kappa;u_0,\tau_0)):(s,\kappa)\in\Omega_{u_0}
\}, \label{fp-pm-form}\ee
with respective envelopes denoted by $\env_P^\pm$.

\begin{figure}
    \centering
    \includegraphics{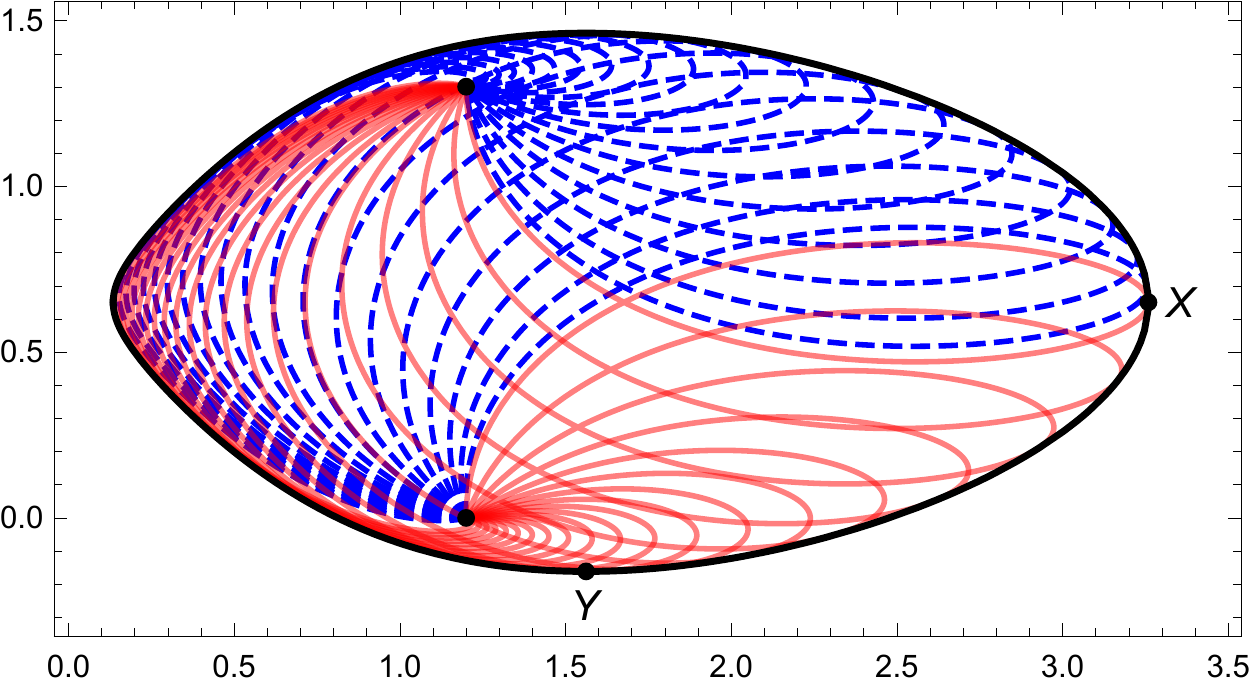}
    \caption{The figure shows (members of) the family of future pointing null geodesics emerging from a point in the $r-\tau$ plane. On each geodesic, the parameter increases in the roughly clockwise sense (see Figure \ref{fig:lem8fig}). The initial point is $P:(r,\tau)=(1.2,0)$. The initially outgoing geodesics are coloured red (solid curves) and the initially ingoing geodesics are coloured blue (dashed curves). Both families reconverge to the future at $(r,\tau)=(1.2,(\sqrt{2}-1)\pi)$. Portions of the envelope $\env_A$ are also shown. The (approximately) elliptical boundary curve comprises the union of $\env_{P,1}^+$ (lower half) and $\env_{P,1}^-$ (upper half). The initial point formally corresponds to $\env_{P,0}$, and the terminal point of the segments is $\env_{P,2}$. See Proposition \ref{prop:env-structure}. The ``south-east" portion of the envelope, $\ese{P}$, corresponds to the segment running clockwise from $X$ to $Y$.}
    \label{fig:envpic}
\end{figure}

A lengthy calculation yields a very satisfying conclusion regarding the description of the envelope, which turns out to be remarkably simple. See Figure \ref{fig:envpic}.

\begin{proposition}\label{prop:env-structure}
The envelope equation $\Delta(s,\kappa)=0$ is equivalent to 
\be
\sigma = \sigma_0 + n\pi,\quad n\in\mathbb{N}, \label{eq:env-equiv} 
\ee
where, as in (\ref{sig-def}) with $\lambda=0$, 
\be 
\sigma = 2\kappa s + \sigma_0 \label{eq:sig-def-state} 
\ee
and $\sigma_0$ is a $\kappa-$dependent constant of integration. 
Thus the envelope $\env_P$ has the form 
\be \env_{P}=\bigcup_{n\in\mathbb{N}}\env_{P,n},
\label{app:env-sol-n-2}
\ee
where 
\be \env_{P,n}=\{(u,\tau)=(\alpha(\frac{n\pi}{2\kappa},\kappa;u_0),\beta(\frac{n\pi}{2\kappa},\kappa;u_0,\tau_0), \kappa\in[\kappa_1(u_0),\kappa_2(u_0)]\}. 
\label{app:env-sol-n-1}
\ee

The branches of the envelope generated by initially outgoing (+) (respectively ingoing (-)) geodesics are defined by
\be \env_{P,n}^\pm=\{(u,\tau)=(\alpha^\pm(\frac{n\pi}{2\kappa},\kappa;u_0),\beta^\pm(\frac{n\pi}{2\kappa},\kappa;u_0,\tau_0), \kappa\in[\kappa_1(u_0),\kappa_2(u_0)]\}. 
\label{app:env-pm-sol-n}
\ee

The branches with $n=2m, m\in\mathbb{N}$ comprise single points:
\be \env_{P,2m} = \{(u_0,\tau_0+(\sqrt{2}-1)m\pi)\},\quad m\in\mathbb{N}. \label{app:env-n-even-1}
\ee

The branches with $n=2m-1, m\in\mathbb{N}$ may be written in the form of the parametrized curves
\be \env_{P,2m-1}^\pm=\{(u,\tau)=(u_{\ce_{2m-1}}^\pm(\kappa;u_0),\tau_{\ce_{2m-1}}^\pm(\kappa;u_0,\tau_0)):\kappa\in[\kappa_1(u_0),\kappa_2(u_0)]\} \label{env1-pm-sol1} \ee
where
\be
u_{\ce_{2m-1}}^\pm(\kappa;u_0) =1+\frac{2\sqrt{2}}{\kappa}-u_0,\quad m\in\mathbb{N}\label{eq:ue-n-odd}
\ee
and 
\be
\tau_{\ce_{2m-1}}^\pm(\kappa;u_0,\tau_0) =\tau_{\ce_1}^\pm(\kappa;u_0,\tau_0) + (\sqrt{2}-1)(m-1)\pi,\quad m\in\mathbb{N}. 
\label{eq:tau-en-odd}\ee
with
\begin{eqnarray}
\tau_{\ce_1}^+(\kappa;u_0,\tau_0) &=& \tau_0-\frac{\pi}{2} + \sqrt{2}\arctan\left(\frac{\sqrt{2}\kappa+1}{(u_0(1-u_0)\kappa^2+2\sqrt{2}u_0\kappa-1)^{1/2}}\right), \label{eq:tau-e1-pos} \\
\tau_{\ce_1}^-(\kappa;u_0,\tau_0) &=&
\tau_0 + \frac{(2\sqrt{2}-1)}{2}\pi-\sqrt{2}\arctan\left(\frac{\sqrt{2}\kappa+1}{(u_0(1-u_0)\kappa^2+2\sqrt{2}u_0\kappa-1)^{1/2}}\right). \label{eq:tau-e1-neg}
\end{eqnarray}
\end{proposition}

\startproof Our starting point is the general solution of the geodesic equations given in Proposition \ref{prop:key-ngs}. The restriction to geodesics with $\lambda=0$ simplifies considerably some terms that arise in the solution, and so we give them here. We have
\begin{eqnarray} 
\alpha(s,\kappa;u_0)&=&\frac12(b-2)+\frac{c}{2}\sin\sigma, \label{app:u-sol}\\
\beta(s,\kappa;u_0,\tau_0) &=& \tau_0+\hat\beta(\sigma,\kappa;u_0)-\hat\beta(\sigma_0,\kappa;u_0),\label{app:tau-sol}
\end{eqnarray}
where 
\be \hat\beta(\sigma,\kappa;u_0) = \sqrt{2}\arctan\left(\frac{c+b\tan\sigma/2}{\sqrt{b^2-c^2}}\right)-\frac{\sigma}{2}+\sqrt{2}\pi\left(1+\floor{\frac{\sigma-\pi}{2\pi}}\right), 
\label{app:be0-def}
\ee
and with
\begin{eqnarray}
{u}_1(\kappa) &=& \frac{2\sqrt{2}+\kappa-(\kappa^2+4\sqrt{2}\kappa+4)^{1/2}}{2\kappa},\label{u1bar-def} \\ 
{u}_2(\kappa) &=& \frac{2\sqrt{2}+\kappa+(\kappa^2+4\sqrt{2}\kappa+4)^{1/2}}{2\kappa}.\label{u2bar-def}
\end{eqnarray}
We also have
\begin{eqnarray}
b &=& 2 + u_1 + u_2 = 3 + \frac{2\sqrt{2}}{\kappa},\label{app:b-def}\\
c &=& u_2-u_1 = \frac{(\kappa^2+4\sqrt{2}\kappa+4)^{1/2}}{\kappa},\label{app:c-def}\\
\sqrt{b^2-c^2} &=& \frac{2(\sqrt{2}k+1)}{k}, \label{app:root-def}\\
\sigma &=& 2\kappa s + \sigma_0. \label{app:sig-def}
\end{eqnarray}
In order to satisfy the initial condition $u(0)=u_0$, we must have
\begin{eqnarray}
\sin\sigma_0 &=& \frac{u_0-u_1-u_2}{u_2-u_1} = \frac{(2u_0-1)\kappa-2\sqrt{2}}{(\kappa^2+4\sqrt{2}\kappa+4)^{1/2}}.\label{app:sig0}
\end{eqnarray}

Now consider the envelope equation
\be \pd{\alpha}{s}\pd{\beta}{\kappa}-\pd{\alpha}{\kappa}\pd{\beta}{s}=0. \label{app:env-eq}
\ee
We can simplify this equation as follows. Define
\begin{eqnarray}
\bar{\alpha}(\sigma,x) &=& \alpha(s,\kappa) = \frac12+\frac{\sqrt{2}}{x}+\frac{h(x)}{2x}\sin\sigma, \label{app:albar-def}\\
\bar{\beta}(\sigma,x) &=& \beta(s,\kappa) = \tau_0+\hat\beta(\sigma,x)-\hat\beta(\sigma_0(x),x), \label{app:bebar-def}
\end{eqnarray}
where $\sigma$ is defined in (\ref{app:sig-def}), $x=\kappa$ and 
\be h(x)=\sqrt{x^2+4\sqrt{2}x+4}.\label{eq:h-def}\ee 
Note then 
\be \sin\sigma_0(x) = \frac{(2u_0-1)x-2\sqrt{2}}{h(x)}.\label{eq:sig0-x}\ee
This amounts to a reparametrisation of the solutions of the null geodesic equations. A straightforward calculation shows that 
\be \pd{\alpha}{s}\pd{\beta}{\kappa}-\pd{\alpha}{\kappa}\pd{\beta}{s}=2\kappa\left( \pd{\bar{\alpha}}{\sigma}\pd{\bar{\beta}}{x}-\pd{\bar{\alpha}}{x}\pd{\bar{\beta}}{\sigma}\right), \label{app:env-alt} \ee
and so the envelope equation is equivalent to vanishing of the right hand side of (\ref{app:env-alt}). This simplifies matters, as it essentially means that we can ignore $\kappa-$derivatives of $\sigma$ when evaluating the left hand side of (\ref{app:env-eq}). Calculating the relevant derivatives of $\bar{\alpha}$ is straightforward. To calculate the $\sigma$ derivative of $\bar{\beta}$, we can use the geodesic equation (\ref{tau-d}) and the definitions above to write down the identity 
\be 2\kappa\pd{\hat\beta(\sigma,x)}{\sigma}=\frac{(1-\bar{\alpha}(\sigma,x))x+\sqrt{2}}{1-\bar{\alpha}(\sigma,x)}.\label{eq:be-d-sig}\ee 
From this we can write down 
\begin{eqnarray}
\pd{\bar{\beta}(\sigma,x)}{\sigma} &=& \frac{(1-\bar{\alpha}(\sigma,x))x+\sqrt{2}}{2x(1-\bar{\alpha}(\sigma,x))},
\label{eq:be-bar-sig}\\
\pd{\bar{\beta}(\sigma,x)}{x} &=& \partial_2\hat{\beta}(\sigma,x)-\partial_2\hat{\beta}(\sigma_0(x),x) - \frac{(1-\bar{\alpha}(\sigma_0(x),x))x+\sqrt{2}}{2x(1-\bar{\alpha}(\sigma_0(x),x))}\sigma_0'(x), \label{eq:be-bar-x}
\end{eqnarray}
where (as usual) $\partial_2A(u(v),v)\equiv \lim_{z\to0}(A(u(v),v+z)-A(u(v),v))/z$. Calculating these derivatives allows us to evaluate the right hand side of (\ref{app:env-alt}). This yields a lengthy expression, but collecting terms that depend only on $\sigma$ reveals an unexpected result: the envelope equation has the essentially explicit form 
\be \tan\sigma = \frac{h}{4}\frac{2\sqrt{2}x\cos\sigma_0-xh(x-h\sin\sigma_0)\sigma_0'}{3x+2\sqrt{2}+h\sin\sig_0}. 
\label{eq:env-sol1}
\ee
Using (\ref{eq:sig0-x}) and its derivative reveals a second unexpected and welcome result: (\ref{eq:env-sol1}) simplifies to yield a remarkably simple form for the envelope equation:
\be \tan\sigma = \tan\sigma_0. \label{eq:env-sol2}\ee
This has the solution $\sigma=\sigma_0+n\pi, n\in\mathbb{Z}$ and so using (\ref{app:sig-def}) and noting that $s\geq 0$, we have 
\be s = \frac{n\pi}{2\kappa},\quad n\in \mathbb{N}. \label{env:s-sol} \ee
This establishes (\ref{eq:env-equiv}), (\ref{app:env-sol-n-2}) and (\ref{app:env-sol-n-1}).

For $n=2m, m\in\mathbb{N}$ (so that $\sigma=\sigma_0+2m\pi$), we can use (\ref{app:u-sol})-(\ref{app:be0-def}) to show that  
\be \env_{P,2m} = \{(u_0,\tau_0+(\sqrt{2}-1)m\pi)\},\quad m\in \mathbb{N}, \label{app:env-n-even}
\ee
so that these branches of the envelope comprise single points as claimed.

For $n=2m-1,m\in\mathbb{N}$, we have $\sigma=\sigma_0+(2m-1)\pi$, and (\ref{eq:ue-n-odd}) is readily established. 

The branches $\env_{P,1}^\pm$ are of particular interest. On these branches, we have $\sigma=\sigma_0+\pi$. To obtain (\ref{eq:tau-e1-pos}) and (\ref{eq:tau-e1-neg}), we note that for initially outgoing geodesics we have
$\cos\sigma_0\geq 0$, while $\cos\sigma_0<0$ for initially ingoing geodesics. Without loss of generality, we an choose $\sigma_0\in[-\pi/2,\pi/2]$ and $\sigma_0\in(\pi/2,3\pi/2)$ for initially outgoing and initially ingoing geodesics respectively. Then using (\ref{app:tau-sol}) and (\ref{app:bebar-def})
we can write 
\be \bar{\beta}(\sigma_0+\pi,x) = \tau_0 - \frac{\pi}{2} + \sqrt{2}\arctan\left(\frac{c-b\cot\sigma_0/2}{\sqrt{b^2-c^2}}\right)-
\sqrt{2}\arctan\left(\frac{c+b\tan\sigma_0/2}{\sqrt{b^2-c^2}}\right)
+\sqrt{2}\pi\theta(\sigma_0),
\label{eq:bebar-env1}
\ee
where $\theta$ is the Heaviside step function. We then use the arctan addition formula, choosing the relevant branches of the tangent functions carefully to ensure that the result does indeed yield points on the envelopes of $\fp_P^\pm$ (which we recall are points on curves of the family). This amounts to making the correct choice of the integer $n=n^\pm$ in the formula
\be \bar{\beta}(\sigma_0+\pi,x) =\tau_0-\frac{\pi}{2}+ \sqrt{2}\arctan\left( \frac{2(\sqrt{2}x+1)}{xh\cos\sigma_0}\right) +\sqrt{2}n\pi. 
\label{eq:bebar-env2}
\ee
For initially outgoing geodesics, we have $\cos\sigma_0\geq 0$ and so (from (\ref{app:sig0}))
\be \cos\sigma_0(x) = \sqrt{1-\sin^2\sigma_0} = \frac{2}{h}(u_0(1-u_0)x^2+2\sqrt{2}u_0x-1)^{1/2},
\label{eq:cos-env-pos}
\ee
and the correct choice is $n=n^+=0$. This yields (\ref{eq:tau-e1-pos}).
For initially ingoing geodesics, the expression above for $\cos\sigma_0$ changes sign, and we require $n=n^-=1$. This yields (\ref{eq:tau-e1-neg}). 

It is straightforward to verify (\ref{eq:tau-en-odd}) given (\ref{eq:tau-e1-pos}) and (\ref{eq:tau-e1-neg}): successive branches of the envelope with $n$ odd are obtained by a translation in the $\tau$ direction. 
\qed

The following technical details relating to $\env_{P,1}^+$ will be of use below; they are easily verified. 

\begin{lemma}\label{lem:env-tech}
Let $u_0\geq1$ and let $\tau_0\in\mathbb{R}$. 
\bi 
\item[(i)] Define
\be \kappa_{\rm{min}}(u_0) = \frac{\sqrt{2}}{u_0}. \label{eq:kap-min-def0} \ee 
Then
\be \kappa_1(u_0) < \kappa_{\rm{min}}(u_0) <\kappa_2(u_0). \label{eq:kap-min}\ee \item[(ii)] The function $\kappa\mapsto \tau_{\env_1}^+(\kappa;u_0,\tau_0)$ is decreasing on $[\kappa_1(u_0),\kappa_{\rm{min}}(u_0))$ and is increasing on 
$(\kappa_{\rm{min}}(u_0),\kappa_2(u_0)]$. 
\item[(iii)] The minimum of $\tau_{\env_1}^+$ on $[\kappa_1(u_0),\kappa_2(u_0)]$ is  
\be \tau_{\env_1,\rm{min}}^+ :=\tau_{\env_1}^+(\kappa_{\rm{min}}(u_0);u_0,\tau_0) = \tau_0-\frac{\pi}{2}+\sqrt{2}\arctan\left(\sqrt{\frac{u_0+2}{u_0}}\right).
\label{eq:tau-env-min}
\ee
This minimum occurs where the geodesic with $\kappa=\sqrt{2}/u_0$ meets the envelope, and at this point, $u=1+u_0$.
\ei
\qed
\end{lemma}

The envelope plays a key role in the construction of the optimal path. This role arises from the following results, the first of which says that we can always find better segments with their endpoints on the ``south-east" portion of the envelope. This is the section of the envelope bounded by the points $X$ and $Y$ in  Figure \ref{fig:envpic}. 

\begin{definition}\label{def:ese} Given a point $P$ with $r(P)\geq1$, we define the \textbf{$\se-$envelope of $P$} to be
\be \ese{P} = \{(u,\tau)=(u_{\env_1}^+(\kappa;u_0,\tau_0),\tau_{\env_1}^+(\kappa;u_0,\tau_0)): \kappa\in[\kappa_1(u_0),\kappa_{\rm{min}}(u_0)]\}. \label{eq:ese-def} \ee
Thus the $\se-$envelope of $P:(u_0,\tau_0)$ is characterised by 
\begin{eqnarray}
u=u(\kappa;u_0,\tau_0) &=& 1+\frac{2\sqrt{2}}{\kappa}- u_0, \label{eq:u-se-env}\\
\tau=\tau(\kappa;u_0,\tau_0) &=& \tau_0-\frac{\pi}{2}+\sqrt{2}\arctan\left(\frac{\sqrt{2}\kappa+1}{(u_0(1-u_0)\kappa^2+2\sqrt{2}u_0\kappa-1)^{1/2}}\right), 
\label{eq:tau-se-env} \\
\end{eqnarray}
where
\be 
\frac{\sqrt{2}u_0-(u_0^2+u_0)^{1/2}}{u_0(u_0-1)} \leq \kappa \leq \frac{\sqrt{2}}{u_0}. 
\label{eq:se-k-bounds}
\ee
\end{definition}

\begin{lemma}\label{lem:envelope-better}
Let $\gamma_a$ be a future pointing null geodesic with $\gamma_a(0)=P$, with parameters $\lambda_a=0$ and (necessarily) $\kappa_a\in[\kappa_1(u_0),\kappa_2(u_0)]$ where $u_0=r_a(0)^2>1$, and with $\dot{r}_a(0)\geq0$. Let $Q=\gamma_a(s),s>0$ be a point to the future of $P$ on $\gamma_a$. Then there exists a future pointing null geodesic $\gamma_b$ with $\gamma_b(0)=P$ and a point $Q'\in \ese{P}$ to the future of $P$ on $\gamma_b$ such that the segment $\gamma_{b_{[P,Q']}}$ is better than the segment $\gamma_{a_{[P,Q]}}$. Furthermore, the geodesic $\gamma_b$ may be chosen with parameters $\lambda_b=0$, $\kappa_b\in[\kappa_1(u_0),\kappa_{\rm{min}}(u_0)]$.
\end{lemma}

\startproof
We refer to Figure \ref{fig:lem8fig} and Figure \ref{fig:envpic}. Since $\dot{r}_a(0)\geq0$ (i.e.\ $\gamma_a$ is an initially outgoing), the initial point $P$ lies on a segment of the form $\gamma_{[A,C]}$. Let $Q=\gamma_a(s), s>0$ with $u|_Q=u_*$ and assume without loss of generality that $Q$ does not lie on the envelope. 

If $u_*<1+u_0$, we can replace the segment $\gamma_{a,[P,Q]}$ with the better segment $\gamma_{b,[P,Q']}$ where we take $\gamma_b$ to be the geodesic with $\kappa=\kappa_{\rm{min}}(u_0)$ and we take $Q'$ to the point where this geodesic meets $\env_{P,1}^+$. From Lemma \ref{lem:env-tech}, this point has coordinates $(u,\tau)=(1+u_0,\tau_{\env_1,\rm{min}}^+)$. The point $Q$ sits above $\env_{P,1}^+$, along which $\tau\geq\tau_{\env_1,\rm{min}}^+$, and $u_*<u_0<1+u_0$. Therefore $\gamma_{b,[P,Q']}$ is indeed better than $\gamma_{a,[P,Q]}$. 

If $u_*\geq 1+u_0$, we drop vertically downwards from $Q$ to the unique point $Q'\in\env_{P,1}^+$ with $u|_{Q'}=u_*$. We need to verify that this point exists and has the properties mentioned. To do so, we define $\kappa_*$ by (\textit{cf.} (\ref{eq:ue-n-odd}) with $m=1$)
\be u_* = 1+\frac{2\sqrt{2}}{\kappa_*}-u_0, \label{eq:u-star-k-star}\ee
so that 
\be \kappa_* = \frac{2\sqrt{2}}{u_*+u_0-1}. \label{eq:k-star}\ee
With a little work we can verify that $\kappa_*\in(\kappa_1,\kappa_{\rm{min}}(u_0)]$, and so there is an initially outgoing future pointing null geodesic from $P$ with parameter $\kappa=\kappa_*$, and with $\kappa$ in the claimed range. By construction, this geodesic meets the envelope at the point $Q'$ and we have $u|_{Q'}=u|_Q$. The inequality $\tau|_{Q'}<\tau|_Q$ follows from the fact that the branch $\env_{P,1}^+$ of the envelope sits below the geodesics forming the family $\fp_P$. By nudging $Q'$ slightly we can produce a point $Q''$ on $\env_{P,1}^+$, which lies on an initially outgoing, future pointing null geodesic $\gamma_b$ from $P$ for which both $u|_{Q''}>u|_Q$ and $\tau|_{Q''}<\tau|_Q$. This yields a segment $\gamma_{b,[P,Q'']}$ that is better (and not just marginally better) than $\gamma_{a,[P,Q]}$.\qed

\begin{figure}
    \centering
    \includegraphics{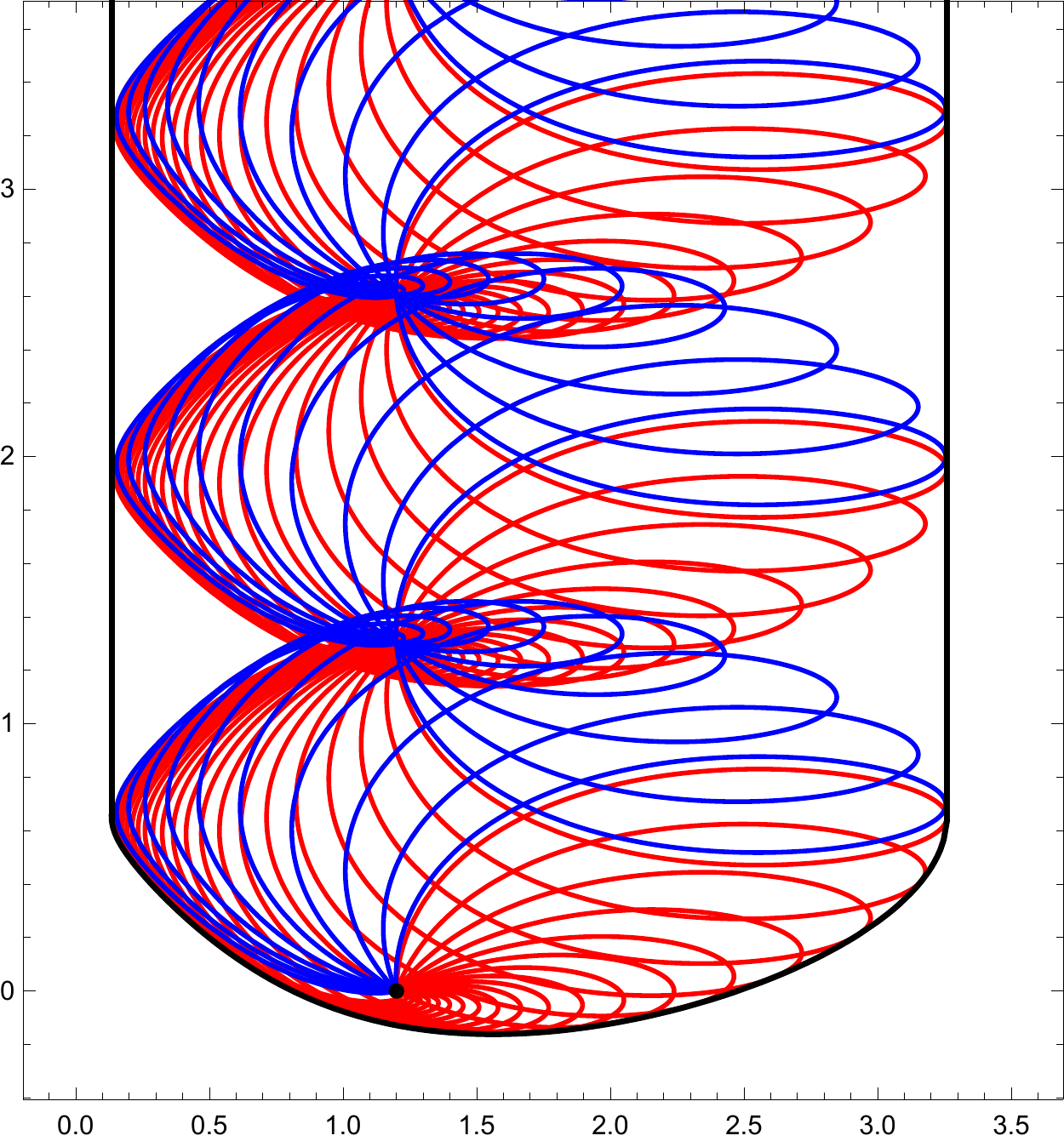}
    \caption{Future directed null geodesic segments from the initial point $P:(r,\tau)=(1.2,0)$. These semi-infinite segments are confined to the region bounded by the left- and right-hand vertical lines (corresponding to the minimum and maximum values of $r$ that can be attained along null geodesics from $P$) and by $\env_{P,1}^+$. The region corresponds to the convex hull of those boundaries. Segments coloured red are initially outgoing; those coloured blue are initially ingoing.} 
    \label{fig:con-hull}
\end{figure}

\begin{comments} We have made reference above to the branch $\env_{P,1}^+$ of the envelope sitting below the geodesics of the family $\fp_P$. This is evident from Figure \ref{fig:envpic}, but for clarity we note the following. Along future pointing null geodesics from $P$, we have 
\be \min_{\kappa\in[\kappa_1(u_0),\kappa_2(u_0)]}u_1(\kappa) \leq u \leq \max_{\kappa\in[\kappa_1(u_0),\kappa_2(u_0)]}u_2(\kappa),
\label{eq:up-max-min} \ee
which gives
\be u_{\rm{min}}(u_0):= u_1(\kappa_2(u_0))\leq u \leq u_2(\kappa_1(u_0))=:u_{\rm{max}}(u_0). \label{eq:up-bounds} 
\ee
 The geodesics share the initial point $P:(u_0,\tau_0)$, reconverge at the later time $\tau_0+(\sqrt{2}-1)\pi$ (corresponding to  $\env_{P,2}$) and each geodesic meets $\env_{P,1}^+$ exactly once: they do not cross and then re-cross the envelope. The geodesic with $\kappa=\kappa_*$ (defined in (\ref{eq:k-star})) meets the envelope at a point $Q_*$ at which $\tau=\tau_*<\tau_0$. So there is certainly one geodesic for which the point at which it meets $\env_{P,1}^+$ lies below $P$ in the $u-\tau$ plane. Appealing to continuous dependence of the geodesics on their parameters $(s,\kappa)$, we can conclude that the entirety of each future-pointing geodesic from $P$ sits above $\env_{P,1}^+$ in the $u-\tau$ plane (strictly above for all points with the exception of that unique point on each geodesic that meets $\env_{P,1}^+$). We can state this formally as follows (see Figure \ref{fig:con-hull}): 
 \end{comments}
 
 \begin{proposition}
 \label{prop:Fp-conv-hull} 
 $\fp_P$ is a subset of the convex hull of the set \be \env_{P,1}^+\cup \{(u_{\rm{min}}(u_0),\tau):\tau\geq T(u_0,\tau_0)\} \cup \{(u_{\rm{max}}(u_0),\tau):\tau\geq T(u_0,\tau_0)\}, 
 \label{eq:conv-hull} 
 \ee
 where $T(u_0,\tau_0)$ is the maximum of $\tau$ on $\env_{P,1}^+$:
 \be T(u_0,\tau_0) = \tau_{\env_1}^+ (\kappa_1(u_0);u_0,\tau_0) = \tau_0+(\sqrt{2}-1)\frac{\pi}{2}. 
 \label{eq:T}
 \ee
 Note that this quantity is independent of $u_0$. 
\end{proposition}

It follows from part (iii) of Lemma \ref{lem:env-tech} and Proposition \ref{prop:Fp-conv-hull} that the minimum of $\tau$ on the envelope is the minimum of $\tau$ taken over all segments from the point $P$ generating that envelope: 

\begin{corollary}
\label{corr:tau-min} Let $P$ satisfy $u(P)=u_0\geq 1$.
The minimum of the elapse of $\tau$ on all future pointing null geodesic segments $\gamma_{[P,Q]}$ is given by 
\be \Delta\tau_{\rm{min}}(u_0) = -\frac{\pi}{2} +\sqrt{2}\arctan\left(\sqrt{\frac{u_0+2}{u_0}}\right). \label{eq:tau-min} 
\ee
This is a negative, decreasing function of $u_0$ with 
\begin{eqnarray}
\lim_{u_0\to 1^+}\Delta\tau_{\rm{min}}(u_0) &=& \sqrt{2}\arctan\sqrt{2}-\frac{\pi}{2} \simeq -0.2198, \\ 
\lim_{u_0\to +\infty}\Delta\tau_{\rm{min}}(u_0) &=& (\sqrt{2}-2)\frac{\pi}{4} \simeq -0.4601.
\end{eqnarray}
\end{corollary}

Two final properties of the envelope are needed before we can give the proof of Proposition \ref{prop:segments-optimal}.

\begin{lemma}\label{lem:env-disjoint}
Let $P, P'$ be points with $\tau(P)=\tau(P')$ and $r(P)<r(P')$. Then the sets $\ese{P}$ and $\ese{P'}$ are disjoint. 
\end{lemma}

\startproof Let $u_0=r(P)^2$ and $u_0'=r(P')^2$, and let $\tau_0=\tau(P)=\tau(P')$. 

To prove the lemma, we must show that there does not exist a pair $k\in[\kappa_1(u_0),\kappa_{\rm{min}}(u_0)]$
and $k'\in[\kappa_1(u_0'),\kappa_{\rm{min}}(u_0')]$ for which 
\be u_{\ce_{1}}^+(k;u_0) = u_{\ce_{1}}^+(k';u_0') \label{eq:env-int1} \ee
and 
\be\tau_{\ce_{1}}^+(k;u_0,\tau_0)=\tau_{\ce_{1}}^+(k';u_0',\tau_0). \label{eq:env-int2}
\ee
From (\ref{eq:ue-n-odd}) and (\ref{eq:tau-e1-pos}), these correspond to, respectively 
\begin{eqnarray}
1+\frac{2\sqrt{2}}{k}-u_0 &=& 1+\frac{2\sqrt{2}}{k'}-u_0',
\label{eq:env-int1a} \\
\frac{\sqrt{2}k+1}{(u_0(1-u_0)k^2+2\sqrt{2}u_0k-1)^{1/2}} &=&
\frac{\sqrt{2}k'+1}{(u_0'(1-u_0'){k'}^2+2\sqrt{2}u_0'k'-1)^{1/2}}
\label{eq:env-int2a}
\end{eqnarray}
These equations have two solutions for $(k',u_0')$: the obvious solution $(k',u_0')=(k,u_0)$ (which contradicts $r(P)<r(P')$), and the solution 
\begin{eqnarray}
k' &=& \frac{\sqrt{2}(1+u_0)k^2}{4-\sqrt{2}(u_0-7)k-4(u_0-1)k^2},
\label{eq:kp-sol}\\
u_0' &=& \frac{8-4\sqrt{2}(u_0-3)k+(u_0^2-7u_0+8)k^2}{(u_0+1)k^2}.
\label{eq:up-sol}
\end{eqnarray}
With a little work, we can show that these expressions for $k'$ and $u_0'$ satisfy the inequality $k'>\sqrt{2}/{u_0'}$. Thus $k'\not\in [\kappa_1(u_0),\kappa_{\rm{min}}(u_0)]$, proving the lemma. \qed

\begin{lemma}\label{lem:se-env-prop} Let $P,P'$ be points with $r(P)<r(P')$ and $\tau(P)>\tau(P')$. If $\gamma_{[P,Q]}$ is an $\se-$segment from $P$, then there exists an $\se-$segment $\gamma'_{[P',Q']}$ such that $r(Q')>r(Q)$ and $\tau(Q')<\tau(Q)$. 
\end{lemma}

\startproof  Let $P,P',\gamma$ and $Q$ be as in the statement of the lemma. Consider the point $P''$ for which $\tau(P'')=\tau(P)$ and $r(P'')=r(P')$. Then by Lemma \ref{lem:env-disjoint}, the $\se-$segments $\ese{P}$ and $\ese{P''}$ are disjoint. They share the common maximum value $\tau(P)+\pi(\sqrt{2}-1)/$ of $\tau$ (see \ref{eq:T}), and by Corollary \ref{corr:tau-min}, the minimum of $\tau$ on $\ese{P''}$ is less than the minimum of $\tau$ on $\ese{P}$. This suffices to prove existence of an $\se-$segment $\gamma_{[P'',Q'']}''$ that is better than $\gamma_{[P,Q]}$. From (\ref{eq:u-se-env}) and (\ref{eq:tau-se-env}), we see that the $\se-$envelope $\ese{P'}$ is obtained by a (downwards) translation in $\tau$ of the $\se-$envelope $\ese{P''}$. The corresponding translation of the geodesic $\gamma''$ and the segment $\gamma_{[P'',Q'']}''$ yields the required geodesic $\gamma'$ and segment $\gamma_{[P',Q']}'$. \qed

Collecting the results of Lemmas \ref{lem:rdot-neg-useless}, \ref{lem:mono-useless-1},  \ref{lem:X} and \ref{lem:envelope-better}, we have the following key result (recall Definition \ref{def:se-seg} above):

\begin{corollary}\label{corr:SE-best} Let $\gamma_{[P,Q]}$ be a segment of a future point null geodesic $\gamma$. Then there exists a future pointing null geodesic $\gamma'$ and an ${\rm{SE}}-$segment $\gamma'_{[P,Q']}$ of $\gamma'$ which is better than $\gamma_{[P,Q]}$. \qed
\end{corollary}


\noindent\textbf{Proof of Proposition \ref{prop:segments-optimal}:} Let $\mu$ be an optimal lightlike path from a point $A_0\in\hor$ with $\tau(A_0)=\tau_* = \pi(\sqrt{2}-1)/2$ to a point $A_M\in\{\tau=0\}$ which comprises $M$ future pointing null geodesic segments. We show that $\gamma$ may be replaced by a lightlike path comprising $M'$ segments (with $M'\leq M$), each of the form described in the statement of the proposition. The proof then follows by considering the observations made in Comment \ref{comm:toS0}. For convenience, we will use $\gamma$ to refer to any future pointing null geodesic that arises in the proof. (Note that since $\mu$ is optimal, we must have $M'=M=N$.) 

Consider the first segment of $\mu$. This has the form $\mu_{[A_0,A_1]}$, and by Corollary \ref{corr:SE-best}, can be replaced by an $\se-$segment $\gamma_{[A_0,A_1']}$ with $r(A_1')>r(A_1)$ and $\tau(A_1')<\tau(A_1)$. Now we can apply Lemma \ref{lem:se-env-prop} to produce an $\se-$segment $\gamma_{[A_1',A_2']}$ with endpoint $A_2'$ satisfying $r(A_2')>r(A_2)$ and $\tau(A_2')<\tau(A_2)$. Iterating, we can produce a sequence of $M$ $\se-$segments with initial points $A_{n-1}'$ and terminal points $A_n', 3\leq n\leq M$, each satisfying $r(A_n')>r(A_n)$ and $\tau(A_n')<\tau(A_n)$. It follows that the lightlike path composed of these $M$ $\se-$segments reaches $\{\tau=0\}$ on or before the $M^{\rm{th}}$ segment.\qed

\subsection{Construction of the optimal path}\label{subsect:optimal-path}

\begin{figure}
    \centering
    {\includegraphics[width=0.45\textwidth]{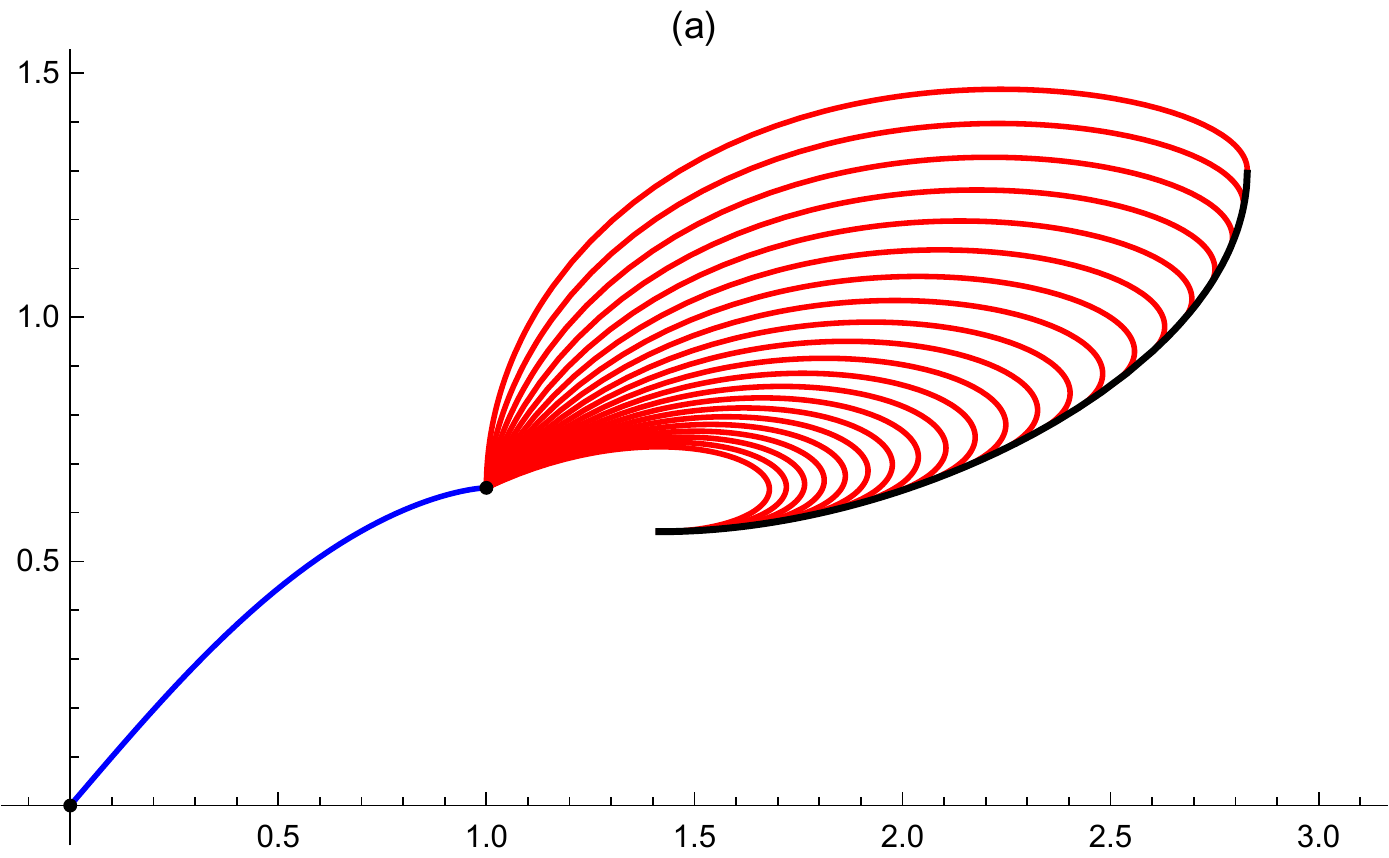}\,\includegraphics[width=0.45\textwidth]{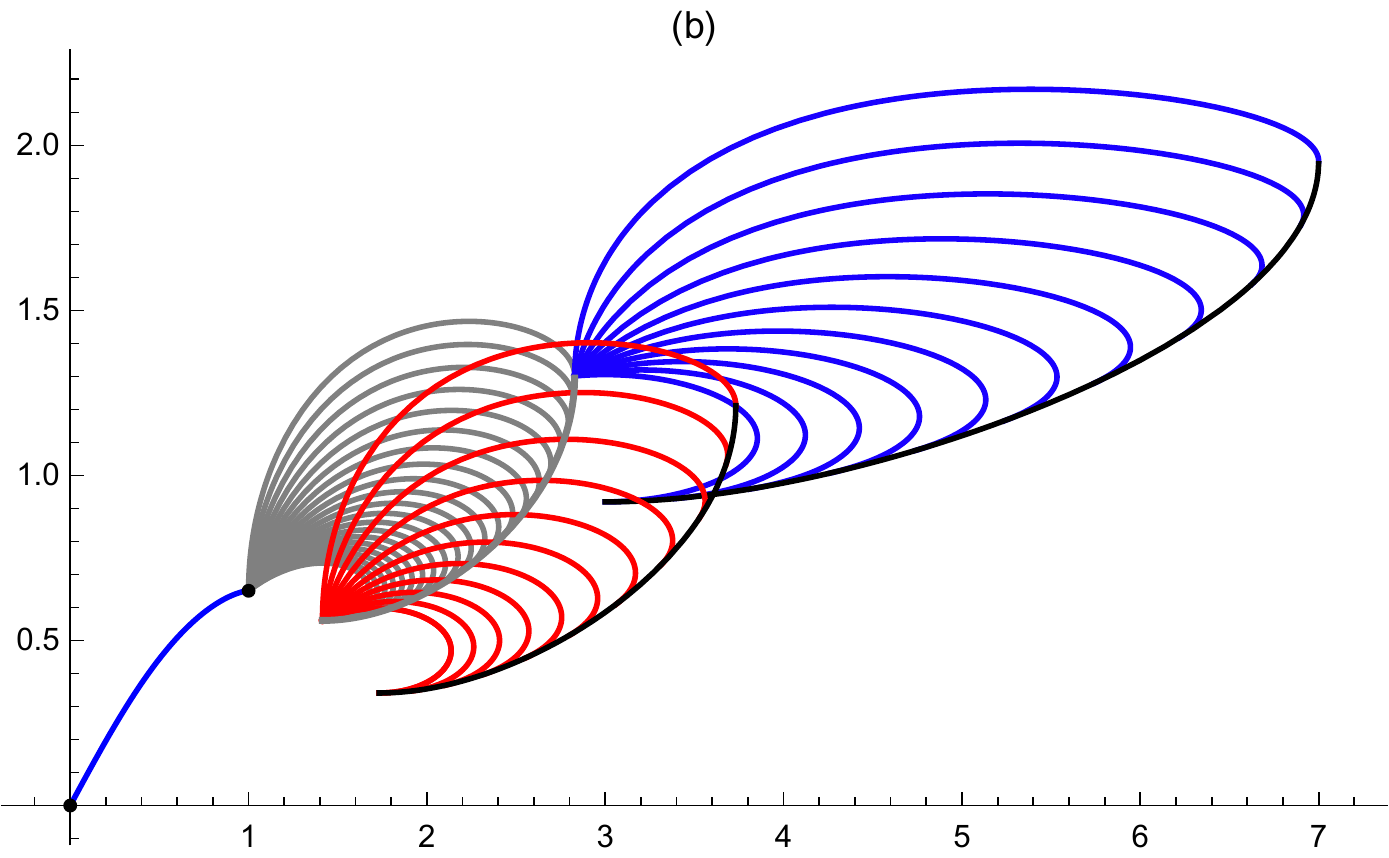}}
    {\includegraphics[width=0.45\textwidth]{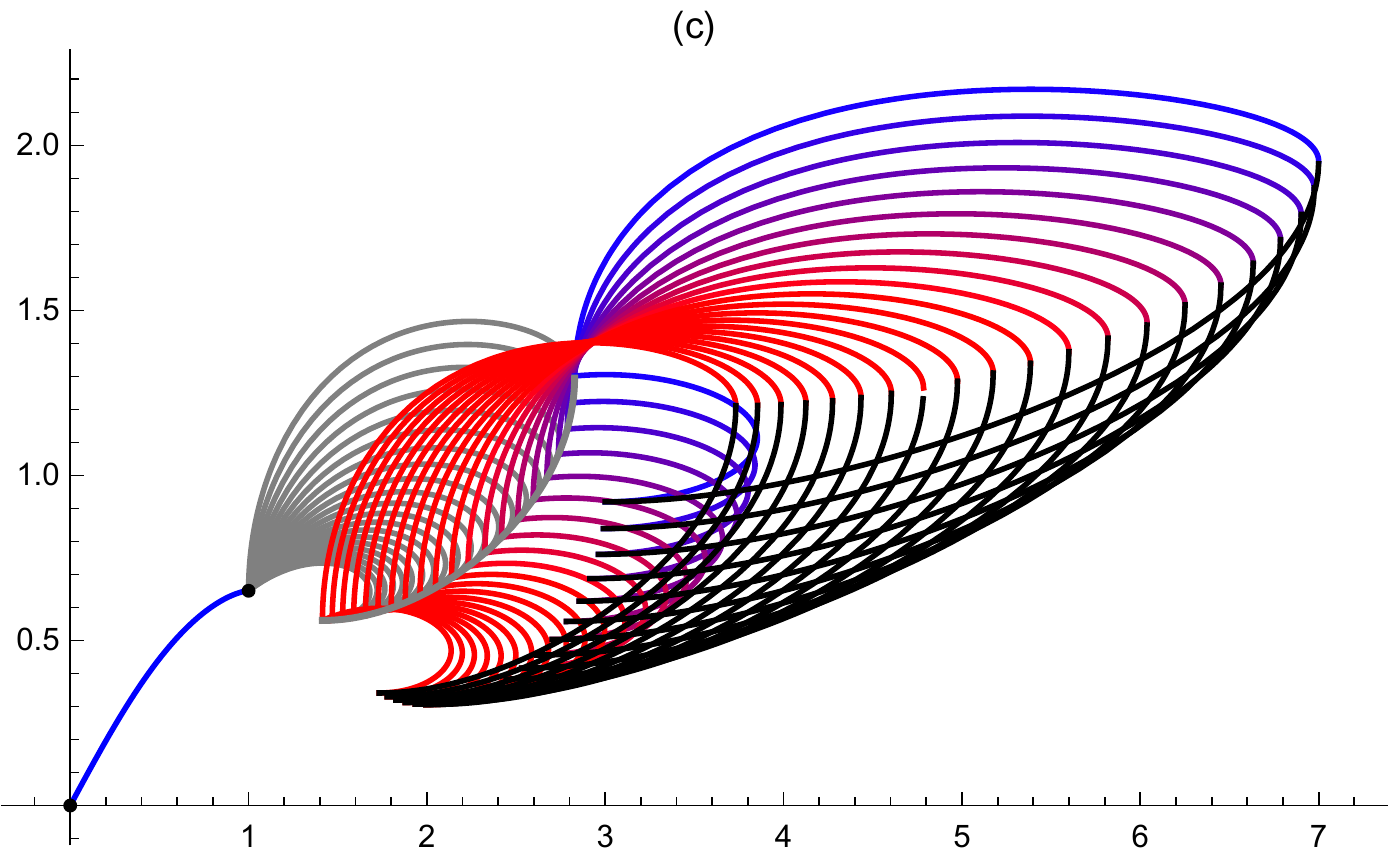}\,\includegraphics[width=0.45\textwidth]{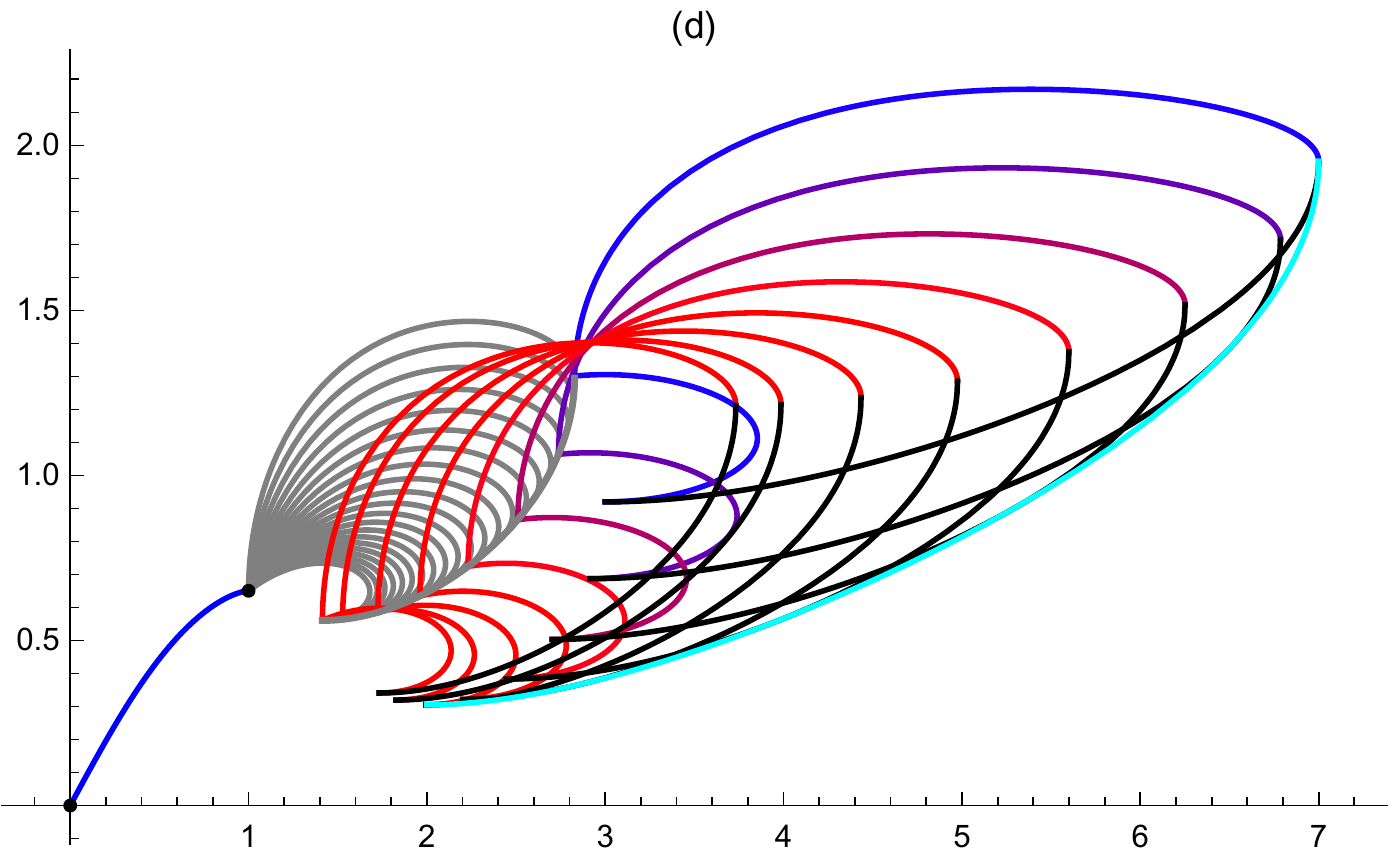}}
    {\includegraphics[width=0.45\textwidth]{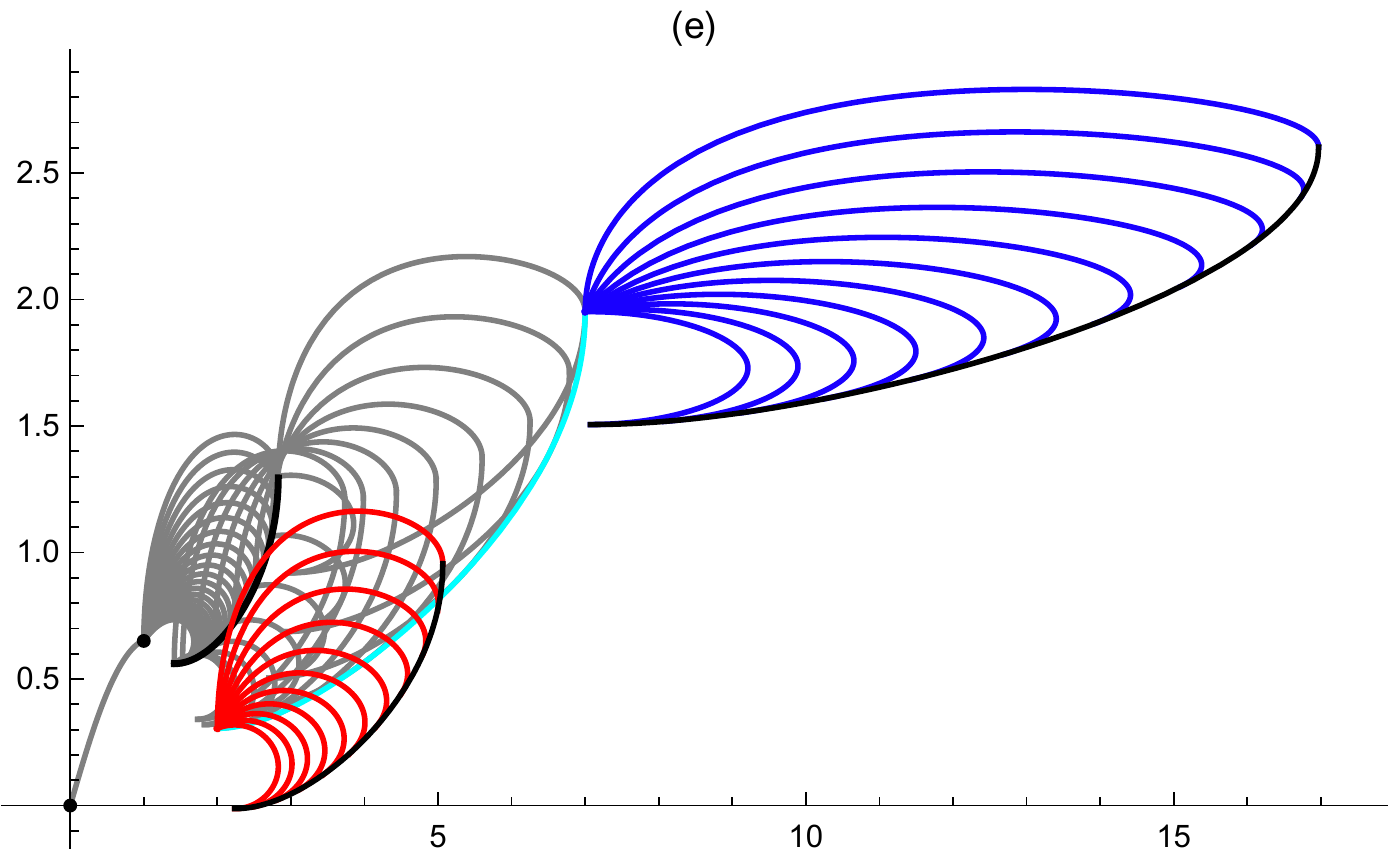}\,\includegraphics[width=0.45\textwidth]{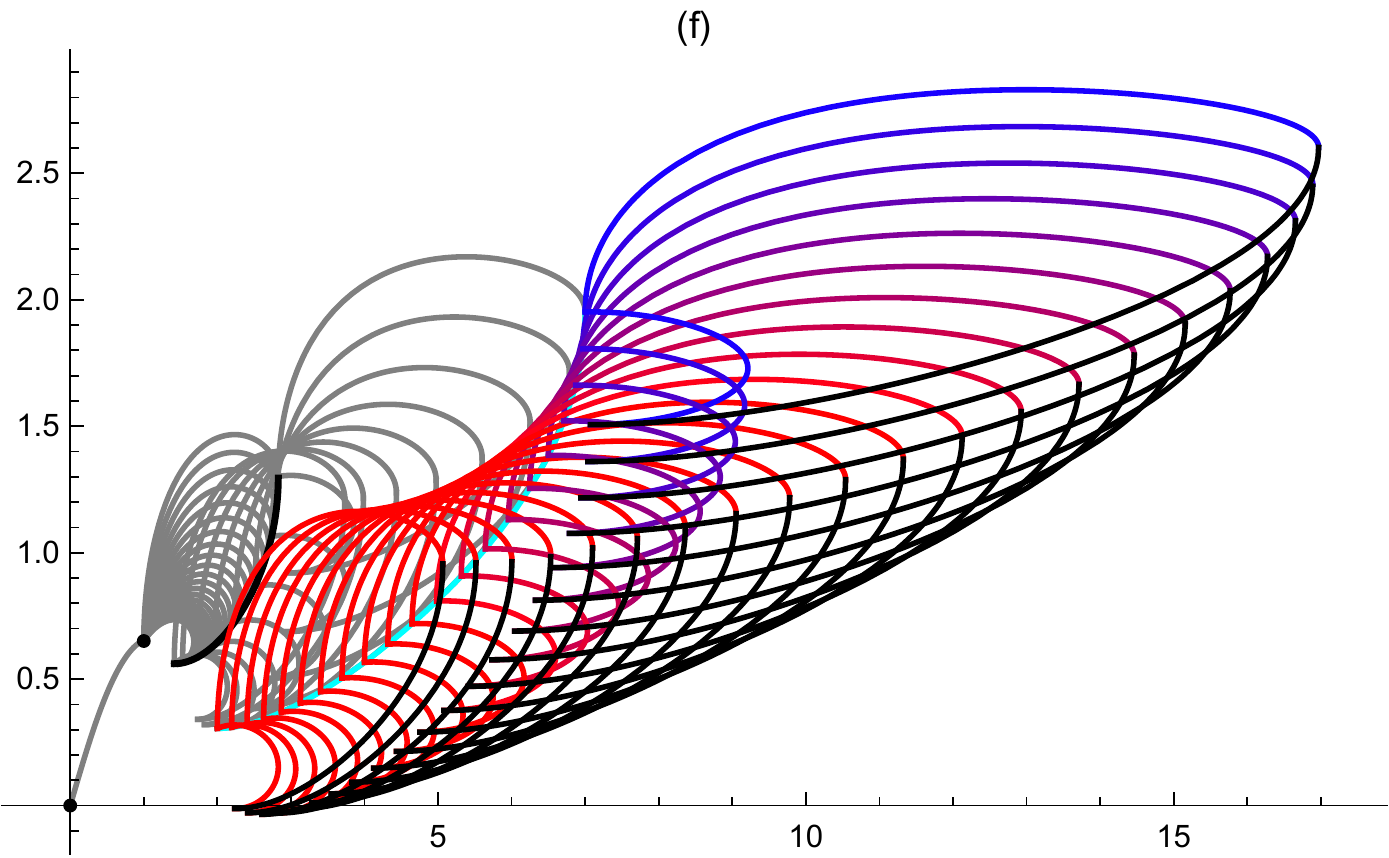}}
    \begin{center}
        {\includegraphics[width=0.45\textwidth]{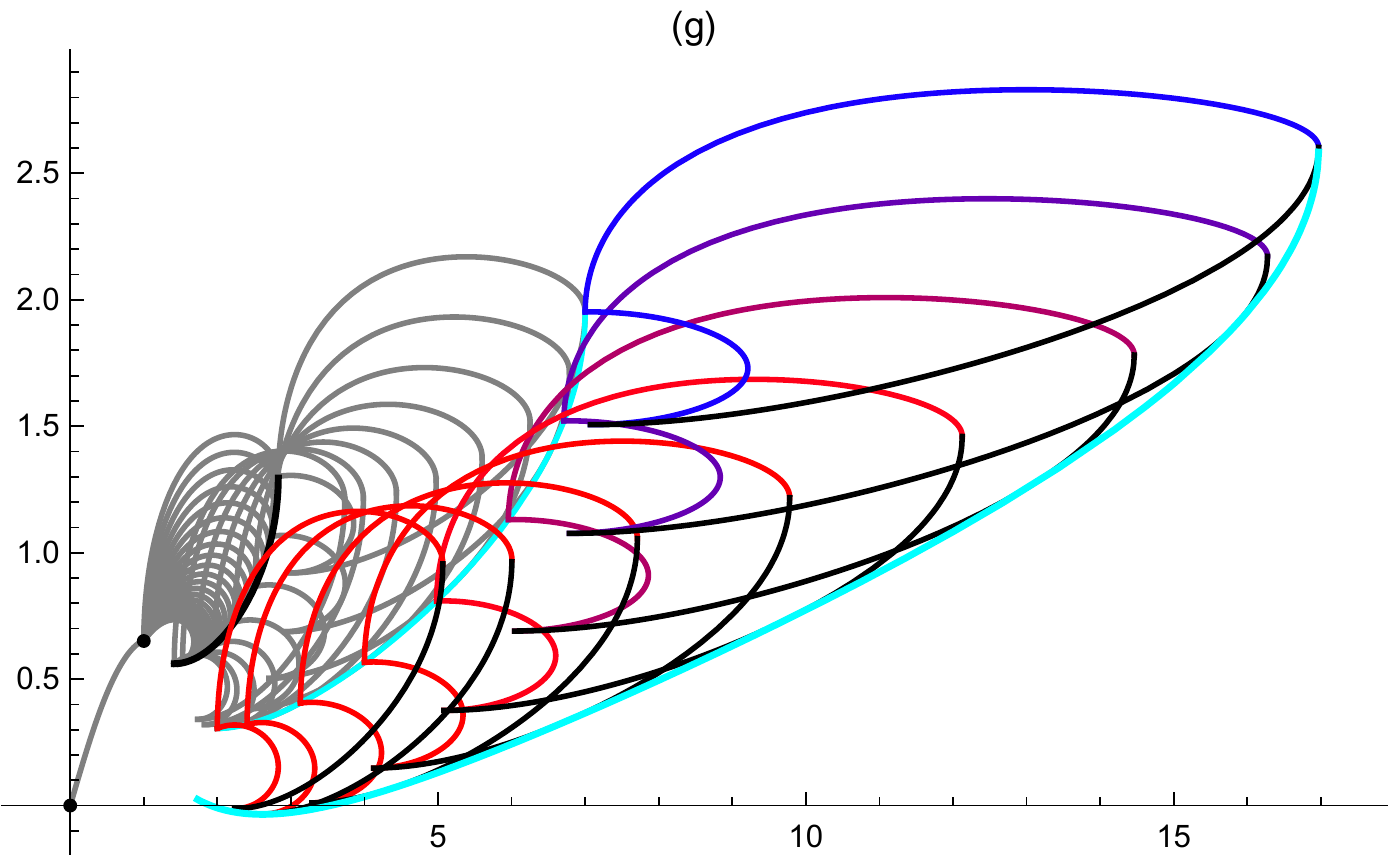}\,\includegraphics[width=0.45\textwidth]{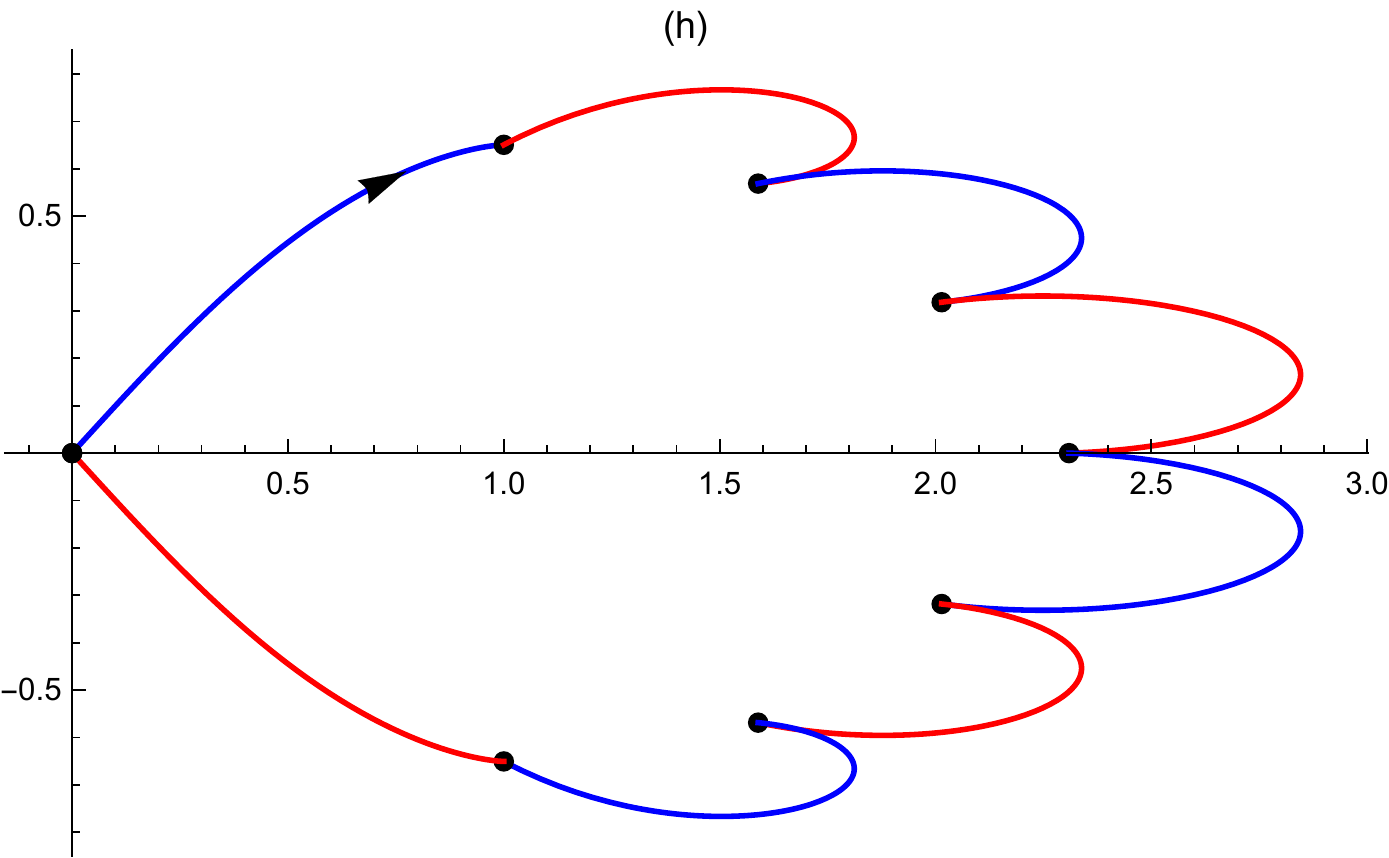}}
    \end{center}
    \caption{Illustration of the construction of an optimal path, and the proof of Theorem \ref{thm-main}. See Comment \ref{comment:optimal} for a description of each panel.}
    \label{fig:construction}
\end{figure}

In Proposition \ref{prop:segments-optimal}, we established the fact that the optimal path can be constructed using $\se-$segments. We have an explicit representation for these, and so it becomes a relatively straightforward task to piece together a sequence of $\se-$segments and so construct the optimal path. The key concern at this stage is to ensure that this is done in such a way as to minimise the number of $\se-$segments involved. Recall also from Comment \ref{comm:toS0} that the object is to produce a lightlike path from the point $A_0:(u,\tau)=(1,\tau_*)$ to the hypersurface $\Sigma_0=\{\tau=0\}$, using the least possible number of future pointing null geodesic segments. In Figure \ref{fig:construction} and the accompanying Comment \ref{comment:optimal}, we give a pictorial account of the argument.

\begin{comments}\label{comment:optimal}
See Figure \ref{fig:construction}. In each figure, $r$ runs along the horizontal axis and $\tau$ along the vertical. Panel (a) shows the first segment extending from $\co:(r,\tau)=(0,0)$ to $A_0:(r,\tau)=(1,\tau_*)$. Also shown is a selection of $\se-$segments from $A_0$ (in red), and the $\se-$envelope $\env_1$ (black; see (\ref{eq:u-env1}) and (\ref{eq:tau-env1})) formed by those segments. These segments provide candidates for the second segment of an optimal path. Points of $\env_1$ provide candidates for the initial point of the third segment of the optimal path. Panels (b) and (c) show candidates for the third segment of the optimal path (with the segments of Panel (a) now shown in grey for clarity). In Panel (b), the segments in blue are $\se-$segments with initial point at the $\tau-$maximising point of $\env_1$, and the segments in red are $\se-$segments from the $\tau-$minimising point of $\env_1$. The $\se-$envelopes of these families are shown in black: these envelopes are members of the 1-parameter family of envelopes $\env_2(\kappa),\kappa\in I_2$ (see (\ref{eq:u-env2}) and (\ref{eq:tau-env2})). Panel (c) shows a different view of this scenario. Here, the `first' and `last' $\se-$segments from a collection of points on $\env_1$ are shown, colour-coded in a fade from blue to red. The envelopes $\env_2(\kappa),\kappa\in I_2$ corresponding to each initial point are also shown (in black). Panel (d) shows a subset of the segments of Panel (c), along with the $\se-$envelope $\env_3$ (cyan) of the 1-parameter family of envelopes $\env_2(\kappa),\kappa\in I_2$; see Lemma \ref{lem:e3-structure}. The fact that the envelope $\env_3$ does not reach $\Sigma_0=\{\tau=0\}$ proves that more than six segments are required to construct a closed lightlike path from $\co$ to $\co$ as per Corollary \ref{corr:more-than-six}. Panel (e) shows a selection of $\se-$segments from the envelope $\env_3$. The segments in blue are $\se-$segments with initial point at the $\tau-$maximising point of $\env_3$, and the segments in red are $\se-$segments from the $\tau-$minimising point of $\env_3$. The envelopes of these families are shown in black. All other segments are greyed-out. Panel (f) shows the `first' and `last' $\se-$segments from a selection of points on $\env_3$ (fading from blue to red), along with (members of) the corresponding  1-parameter family of envelopes, $\env_4(\kappa), \kappa\in I_4$. Finally, panel (g) shows a subset of these $\se-$segments and their envelopes, along with the envelope $\env_5$ (cyan) of the 1-parameter family $\env_4(\kappa), \kappa\in I_4$. As panels (e)-(g) show, a fourth segment from $\co$ can be found which crosses $\Sigma_0$. The minimum of $\tau$ on a fourth segment corresponds to the minimum of $\tau$ on $\env_5$. Since this minimum is clearly greater than $-\tau_*$, a further four future pointing null geodesic segments are required to return to $\co$. Panel (h) shows the projection into the $r-\tau$ plane of an optimal closed lightlike path from $\co$ to $\co$ (as illustrated in Figure \ref{fig:CTC}). The relevant parameters of each segment of this path are given in Appendix \ref{app:params}. The first segment indicates the direction of increase of proper time. The second, third and fourth segments are $\se-$segments from the endpoint of the previous segment. Parameters are chosen to ensure that the fourth segment terminates on $\Sigma_0$. The fifth to eight segments retrace the fourth down to first respectively, as described in  Lemma \ref{lem:retrace}.
\end{comments}

We now provide the analysis that underpins the account just given - that is, we prove Theorem \ref{thm-main}.

The first $\se-$segment of the path extends from $A_0$ to the $\se-$envelope of this point, which (using (\ref{eq:u-se-env}) and (\ref{eq:tau-se-env})) is given by 
\begin{eqnarray} 
u=u_{\env_1}(\kappa) &:=& \frac{2\sqrt{2}}{\kappa},\label{eq:u-env1}\\
\tau=\tau_{\env_1}(\kappa) &:=& \tau_*-\frac{\pi}{2}+\sqrt{2}
\arctan
\left(\frac{\sqrt{2}\kappa+1}{(2\sqrt{2}\kappa-1)^{1/2}}\right), \label{eq:tau-env1} 
\end{eqnarray}
with 
\be
\frac{1}{2\sqrt{2}}\leq\kappa \leq \sqrt{2}. \label{eq:k-env1} 
\ee

We refer to this envelope as $\env_1$. Each point of $\env_1$ (which is parametised by $\kappa$) generates its own envelope $\env_2(\kappa)$, and we know that the second segment of the optimal path terminates on one of these envelopes. These envelopes are described by the 2-parameter family of curves (found by using (using (\ref{eq:u-se-env}) and (\ref{eq:tau-se-env}) with $(u_0,\tau_0)=(u_{\env_1}(\kappa),\tau_{\env_1}(\kappa))$)
\begin{eqnarray}
u = u_{\env_2}(\kappa,\rho) &:=& 1+2\sqrt{2}\left(\frac{1}{\rho}-\frac{1}{\kappa}\right), \label{eq:u-env2} \\
\tau = \tau_{\env_2}(\kappa,\rho) &:=&
\tau_{\env_1}(\kappa) - \frac{\pi}{2} +\sqrt{2}\arctan\left(\frac{\sqrt{2}\rho+1}{q^{1/2}(\rho,u_{\env_1}(\kappa))}\right), \label{eq:tau-env2} 
\end{eqnarray}
where $1/(2\sqrt{2})\leq\kappa\leq\sqrt{2}$,  
\be \frac{\sqrt{2}\kappa-\kappa(1+\frac{\kappa}{2\sqrt{2}})^{1/2}}{2\sqrt{2}-\kappa} =  \kappa_1(u_{\env_1}(\kappa)) \leq \rho \leq \frac{\sqrt{2}}{u_{\env_1}(\kappa)} = \frac{\kappa}{2}, 
\label{eq:rho-env2}
\ee
and
\begin{eqnarray} 
q(\rho,u_{\env_1}(\kappa)) &=& \ueone(1-\ueone)\rho^2+2\sqrt{2}\ueone \rho - 1 \nonumber \\
&=& 2\sqrt{2}(\kappa-2\sqrt{2})\frac{\rho^2}{\kappa^2}+8\frac{\rho}{\kappa} - 1. 
\label{eq:q-env2} 
\end{eqnarray}

The optimal $\se-$segment from a point on $\env_1$ to the corresponding $\env_2(\kappa)$ must terminate on a point on the boundary of the region filled by the family of envelopes $\env_2(\kappa), \kappa\in[1/(2\sqrt{2}),\sqrt{2}]$. The boundary of this region is a subset of the envelope of the parametrised cuves $(u,\tau)=(u_{\env_2}(\kappa,\rho),\tau_{\env_2}(\kappa,\rho))$ (see \S 5.16 of \cite{bruce1992curves}), and so we require the envelope of the family $\env_2(\kappa)$. This is determined by the solutions of the equation
\be \Delta_2(\rho,\kappa) = \pd{u_{\env_2}}{\kappa}\pd{\tau_{\env_2}}{\rho}-\pd{u_{\env_2}}{\rho}\pd{\tau_{\env_2}}{\kappa} = 0. \label{eq:env2-eqn} 
\ee
With some work, we can show that this equation has the solutions $\rho=\kappa$ (which is ruled out by the second inequality in (\ref{eq:rho-env2})) and 
\be 
\rho = \frac{\kappa^2}{2\sqrt{2}-\kappa}. \label{eq:rho-env2-sol}\ee
Substituting into (\ref{eq:u-env2}) and (\ref{eq:tau-env2}) yields the following lemma:

\begin{lemma}\label{lem:e3-structure} The envelope $\env_3$ of the 1-parameter family $\env_2(\kappa), \kappa\in I_2=[1/(2\sqrt{2}),\sqrt{2}]$ of envelopes of families of $\se-$segments emanating from the envelope of the family of $\se-$segments emanating from $A_0$ is given by 
\begin{eqnarray}
u = u_{\env_3} & := & \frac{\kappa^2-4\sqrt{2}\kappa + 8}{\kappa^2} \label{eq:u-env3} \\
\tau = \tau_{\env_3} & :=& \frac32(\sqrt{2}-1)\pi - \sqrt{2}\arctan\left(\frac{(2\sqrt{2}\kappa-1)^{1/2}}{\kappa^2-\sqrt{2}\kappa+1}\right),\label{eq:tau-env3} 
\end{eqnarray}
with $\kappa\in I_2$.\qed
\end{lemma}

It is straightforward to show that $u_{\env_3}$ is decreasing on $[\frac{1}{2\sqrt{2}},\sqrt{2}]$, and that $\tau_{\env_3}(\kappa)$ attains its minimum at $\kappa  =2\sqrt{2}/3$. It follows that an $\se-$segment arriving at the portion of $\env_3$ with $\kappa\in (\frac{2\sqrt{2}}{3},\sqrt{2}]$ may be replaced by a better segment arriving at the portion of $\env_3$ with $\kappa\in[\frac{1}{2\sqrt{2}},\frac{2\sqrt{2}}{3}]$. We also have the following implications:

\begin{proposition}
\label{prop:min-tau-3-segments}
The minimum of $\tau$ on a 3-segment lightlike path from $\co:(u,\tau)=(0,0)$ is 
\be \tau_{\env_3}(\frac{2\sqrt{2}}{3}) = \frac32(\sqrt{2}-1)\pi - \sqrt{2}\arctan\left(3\sqrt{\frac35}\right) \simeq 0.3052. 
\label{eq:min-tau-3-segments}
\ee
\qed
\end{proposition}

Since this minimum is positive, we can state the following: 

\begin{corollary}\label{corr:more-than-six}
$N>6.$\qed
\end{corollary}

The next segment of an optimal path (the third from $A_0$ and the fourth overall) emanates from a point on $\env_3$ corresponding to $\kappa\in[\frac{1}{2\sqrt{2}},\frac{2\sqrt{2}}{3}]$. By Proposition \ref{prop:segments-optimal}, we can assume without loss of generality that this is an $\se-$segment, and so it terminates on the $\se-$envelope of the point of $\env_3$ from which it emanates. As above, there is a 1-parameter family of these envelopes, and combining Lemma \ref{lem:e3-structure} with (\ref{eq:u-se-env}) and (\ref{eq:tau-se-env}) allows us to describe them as follows:
\begin{eqnarray}
u = u_{\env_4}(\kappa,\rho) &:=& 1+\frac{2\sqrt{2}}{\rho}-u_{\env_3}(\kappa), \label{eq:u-env4} \\
\tau = \tau_{\env_4}(\kappa,\rho) &:=&
\tau_{\env_3}(\kappa) - \frac{\pi}{2} +\sqrt{2}\arctan\left(\frac{\sqrt{2}\rho+1}{q^{1/2}(\rho,u_{\env_3}(\kappa))}\right), \label{eq:tau-env4} 
\end{eqnarray}
where $1/(2\sqrt{2})\leq\kappa\leq 2\sqrt{2}/3$ and   
\be  \kappa_1(u_{\env_3}(\kappa)) \leq \rho \leq \frac{\sqrt{2}}{u_{\env_3}(\kappa)} = \frac{\sqrt{2}\kappa^2}{(2\sqrt{2}-\kappa)^2}. 
\label{eq:rho-env4}
\ee
For completeness, we note that 
\be \kappa_1(u_{\env_3}(\kappa)) =\frac{\kappa^2}{4}\left(\frac{-\kappa^2+4\sqrt{2}\kappa-8+(2\sqrt{2}-\kappa)(\kappa^2-2\sqrt{2}\kappa+4)^{1/2}}{(\sqrt{2}-\kappa)(-\kappa^2+4\sqrt{2}\kappa-8)}\right). 
\label{eq:kleft-env4}
\ee

Repeating the argument above, we seek the envelope of \textit{this} 1-parameter family of curves by solving 
\be \Delta_4(\rho,\kappa) = \pd{u_{\env_4}}{\kappa}\pd{\tau_{\env_4}}{\rho}-\pd{u_{\env_4}}{\rho}\pd{\tau_{\env_4}}{\kappa} = 0. \label{eq:env4-eqn} 
\ee

Remarkably, it is possible to solve this equation in closed form. We find three solutions in the form $\rho=\rho_i(\kappa), i=1,2,3$. Only one of these corresponds to values of $\rho$ in the permitted interval (\ref{eq:rho-env4}). This solution is 
\be \rho = \frac{\kappa^3}{2(2\sqrt{2}-\kappa)(\sqrt{2}-\kappa)}.
\label{eq:rho-env4-sol}
\ee

Substituting into (\ref{eq:u-env4}) and (\ref{eq:tau-env4}) yields the following result:

\begin{lemma} Given $A_0:(u,\tau)=(1,\tau_*)$, let $\env_1$ be the $\se-$envelope of the family of $\se-$segments from $A_0$. Let $\env_2(\kappa), \kappa\in I_2$ be the 1-parameter family of $\se-$envelopes of $\se-$segments from points on $\env_1$, and let $\env_3$ be the envelope of the 1-parameter family $\env_2(\kappa), \kappa\in I_2$. Let $\env_4(\kappa)$ be the 1-parameter family of $\se-$envelopes of $\se-$segments from points on $\env_3$ and let $\env_5$ be the envelope of the 1-parameter family $\env_4(\kappa), \kappa\in I_4=[\frac{1}{2\sqrt{2}},\frac{2\sqrt{2}}{3}]$. Then $\env_5$ is described by the parametrised curve
\begin{eqnarray} 
u=u_{\env_5}(\kappa) &:=& \frac{8\sqrt{2}}{\kappa^3}(2-2\sqrt{2}\kappa+\kappa^2)
\label{eq:u-env5} \\
\tau = \tau_{\env_5}(\kappa) &:=& (3\sqrt{2}-4)\frac{\pi}{2}-\sqrt{2}\arctan\left(
\frac{\sqrt{2}(2\sqrt{2}-7\kappa+2\sqrt{2}\kappa^2+\kappa^3)}{\sqrt{-1+2\sqrt{2}\kappa}(-4+3\sqrt{2}\kappa-2\kappa^2)}\right),
\label{eq:tau-env5} 
\end{eqnarray}
with $\kappa\in I_4$.\qed
\end{lemma}

We can now write down the proof of the main theorem. 

\noindent\textbf{Proof of Theorem \ref{thm-main}:} It is straightforward to show that the minimum of $\tau$ on $\env_5$ occurs at $\kappa=2(\sqrt{2}-1)\in (\frac{1}{2\sqrt{2}},\frac{2\sqrt{2}}{3})$, and the minimum value is 
\be \tau_{\env_5,\rm{min}} = (3\sqrt{2}-4)\frac{\pi}{2}-\sqrt{2}\arctan\left(\frac{\sqrt{95-64\sqrt{2}}}{7}\right) \simeq - 0.0346. \label{eq:tau-min-fourth} \ee
The construction above proves that this is also the minimum of $\tau$ over all four-segment lightlike paths from $\co$ (or three-segment lightlike paths from $A_0$). This proves that there is a sequence of eight future pointing null geodesic segments forming a closed lightlike path from $\co$ to $\co$: We take the fourth segment to be an $\se-$segment from $\env_4(\kappa)$ which has its endpoint on $\{\tau=0\}$. Such a segment exists since $\tau_{\env_5,\rm{min}}$ is negative. We take the fifth to eight segments to retrace the paths of the first four in the sense of Comment \ref{comm:toS0} and Lemma \ref{lem:retrace}. 

Furthermore, since 
\be (3\sqrt{2}-4)\frac{\pi}{2}-\sqrt{2}\arctan\left(\frac{\sqrt{95-64\sqrt{2}}}{7}\right) > 
-\left(\frac32(\sqrt{2}-1)\pi - \sqrt{2}\arctan\left(3\sqrt{\frac35}\right)\right), \label{eq:compare-mins} 
\ee
we see that we cannot construct a closed lightlike path with seven segments. This follows by considering the time-reversed version of Proposition \ref{prop:min-tau-3-segments}. The quantity on the right hand side of (\ref{eq:compare-mins}) is the \textit{maximum} value of $\tau$ from which a three-segment lightlike path can reach $\co$. \qed

\begin{comments}\label{comm:phi}
We have the good fortune not to have to consider the angular coordinate $\phi$ in the analysis. The value of $\phi$ varies on each segment of the optimal path. The seventh (and penultimate) segment terminates on $\hor$ at some value $\phi_7$ of $\phi$. But since the last segment terminates at $r=0$, whereat the value of $\phi$ is unimportant, the value of $\phi_7$ is likewise irrelevant, as are the values of $\phi$ along all segments considered. 
\end{comments}

\section{Conclusions}\label{sect:conclusions}

In the context of spacetime geometry, time travel requires two things. First, the spacetime must admit the appropriate closed causal curves. Secondly, there must be a means of travelling along these closed curves. In \godel's spacetime, the latter requirement appears to present insurmountable difficulties, as outlined in the introduction above. Thus, on a purely conceptual basis, violating causality by means of a sequence of signals appears to provide an interesting alternative. But it should be noted that this only provides an alternative to the second necessary element for time travel. Theorem \ref{thm-lightlike-trips} shows that a CTC may be replaced by a closed lightlike path. But this theorem also shows that a closed lightlike path can exist only when a CTC is present. There is one exception: this is when the spacetime admits a closed null geodesic, but no closed timelike curves, and so is chronological but non-causal \cite{minguzzi2019lorentzian}. Examples of such spacetimes exist; see e.g.\ Figure 11 and the associated Proposition 4.32 of  \cite{minguzzi2019lorentzian}. So the underlying spacetime geometries that support closed timelike curves and closed lightlike paths appear to be by and large the same. The question then arises as to whether the lightlike path option is indeed favourable. It must be noted that we need to rely on what we have referred to as cooperative agents in remote (and possible unpopulated) regions of the universe, or a system of mirrors. Placing these mirrors in the appropriate locations will of course incur a fuel bill of some sort. It would be interesting to know what this fuel bill would be, but we have not pursued this in the present paper, which provides the `proof of concept' for this type of causality violation. 

It is also worth commenting on the nature of these mirrors. The term is used somewhat analogously, as there is a deflection in spacetime rather than (just) in space at each non-smooth junction of the closed lightlike path. According to the geodesic equations (\ref{tau-d})-(\ref{z-d}), the tangent to a light ray at a fixed point of spacetime is determined by the value of $\kappa$ and a choice of sign of $\dot{r}$ (recall that $\lambda=0$ on all relevant segments). From panel (h) in Figure \ref{fig:construction}, it is evident that $\dot{r}$  changes sign at each junction. Table \ref{table:params} of Appendix B provides the details of the values of $\kappa$ required along each segment. Recall that $\kappa=L_1/L_2$ where $L_1$ and $L_2$ are respectively energy and angular momentum constants associated with the Killing vectors $\frac{\partial}{\partial \tau}$ and $\frac{\partial}{\partial\phi}$. On the second to seventh segment, we can set $L_2=1$ and identify $L_1$ with $\kappa$, and on the first and eighth segments, we have $L_1=1$ and $L_2=0$. Thus the reflection in the $r-$direction is accompanied by a jump in the energy $\kappa$ of the light ray. So perhaps the term `acousto-optic modulator' would be more appropriate than `mirror' as a description of the device needed at each junction of the lightlike path \cite{scrubylaser}. Note the implication of a further fuel cost. 

Having constructed a closed lightlike path, it is evident that it is possible to send a signal strictly into one's own past in \godel's universe. There is no limit to how far into one's own past such a signal can be sent. There exist future-directed timelike curves connecting any two points of \godel's spacetime, including the case where the second point $Q$ lies on the past world-line of an observer at the first point $P$ (see Proposition 2 of \cite{franchi2009relativistic}). Then Theorem \ref{thm-lightlike-trips} applies to prove the existence of a lightlike path from $P$ to $Q$. But given that we have demonstrated that at least eight segments are required to close a lightlike path, it is of interest to consider how far into one's past such a signal (i.e.\ and eight-segment lightlike path) may be sent. Equation (\ref{eq:tau-min-fourth}) shows that we can construct a seven-segment lightlike path that originates at $\co$ and terminates at a point $P\in\hor$ with $\tau(P)= -\tau_*+2\tau_{\env_5,{\rm{min}}}\simeq -0.7918$. The future directed, planar, ingoing null geodesic from $P$ meets the axis at $\tau=2\tau_{\env_5,{\rm{min}}}\simeq -0.0691$. Since $\tau$ is proper time along the worldline along the axis, an eight-segment lightlike path can travel this far into this observer's past. Now recall that we have been working in the spacetime with line element rescaled by the factor $\alpha$ - see (\ref{metric-cyl1}). Thus in the ``physical" universe, this corresponds to an elapse of proper time of the order $\Delta s \simeq 0.07 \alpha^{-1}$. For illustrative purposes, consider a \godel\ universe with $\Lambda=0$, and with a value of $\rho$ corresponding to the average density of our universe $O(10^{-30})\ g/cm^3$. This yields $\Delta s = O(10^{16}) s$, giving plenty of time to send oneself winning lottery numbers on a useful time scale. A major drawback is that the ``backwards in $\tau$" segments (which require mirrors and/or the cooperation of locals) are all located outside the \godel\ horizon $\hor$, which in the physical spacetime is located a distance $r=\alpha^{-1}$ away: this is $O(10^{24})\ m$ away using the numbers above - just a couple of orders of magnitude away from estimates of the current size of the universe. 

We conclude by noting that the results above rely on some very nice analytic properties of \godel's spacetime. It is rare that one has access to a complete closed form general solution of the geodesic equations in a spacetime which has a clear physical and geometric interpretation. Furthermore, as we have seen, it is possible to describe in closed form the envelope of future pointing null geodesics from a point of the spacetime - and to then describe in closed form the iterated envelopes that were required in Section IV-C. It would be of interest to know if there is any underlying geometric reason for the observed simplicity of the envelope structure.  

\acknowledgements
I thank Abraham Harte, Ko Sanders and Peter Taylor for useful conversations.

\appendix
\section{Proofs deferred from the main text.}

\noindent\textbf{Proof of Theorem \ref{thm-lightlike-trips}:}

The proof of part (i) of the theorem is trivial. If $A\lll B$, then immediately $A\prec B$, and so either $A\ll B$ or there exists a null geodesic from $A$ to $B$ by Proposition 3. In the former case, Proposition 4 applies. 

The proof of part (ii) is more involved, but is conceptually straightforward: we construct the lightlight path using a sequence of (short) `outgoing' and `ingoing' null geodesic segments that respectively originate at and terminate at points of the timelike curve from $A$ to $B$. 

So assume that there is a future-pointing timelike curve $\gamma$ from $A$ to $B$. We apply Proposition 4 to deduce the existence of a set of points $A_0=A, A_1,A_2,\dots,A_n=b$ and a set of future-pointing timelike geodesics $\gamma_i$ from $A_{i-1}$ to $A_i$, $1\leq i\leq n$. We show that each pair of points satisfies $A_{i-1}\lll A_i$: this proves the result by taking the lightlike path from $A$ to $B$ to be the union of the lightlike paths from each $A_{i-1}$ to $A_i$. Thus we can assume (without loss of generality) that the future-pointing timelike curve from $A$ to $B$ is in fact a future-pointing timelike geodesic. So we consider the geodesic
\be \gamma:[0,1]\to\cm,\quad s\mapsto \gamma(s)\label{ab-geo}\ee
with $\gamma(0)=A, \gamma(1)=B$. We introduce a tetrad $\vec{e}_i, i=0,1,2,3$ with $\vec{e}_0=\gamma'$ and with $\vec{e}_i, i=1,2,3$ parallel transported along $\gamma$ so that $\nabla_{\gamma'}\vec{e}_i=0$. Let $P=\gamma(s)$ for some $s\in [0,1]$ and consider the exponential map at $P$:
\be \exp_P : U_s \to \cm,\label{expp}\ee
where $U_s$ is an open neighbourhood of the origin in $T_{\gamma(s)}\cm$, and where 
\be \exp_P(\vec{v}) =G^\alpha(1;P,\vec{v}) \label{exp-eval}\ee
where
\be x^\alpha(u) = G^\alpha(u;P,\vec{v}) \label{exp-geo}\ee
is the unique solution of the geodesic equations with $x^\alpha(0)=x^\alpha|_{P}$ and $\frac{dx^\alpha}{du}(0)=v^\alpha$ and where $u$ is an affine parameter along the geodesic. The exponential map at $P$ is defined for those $\vec{v}\in T_P(\cm)$ for which the solution (\ref{exp-eval}) of the geodesic equations extends to the parameter value $u=1$. We define $N_P$ to be the maximal normal neighbourhood of $P$, so that $N_P$ is the image of $\exp_P$ on the maximal domain $U_s$. 

We recall that Riemann normal coordinates are defined on $N_P$ by 
\be X^\alpha_R(Q)=v^\alpha, \quad Q\in N_P, \label{rnc}\ee
where $\vec{v}\in T_P(\cm)$ is the vector at $P$ for which 
\be G^\alpha(1;P,\vec{v}) = x^\alpha|_Q.\label{rnc-con}\ee
By uniqueness of the solutions of the geodesic equations, we have 
\be G^\alpha(u;P,\vec{v}) = G^\alpha(1;P,u\vec{v}),\label{geo-uni}\ee
where $u$ is an affine parameter. Thus
\be X^\alpha_R(Q)=u v^\alpha \ee
are the Riemann normal coordinates of a point $Q\in N_P$ at affine distance $u$ from $P$ along the geodesic with tangent $\vec{v}$ at $P$. 

Now return to the tetrad along $\gamma$, and take this (without loss of generality) to be orthonormal, so that 
\be g(\vec{e}_i(s),\vec{e}_i(s))|_{\gamma(s)}=\eta_{ij} \quad \forall s\in [0,1], \label{ortho-normal}
\ee
where $\eta$ is the unit Minkowski tensor. Then $\vec{v}\in T_{\gamma(s)}(\cm)$ may be written as 
\be \vec{v}=v^i(s)\vec{e}_i(s). \ee
This gives rise to \textit{Minkowski normal coordinates} (MNCs) by taking tetrad components of (\ref{rnc}): for $Q\in N_P, P=\gamma(s)$,
\be X^i(Q) = u v^i. \label{mnc}\ee
It follows that the geodesic $PQ$ is timelike (null, spacelike) if and only if $v^i$ is timelike (null, spacelike) with respect to $\eta$, and the causal geodesic $PQ$ is future-pointing if and only if $v^0>0$. 

Now consider $\overline{X}=(X^0,X^1,X^2,X^3)$ as elements of $\mathbb{R}^4$ with the standard Euclidean norm:
\be |\overline{X}|=\left(\sum_{i=0}^3 |X^i|^2\right)^{1/2}. \label{euclid} 
\ee
The existence of the open neighbourhood $U_s$ of the origin $T_{\gamma(s)}(\cm)$ on which $\exp_P$ is defined implies the existence of a closed ball centred at the origin of $\mathbb{R}^4$, $B(O,\delta_s)$ such that
\be \{\overline{X}\in\mathbb{R}^4, \overline{X}\in B(O,\delta_s)\} \subset N_{\gamma(s)}. 
\label{ball_in_n}
\ee
By continuous dependence of geodesics on their initial values and initial tangents, it follows that the mapping
\be s\mapsto \Delta_s = \sup_{\delta_s>0}\{\delta_s:B(O,\delta_s)\subset N_{\gamma(s)}\} 
\ee
is continuous on $[0,1]$. This function is strictly positive on the closed interval $[0,1]$, and thus attains a positive minimum. So we have:

\begin{lemma}\label{lem:delta-mnc} There exists $\delta>0$ such that for all $s\in[0,1]$, the points $Q$ with Minkowski normal coordinates $\overline{X}$ at $P$ lie in $N_P$ for all $\overline{X}\in B(O,\delta)$. \qed
\end{lemma}

We can now construct a lightlike path from $A$ to $B$. We do this by building a chain of `outgoing' and `ingoing' future pointing null geodesics along $\gamma$, constructed explicitly in MNCs, and staying within $B(O,\delta)$ at each point. 

So let $\overline{n}\in\mathbb{R}^4$ with $n^0>0$, $\eta_{ij}n^in^j=0$ and $|\overline{n}|=1$ (e.g.\ $\overline{n}=(1,1,0,0)$). Consider the path in $\cm$ whose Minkowski normal coordinates at $P$ are given by 
\be X^i = u n^i,\quad 0\leq u\leq \delta.\label{out-path}\ee
This is a future pointing null geodesic from $P$ to $Q_1$, where $X^i|_{Q_1}=\delta n^i$. The point $Q_2$ with MNCs 
\be X^i|_{Q_2}=\frac12\delta n^i \label{mnc-q2} \ee
lies on this geodesic. Take $Q_3$ to be the point with MNCs given by 
\be X^i|_{Q_3}=\delta(1,0,0,0)\in\gamma, \label{mnc-q3} \ee and consider the path in $\cm$ with MNCs at $P$ given by 
\be X^i(t) = t X^i|_{Q_3} +(1-t)X^i|_{Q_2},\quad t\in[0,1]. \label{in-path} \ee
This null path must be a null geodesic (\textit{cf}.\ Proposition 2.20 of \cite{penrose1972techniques}; there is no timelike trip from $Q_2$ to $Q_3$). Then the path $PQ_2\cup Q_2Q_3$ is a lightlike path from $P$ to $Q_3$. 

This path exhausts a \textit{finite} portion of the timelike geodesic $\gamma$: we have $Q_3=\gamma(\epsilon)$ for some $\epsilon$ that is bounded away from 0. Therefore a finite number of paths constructed in this way yields a path from $A$ to $B$. We note that the last leg of this path must be adjusted (by choosing the point corresponding to $Q_2$ above appropriately) to ensure that the point corresponding to the $Q_3$ point on this final segment is indeed $B$.
\qed

\noindent\textbf{Proof of Lemma \ref{lem:rdot-neg-useless}:}

Let $\gamma_b$ be the complete null geodesic with the same values of $\kappa,\lambda$ as $\gamma_a$, with $\gamma_b(0)=P$ (so that $r_b(0)=r_a(0), \tau_b(0)=\tau_a(0)$), but with $\dot{r}_b(0)=-\dot{r}_a(0)>0$. Referring to Figure 1, $\gamma_a(0)=P$ must lie on a segment equivalent to $\gamma_{(Y,A)}$.

Consider first the case where $\gamma_a(0)=P\in\gamma_{(Y,W_+]}$. Let $s_-<0$ be the greatest negative value of $s$ for which $r_a(s_-)=r_a(0)$ (in Figure \ref{fig:lem8fig}, $\gamma(s_-)$ is the point on $\gamma_{[W_-,Y)}$ lying vertically above $P$). Then we claim that the unique solution of the geodesic equations for $\gamma_b$ for $r,\tau$ is given by 
\begin{eqnarray}
\tau_b(s)&=&\tau_a(s+s_-)+\tau_a(0)-\tau_a(s_-),\label{tau1solminus} \\
r_b(s) &=& r_a(s+s_-). \label{r1sol}
\end{eqnarray}
So define $r_b(s)=r_a(s+s_-)$. Then we easily find that $r_b(s)$ satisfies (\ref{r-dot}) and that $r_b(0)=r_a(s_-)=r_a(0)$, by the definition of $s_-$. From (\ref{r-dot}), we also have $\dot{r}_b^2(0)=\dot{r}_a^2(s_-)$. Since $\dot{r}_a(s_-)$ must have the opposite sign to that of $\dot{r}_a(0)$, this yields $\dot{r}_b(0)=-\dot{r}_a(0)$. This shows that $r_b(s)=r_a(s+s_-)$ satisfies the differential equation and initial conditions for $r_b$ along $\gamma_b$. It is then straightforward to show that the same holds for $\tau_b$, and so the claim is proven. 

Now let $Q=\gamma_a(s), s>0$ be any point on $\gamma_a$ to the future of $P$, and take $Q'=\gamma_b(s-s_-)$, so that $Q'$ lies to the future of $P$ on $\gamma_b$ (recall that $s_-<0$). Then by construction, $r|_{Q'}=r|_{Q}$, and 
\begin{eqnarray}
\tau|_{Q'}&=&\tau_b(s-s_-) \nonumber \\
&=&\tau_a(s)+\tau_a(0)-\tau_a(s_-) \nonumber \\
&=&\tau|_{Q} + \tau_a(0)-\tau_a(s_-) < \tau|_{Q},
\end{eqnarray} 
the final inequality following from the definition of $s_-$. This completes the proof for this case. 

Next, we consider the case where $\gamma_a(0)=P\in\gamma_{(W_+,A)}$. In this case, we define $s_+$ to be the least positive value of $s$ for which $r_a(s_+)=r_a(0)$ (in Figure \ref{fig:lem8fig}, $\gamma(s_+)$ is the point on $\gamma_{(A,E_-)}$ lying vertically above $P$). As above, we can show that the solutions for $\tau_b,r_b$ of the geodesic equations for $\gamma_b$ are given by 
\begin{eqnarray}
\tau_b(s)&=&\tau_a(s+s_+)+\tau_a(0)-\tau_a(s_+),\label{tau1solplus} \\
r_b(s) &=& r_a(s+s_+). \label{r1solplus}
\end{eqnarray}
Now consider a point $Q=\gamma_a(s), s>0$ to the future of $P$ on $\gamma_a$. For $s>s_+$, we take $Q'=\gamma_b(s-s_+)$. As above, this fulfills the conditions of the lemma (note that we require $s>s_+$ to ensure that $Q'$ lies to the future of $P$ on $\gamma_b$). For $0<s\leq s_+$, we define $s_2$ to be the first positive value of $s$ at which $\gamma_a$ meets the maximum value of $r$, so that $r_a(s)$ is increasing on the interval $(0,s_2)$. There are two cases to consider. 

Case (i): If $s_2\geq s_+$, we define $Q'=\gamma_b(s)$. Then $r|_{Q'}=r_b(s)=r_a(s+s_+)>r(s), 0<s\leq s_+$. The inequality follows from the fact that $r_a(s)\leq r_a(0)=r_a(s_+)$ for $0<s<s_+$, whereas $r_b(s)$ is increasing on $(0,s_+)\subset (0,s_2)$, and has initial value $r_b(0)=r_a(0)$. Next, we note that from (\ref{tau-dot}) we have 
\be \dot{\tau}=f(r),\quad f(r)=(1+r^2)^{-1}((1-r^2)\kappa+\sqrt{2}), \label{tau-dot-again} \ee
so that $f(r)$ is a decreasing function of $r$. Since $r_b(s)>r_a(s)$, this gives $\dot{\tau}_b(s)<\dot{\tau}_a, 0<s\leq s_+$. Since $\tau_b(0)=\tau_a(0)$, this yields 
\be \tau|_{Q'}=\tau_b(s)<\tau_a(s)=\tau|_Q,\quad 0<s\leq s_+, \label{tau-q1-q-case1} \ee
completing the proof of the lemma in this case. 

Case (ii): If $s_2<s_+$, then $r_b(s)$ reaches its maximum before $r_a(s)$ reaches $r_a(s_+)=r_a(0)$. Then $r_a(s)$ first reaches its maximum on $\gamma$ at $s=s_3=s_2+s_+$ ($\gamma(s_3)=C$ in Figure \ref{fig:lem8fig}). Then we define
\be
Q' = \left\{ \begin{array}{ll}
    \gamma_b(s),  & 0<s<s_2;  \\
    \gamma_b(s_2), & s_2\leq s \leq s_3; \\
    \gamma_b(s), & s_3< s. 
    \end{array}\right. 
    \label{Q1-case2-def} 
\ee

Repeating the proof of Case (i) verifies that $\tau|_{Q'}<\tau|_Q$, $r|_{Q'}>r|_Q$ while $s<s_2$. On $\gamma_a$, $\tau_a$ is initially increasing from $\gamma_a(s_2)$ to $E_-$, and $\tau_a(s)\geq\tau|_{E_-}=\tau|_C$ (see (\ref{tauE-tauC})). Thus $\tau_a(s)>\tau_a(s_2)\geq \tau_b(s_2)$ for $s\in[s_2,s_3]$. $r_a(s)\leq r_b(s_2)$ for $s\in[s_2,s_3]$ is immediate since $r_b(s_2)$ is the global maximum of $r$ on both $\gamma_a$ and $\gamma_b$. Repeating once again the proof of Case (i) verifies that $\tau|_{Q'}<\tau|_Q$ for $s>s_3$, and $r|_{Q'}=r|_Q$ for $s>s_3$ by construction. This completes the proof. \qed

\noindent\textbf{Proof of Lemma \ref{lem:mono-useless-1}:} 

We take $\gamma_a$ and $\gamma_b$ be as defined in the statement of the lemma. (Note that the third part of Lemma \ref{lem:ri-l3-dep} guarantees existence of $\gamma_b$, and the assumed data yield uniqueness.) With this choice and using (\ref{r-d}), 
which yields 
\be \dot{r}_b^2(0) > \dot{r}_a^2 (0),\quad s\geq 0, \label{rdotsq-ab-all}
\ee
we see that 
\be \dot{r}_b(0) > \dot{r}_a(0),\label{rdot-ab-init} \ee
and so there exists $s_*>0$ such that 
\be r_b(s) > r_a (s),\quad 0<s<s_*. \label{r-ab-interval} \ee
It follows from (\ref{tau-dot}) that 
\be \dot{\tau}_b(s) < \dot{\tau}_a(s),\quad 0<s<s_*, \label{tdot-ab-interval} \ee
and so, since $\tau_b(0)=\tau_a(0)$,
\be {\tau}_b(s) < {\tau}_a(s),\quad 0<s<s_*.
\label{tau-ab-interval}
\ee

Next, we establish that $r_b$ reaches its maximum value $r_2(\kappa,0)$ along $\gamma_b$ before any subsequent crossing of the geodesics $\gamma_a$ and $\gamma_b$. So suppose that a crossing occurs before $r_b$ has reached its maximum. Then $\dot{r}_b>0$ at the point of intersection. Considering the relevant graphs in the $s-r$ plane, $r_b(s)$ is initially above $r_a(s)$.  It follows that the graph of the increasing function $s\to r_b(s)$ meets the graph of the function $s\to r_a(s)$ from above at the point of intersection (at $s=s_1$ say), which leads to 
\be 0 < \dot{r}_b(s_1) < \dot{r}_a(s_1), \label{rdot-inter} \ee
giving 
\be 0 < \dot{r}_b^2(s_1) < \dot{r}_a^2(s_1). \label{rdotsq-inter} \ee
But (\ref{r-d}) shows that we must have the opposite inequality at the point of intersection, yielding a contradiction. 

It follows that the inequalities (\ref{r-ab-interval}) and (\ref{tau-ab-interval}) hold in the case where $s_*>0$ is the parameter value corresponding to the first occurrence on $\gamma_b$ of the global maximum $r_2(\kappa,0)$ of $r_b$. Fix $s_*$ to be this value.

To complete the proof, let $Q=\gamma_a(s), s>0$. The geodesic $\gamma_b$ provides the future pointing null geodesic whose existence is claimed in the statement of the lemma: it remains to identify a suitable point $Q'$. If $\tau_b$ is initially increasing, say on $s\in(0,s_3)$ (where we must have $s_3<s_*$ by virtue of part (iii)-(b) of Proposition \ref{prop:key-ngs}), then $Q'=\gamma_b(s_{**})$ fulfills the requirements of the lemma. 

Now consider $s\geq s_3$. As $\tau_b$ is decreasing on $(s_3,s_*)$, we have
\be \tau_b(s_*) < \tau_a(s),\quad s\geq s_3. \ee
Since the maximum of $r$ on $\gamma_a$ is $r_2(\kappa,\lambda)<r_2(\kappa,0)$, we have $r_a(s) < r_2(\kappa,0)=r_b(s_*)$ for all $s\geq 0$. It follows that the point $Q'=\gamma_b(s_*)$ fulfills the requirements of the theorem in this case. 

If $\tau_b$ is not initially increasing, then the initial point $P$ lies on a segment of the form $\gamma_{[B,C]}$ in Figure \ref{fig:lem8fig} (where it should be understood that $\gamma=\gamma_b$ here). In this case, we take $Q'=\gamma_b(s_*)$ and note that $r|_{Q'}$ is the maximum of $r$ on $\gamma_b$, which exceeds the maximum of $r$ on $\gamma_a$, and that $\tau|_{Q'}<\tau|_P$, which is less that the value of $\tau_a(s)$ for all $s>0$, given monotonicity of $\tau$ on $\gamma_a$. This completes the proof. \qed

\noindent\textbf{Proof of Lemma \ref{lem:X}:} 

We use the labelling of  Figure \ref{fig:lem8fig} throughout the proof. Three cases arise. 

\begin{figure}
    \centering
    \includegraphics{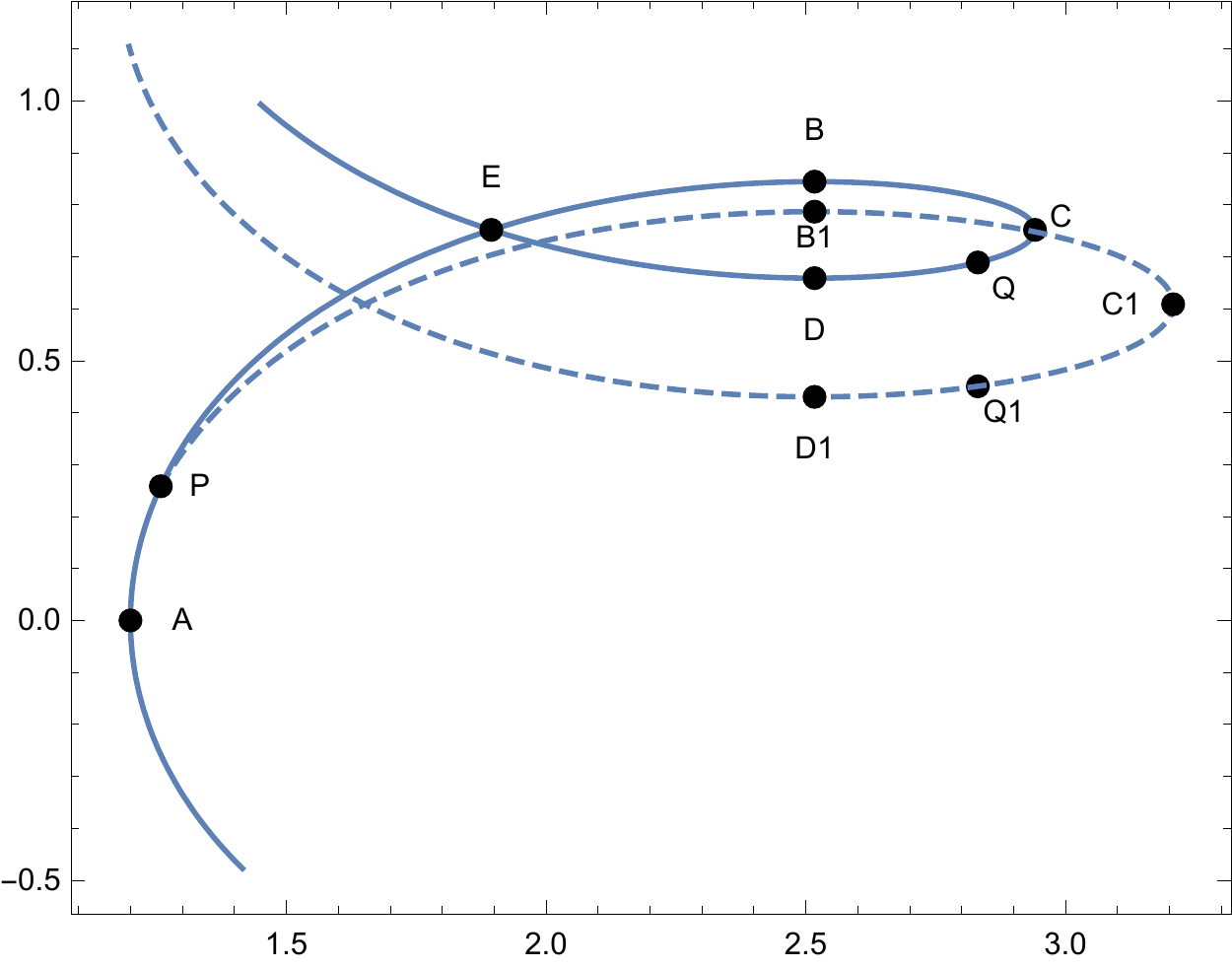}
    \caption{Relevant portions of $\gamma_a$ (solid; $\lambda>0$) and $\gamma_b$ (dashed; $\lambda=0$) for Case (i) of the proof of Lemma \ref{lem:X}, along with some of the critical points of each. $P$ is the initial point of both geodesics ($\gamma_a(0)=\gamma_b(0)=P$). The points $Q,Q_1$ correspond to the points introduced in the fourth bullet point of the proof of Case (i).} 
    \label{fig:lemXfig}
\end{figure}

\noindent\textit{Case (i):} $\dot{\tau_a}(0)>0$. Then we can consider that $P\in\gamma_{[A,B)}$ - with the understanding that $\gamma=\gamma_a$. From (\ref{r-dot}) and the properties of $\gamma_a$ and $\gamma_b$, we have 
\be 0\leq \dot{r}_a(0) < \dot{r}_b(0) \label{X-rdot-init}\ee
and 
\be \dot{\tau}_a(0)=\dot{\tau}_b(0)>0. \label{X-tdot-init}\ee
Thus (considering the projections of the geodesics into the $r-\tau$ plane), $\gamma_b$ crosses $\gamma_a$ from above at $P$, and (initially) site below $\gamma_a$. That is, there exists $s_*>0$ such that 
\be r_a(s) < r_b(s),\quad 0<s<s_*. \label{X-r-below} \ee
From (\ref{tau-dot-again}), we have $\partial_rf<0$, which yields
\be \tau_a(s)>\tau_b(s),\quad 0<s<s_*. \label{X-t-below} \ee
Define $B_1, C_1$ and $D_1$ to be the points on $\gamma_b$ corresponding to the points $B, C, D$ on $\gamma_a$ (see Figure \ref{fig:lemXfig}). Since both $r$ and $\tau$ are increasing on $\gamma_{a(A,B)}$, we can repeat the argument above to conclude that $\gamma_{b(A,B_1)}$ sits below $\gamma_{a(A,B)}$, and so $\tau|_B>\tau|_{B_1}$. Note that $r(B)=r(B_1)=r(D)=r(D_1)$. By Lemma \ref{lem:back-c-shaped} on \backc segments, 
\be \tau(D_1)-\tau(B_1) < \tau(D) - \tau(B), \label{tau-compare-1} \ee
and so 
\be \tau(D_1) < \tau(D) \label{tau-compare-2} \ee
and
\be \tau(C_1)-\tau(B_1) < \tau(C) - \tau(B) \label{tau-compare-3} \ee
since (as shown in the proof of Lemma \ref{lem:back-c-shaped}), the left and right hand sides here equate to one half of the left and right hand sides of (\ref{tau-compare-1}). So 
\be \tau(C_1) < \tau(C)+\tau(B_1)-\tau(B) <\tau(C). \label{tau-compare-4}\ee
Applying Lemma \ref{lem:ri-l3-dep}, we also have
\be r(C_1) > r(C), \label{r-compare-1}\ee
and consequently, since $r(C)$ is the global maximum of $r$ on $\gamma_a$, 
\be r(C_1) > r_a(s),\quad s\geq0. \label{r-compare-2}\ee
Then:
\begin{itemize}
    \item If $Q\in\gamma_{a(P,C]}$, it is immediate that $\gamma_{b[P,C_1]}$ is better than $\gamma_{a[P,Q]}$.
    \item If $Q\in\gamma_{a[D,E_+)}$, then $\tau(Q)\geq\tau(D)$ by an earlier lemma, and $r(Q)\leq r(D)$. In this case, $\gamma_{b[P,D_1]}$ is better than $\gamma_{a[P,Q]}$.
    \item If $Q\in \gamma_{a[E_+,+\infty)}$, then $\tau(Q)>\tau(E_+)=\tau(C)>\tau(C_1)$ and so $\gamma_{b[P,C_1]}$ is better than $\gamma_{a[P,Q]}$.
    \item If $Q\in\gamma_{a[C,D]}$, we choose $Q_1$ to be the point on $\gamma_{b[C_1,D_1]}$ that sits vertically below $Q$ in the $r-\tau$ plane (see Figure \ref{fig:lemXfig}). Then $\tau(Q_1)<\tau(Q)$ and $r(Q_1)=r(Q)$, so that $\gamma_{b[P,Q_1]}$ is marginally better than $\gamma_{a[P,Q]}$. 
\end{itemize}
This completes the proof for Case (i). 

\noindent\textit{Case (ii):}$\dot{\tau}_a(0)< 0$. In this case, we have $P\in\gamma_{[B,C]}$. As above, $0\leq\dot{r}_a(0)<\dot{r}_b(0)$, but $\dot{\tau}_a(0)=\dot{\tau}_b(0)<0$. Then 
\be \left.\frac{d\tau}{dr}\right|_{\gamma_a(0)}=\frac{\dot{\tau}_a(0)}{\dot{r}_a(0)}<\frac{\dot{\tau}_b(0)}{\dot{r}_b(0)}=\left.\frac{d\tau}{dr}\right|_{\gamma_b(0)}<0. \label{X-slopes} \ee
Thus $\gamma_b$ crosses $\gamma_a$ from below in the $r-\tau$ plane at $P$. Defining $B_1, C_1, D_1$ as in Case (i), it follows that 
\be \tau(B_1)<\tau(B),\quad \tau(C_1)<\tau(C),\quad \tau(D_1)<\tau(D). \label{X-tau-compare} \ee
Then: 
\begin{itemize}
    \item If $Q\in\gamma_{a(P,C)}$, then $\gamma_{b[P,C_1]}$ is better than $\gamma_{a[P,Q]}$.
    \item If $Q\in\gamma_{a[D,E]}$, then $\gamma_{b[P,D_1]}$ is better than $\gamma_{a[P,Q]}$.
    \item If $Q\in\gamma_{a(E,+\infty)}$, then $\gamma_{b[P,C_1]}$ is better than $\gamma_{a[P,Q]}$.
    \item If $Q\in\gamma_{a[C,D]}$, then take $Q_1$ as in the proof of Case (i). It follows that $\gamma_{b[P,Q_1]}$ is better than $\gamma_{a[P,Q]}$.
\end{itemize}

\noindent\textit{Case (iii):} Here, $\dot{\tau}_a(0)=\dot{\tau}_b(0)=0$, and so $P=B$. From (\ref{tau-dot}), we have $\ddot{\tau}=\partial_rf\dot{r}$. Then using (\ref{tau-dot-again}), (\ref{r-dot}) and (\ref{l3-bound}), we can show that as in Case (ii), $\gamma_b$ crosses $\gamma_a$ from below in the $r-\tau$ plane at $P$. The proof of Case (ii) then carries over. \qed

\section{Parameters for an optimal path.}\label{app:params}
As shown in Lemma \ref{lem:axis-to-hor}, the first segment of the optimal path extends from $\co:(r,\tau)=(0,0)$ to $A_0:(r,\tau)=(1,\tau_*)$. Recall that $\tau_*=\pi(\sqrt{2}-1)/2\simeq 0.65065$.  
This segment corresponds to a solution of (\ref{tau-dot})-(\ref{r-dot}) with $L_2=L_3=\epsilon=0$, and (without loss of generality) $L_1=1$.

The table below describes the parameters used for the construction of the next six segments of the optimal path as shown in the right hand image of Figure \ref{fig:CTC} and in panel (h) of Figure \ref{fig:construction}. The second and third segments terminate on $\env_1$ and $\env_3$ respectively, and so the range of the affine parameter is $s\in[0,\frac{\pi}{2\kappa}]$ on each: see (\ref{env:s-sol}). We reset $s$ to zero at the initial point of each segment. On the fourth segment, we solve numerically to find that the segment meets $\Sigma_0$ at $s\simeq4.39302$. The final segment extends from $A_7:(r,\tau)=(1,-\tau_*)$ to $\co$, with affine parameter $s\in[0,\frac{\pi}{2}]$.

\begin{table}
 \begin{tabular}{||c | c | c | c||} 
 \hline
 Segment & Initial value of $(r,\tau)$ & Final value of $(r,\tau)$ &  $\kappa$  \\ [0.5ex] 
 \hline\hline
 1 & $(0,0)$ & $(1,\tau_*)$ & NA \\ 
 \hline
 2 & $(1,\tau_*)$ & $(\sqrt{2},0.56081)$ & $\sqrt{2}$ \\
 \hline
 3 & $(\sqrt{2},0.56081)$ & (2.01427, 0.31845) & 0.55994 \\
 \hline
 4 & (2.01427, 0.31845) & (2.30928,0) & 0.34856 \\
 \hline
 5 & (2.30928,0) & (2.01427, -0.31845) & 0.34856 \\ 
 \hline 
 6 & (2.01427, -0.31845) & $(\sqrt{2},-0.56081)$ & 0.55994 \\
 \hline
 7 & $(\sqrt{2},-0.56081)$ & $(1,-\tau_*)$ & $\sqrt{2}$ \\
 \hline 
 8 & $(1,-\tau_*)$ & (0,0) & NA \\ [0.25ex]
 \hline 
\end{tabular}
\caption{\label{table:params}Parameters for an optimal closed lightlike path.}
\end{table}

\section*{References}
\bibliographystyle{unsrtnat}
\bibliography{mybib}

\end{document}